\documentclass[fleqn,usenatbib]{mnras}

\usepackage{mathptmx}

\usepackage[T1]{fontenc}

\DeclareRobustCommand{\VAN}[3]{#2}
\let\VANthebibliography\thebibliography
\def\thebibliography{\DeclareRobustCommand{\VAN}[3]{##3}\VANthebibliography}

\usepackage{graphicx}	
\usepackage{amsmath}	
\usepackage{amssymb}	
\usepackage{url}
\usepackage{adjustbox}
\usepackage{caption}
\usepackage{subcaption}
\usepackage{mathtools}
\usepackage{amsfonts}
\usepackage{todonotes}
\usepackage[capitalize]{cleveref}
\usepackage{multirow}
\usepackage{array}
\usepackage{rotating}
\usepackage{hyperref}
\usepackage{ulem}
\newcolumntype{H}{>{\setbox0=\hbox\bgroup}c<{\egroup}@{}}

\title[Study of photochemical hazes in 10 hot-Jupiter atmospheres]{A large range of haziness conditions in hot-Jupiter atmospheres}

\author[A. Arfaux et al.]{
Anthony Arfaux,$^{1}$\thanks{E-mail: anthony.arfaux@univ-reims.fr}
and Panayotis Lavvas,$^{1}$
\\
$^{1}$Groupe de spectrom\'erie mol\'eculaire et atmosph\'erique, Universit\'e de Reims Champagne Ardenne, Reims, France\\
}

\date{Accepted 2022 June 21. Received 2022 June 9; in original form 2022 April 14}

\pubyear{2021}

\begin{document}
\label{firstpage}
\pagerange{\pageref{firstpage}--\pageref{lastpage}}
\maketitle

\begin{abstract}

We present a study of photochemical hazes of exoplanet atmospheres based on a self-consistent model including haze microphysics, disequilibrium chemistry, and radiative feedbacks. 
We derive the haze properties required to match HST observations of ten hot-Jupiters. 
HAT-P-12b, HD-189733b, HD-209458b and WASP-6b require haze mass fluxes between 5x10$^{-15}$ and 9x10$^{-12} g.cm^{-2}.s^{-1}$ to match the observations. 
WASP-12b and WASP-19b with equilibrium temperatures above 2000 K are incompatible with the presence of haze and are better fitted by heavy metals. 
HAT-P-1b and WASP-31b do not show clear evidence for the presence of hazes with upper mass fluxes of 10$^{-15}$ and 10$^{-16}g.cm^{-2}.s^{-1}$, respectively, while WASP-17b and WASP-39b present an upper mass flux limit of 10$^{-16} g.cm^{-2}.s^{-1}$.
We discuss the implications of the self-consistent model and we derive upper limits for the haze abundances based on photochemistry results. 
Our results suggest HCN as the main haze precursor up to 1300 K effective temperatures and CO above. 
Our derived haze mass fluxes based on the fit to the observations are consistent with the photochemistry with formation yields up to $\sim$6.4\%. 
Disequilibrium chemistry has negligible impact on the spectra considering the low resolution observations used but impacts the chemical composition and temperature profiles. 
We find that hazes produce hotter upper atmosphere temperatures with a detectable impact on the spectra. 
Clouds may have implications for interpreting the transit spectra of HD-209458b, WASP-31b and WASP-39b. 
Nevertheless, the presence of silicate and iron clouds is expected in all studied atmospheres except WASP-12b and WASP-19b. 
\end{abstract}




\section{INTRODUCTION}

\begin{table*}
\caption{Physical properties of the studied exoplanets. Equilibrium temperatures are computed assuming a zero albedo. The planets radius, mass, surface gravity and star distance are taken from \protect\cite{Sing16}.}
\begin{tabular}{c|ccccccc}
\hline
Planet 		& Radius 	& Mass 		& Surface gravity 	& Equilibrium 		& Semi-major 		& Host star 	& Scale heigh	\\
			& ($R_J$) & ($M_J$) 	& ($ m.s^{-2} $) 	& Temperature (K) 	& axis (A.U.) 		& stellar type 	& (km)		\\
\hline
HAT-P1b 		& 1.32	& 0.53 		& 7.5 			& 1,320 			& 0.05561 		&  G0V		&  636.2	 	\\
HAT-P-12b 	& 0.96 	& 0.21 		& 5.6 			& 960 			& 0.0384 			& K5			&  619.7		\\
HD-189733b 	& 1.14	& 1.14 		& 21.4 			& 1,200 			& 0.03142 		& K1-K2		&  202.7		\\
HD-209458b 	& 1.36 	& 0.69 		& 9.4 			& 1,450 			& 0.04747 		& G0V		&  557.6		\\
WASP-6b 		& 1.22 	& 0.50 		& 8.7 			& 1,150 			& 0.0421 			& G8V		& 477.8		\\
WASP-12b 	& 1.73 	& 1.40 		& 11.6 			& 2,510 			& 0.0234 			& G0V		& 782.2		\\
WASP-17b 	& 1.89 	& 0.51 		& 3.6 			& 1,740 			& 0.0515  			& F4			& 1747.1		\\
WASP-19b 	& 1.41 	& 1.14 		& 14.2 			& 2,050 			& 0.01616 		& G8V		& 521.9		\\
WASP-31b 	& 1.55 	& 0.48 		& 4.6 			& 1,580 			& 0.04657 		& F			& 1241.6		\\
WASP-39b 	& 1.27 	& 0.28 		& 4.1 			& 1,120 			& 0.0486 			& G8			& 987.5		\\
\hline
\end{tabular}
\label{Tab:properties}
\end{table*}

The comprehension of exoplanet atmospheres composition might give precious clues in the understanding of their formation and therefore on the mechanisms underlying the evolution of stellar systems.
The study of exoplanets is a complicated task as they are small and faint objects hardly observable and too far away for \it in situ \rm exploration.
Currently, the most common way for probing their atmosphere uses the transit phenomenon.
Indeed, as the planet passes in front of its host star, crossing the line of sight between the star and the observer, a part of the starlight is occulted, leading to a variation in the measured transit depth.
According to the atmospheric composition, such variation is dependent on the wavelength of observation.
The resulting transit spectrum may therefore present clues about the composition of the transiting planet.
The probability of such phenomenon increases as the planet to star distance decreases \citep{Brown01}, therefore, exoplanets in close-in orbit are the most likely to present a transit.
On the other hand, the larger the planet, the more important the occultation, thus providing an enhanced detection probability.
Hot-Jupiters being large planets with small semi-major axes, they are the best suited for such an investigation.

Multiple studies about hot-Jupiter atmospheres have been released \citep[{e.g. }][]{Charbonneau02,Pont08,Sing16,Barstow17}, revealing interesting features within their transit spectra, especially a UV decreasing slope whose origin has been the subject of multiple investigations.
\cite{Lecavelier08} point out the role of H$_2$ Rayleigh scattering to explain such a slope in HD-209458b transit spectrum.
Indeed, Rayleigh scattering is known to present a $1/\lambda^4$ behavior leading to a similar feature.
However, this single phenomenon, although sufficient for some cases, does not provide a match for others \citep{Pont08,Sing16}.

Another hypothesis explaining the presence of this UV slope are photochemical hazes, that is particules suspended in the atmosphere created by photochemistry.
The hypothesis of absorption by haze has yet been suggested for some of the planet we are investigating \citep{Pont08,Pont13,Sing09,Sing11,Sing13,Sing15,Nikolov14,Nikolov15,Wong20}.
\cite{Sing16,Barstow17} observed the presence of haze in the atmosphere of ten hot-Jupiters based on Hubble Space Telescope (HST) and Spitzer Space Telescope (SST) observations and they derived \it ad hoc \rm values of the particle density to match the measured spectra using a $\lambda^{-4}$ parameterization for photochemical haze, implicitly assuming non-absorbing, small particles.

The use of haze microphysics models \citep{Lavvas17,Lavvas19,Ohno18,Ohno20,Kawashima19,Adams19,Gao20} brought additional insight in the comprehension of haze formation and properties allowing to describe the haze particle size distribution along with their number density.
The average particle size for HD-189733b was found to be quite small around 1 to 2 nm \citep{Lavvas17}, while \cite{Lavvas21} demonstrate the impact of haze radiative feedback on the temperature profile and the consequent enhancement of the UV slope.

Although the possibility of silicate clouds to explain the observed UV slope has been rejected for the specific case of HD-189733b \citep{Lee15,Pinhas17}, clouds remain a viable option for other planets.
Also, other types of clouds can lie in hot gaseous giant atmospheres and may lead to additional opacities that can match the observations in the UV region \cite[][who proposed sulphide clouds to be a likely solution]{Wakeford15,Pinhas17}.
Therefore, although the study of clouds is out of the scope of this investigation, we keep in mind their possible implication in the spectra.

Another possibility for explaining the observed UV slope arises from heavy atom absorption.
The studies of \cite{Barman07,Lavvas14,Salz19} and \cite{Lothringer20} show the possible influence of metal absorption on this UV slope.
The overlapping features of metals in the UV are likely to result in a similar behavior.
However, the heavy metal elements can be depleted from the upper atmosphere due to condensation thus have negligible impact on the spectra with decreasing equilibrium temperatures \citep{Lothringer20}. Therefore, they can not explain the spectral slope observed for most of the studied planets.

Finally, stellar heterogeneities might let prints in planet transit spectra as found by \cite{Pinhas18}, who derived both star and planet properties through a coupled retrieval model and found evidences for inhomogeneities for WASP-6 and WASP-39.
However, this explanation does not apply for every planet.

Here, we focus on the photochemical haze hypothesis and extend the study of \cite{Lavvas17,Lavvas21} to a larger set of hot-Jupiters based on the target list from \cite{Sing16}.
Our goal is to evaluate if these planets can be fitted by photochemical hazes and what are the required haze properties.
We use a self-consistent model that couples disequilibrium chemistry with the haze microphysics and the radiative transfer in the atmosphere in 1D.
Photochemical mass fluxes for haze precursors are derived in a disequilibrium chemistry model providing additional constraints on the haze amount possibly available, while the haze radiative feedback and the disequilibrium composition are accounted in the temperature profiles.
These calculations aim to assess if photochemistry can actually support the amount of haze required for matching the observations and if the haze radiative feedback has a positive influence on the fit of the observations.

3D effects have been suggested to have an important impact on the transit spectra.
\cite{Caldas19} found that the temperature and composition gradient around the terminators could drastically modify the transit spectra.
However, this effect is limited and mostly effective for very hot exoplanets as shown by \cite{Komacek16,Komacek17,Pluriel21} that find the dayside-nightside temperature difference to decrease with decreasing planet effective temperature.
\cite{Steinrueck21} did not demonstrate important modifications between morning and evening terminators in the spectrum of HD-189733b.
Thus, we do not include these effects through our 1D simulation but we keep in mind their possible impact for the very hot exoplanets.

In \cref{Sec:Model}, we present the planet and star properties as well as the details of our approach and of the microphysics, disequilibrium chemistry and radiative-convective models used.
In \cref{Sec:GenResults} we present the general results detailing the phenomena impacting our study before to further describe the planet results individually in  \cref{Sec:PlanResults}.
A discussion of the model results is presented in \cref{Sec:Discussion}.

\section{METHODS}
\label{Sec:Model}

Here we provide an overview of the exoplanets studied and of the approach followed in our analysis.

\begin{table*}
\caption{Table of the observations used in the work.
Sing16 stands for \protect\cite{Sing16};
Wong20 for \protect\cite{Wong20};
Yan20 for \protect\cite{Yan20};
Carter20 for \protect\cite{Carter20};
Sedaghati16 for \protect\cite{Sedaghati16};
Espinoza19 for \protect\cite{Espinoza19};
Gibson17 for \protect\cite{Gibson17};
Gibson19 for \protect\cite{Gibson19};
Fischer16 for \protect\cite{Fischer16};
Nikolov16 for \protect\cite{Nikolov16};
Saba21 for \protect\cite{Saba21};
Sedaghati21 for \protect\cite{Sedaghati21}
.}
\begin{adjustbox}{width=\textwidth}
\begin{tabular}{c|c|c|c|c|c|c|c|c}
\hline
Planet 		& \multicolumn{2}{c|}{HST} 					& SST 			& \multicolumn{3}{c|}{VLT} 				& LBT 	& Magellan 	\\
 			& STIS 				& WFC3 				& IRAC 			& FORS2 		& ESPRESSO	& 	UVES	& MODS 	& IMACS 		\\
\hline
HAT-P-1b 		& Sing16 				& Sing16 				& Sing16 			& 			& 			&			&		& 			\\
HAT-P-12b 	& Wong20 / Alexoudi18 	& Wong20 / Alexoudi18 	& Sing16 			& 			& 			&			& Yan20 	&  			\\
HD-189733b 	& Sing16 				& Sing16 				& Sing16 			& 			&			&			& 		&  			\\
HD-209458b 	& Sing16 				& Sing16 				& Sing16 			&  			& 			&			&		&  			\\
WASP-6b 		& Sing16 / Carter20 	 	& Carter20 			& Sing16 / Carter20 	& 			&			&			& 		&   			\\
WASP-12b 	& Sing16 				& Sing16 				& Sing16 			& 			& 			&			&		&  			\\
WASP-17b 	& Sing16 / Saba21		& Sing16 / Saba21		& Sing16 			& Sedaghati16 & 			&			&		&  			\\
WASP-19b 	& Sing16 				& Sing16				& Sing16 			& 			& Sedaghati21	& 			&		& Espinoza19	\\
WASP-31b 	& Sing16 / Gibson17 	&  Sing16 / Gibson17	& Sing16 			&  Gibson17 	& 			& Gibson19 	&		&  			\\
WASP-39b 	& Sing16  / Fischer16	 & Sing16 / Fischer16	& Sing16 / Fischer16	& Nikolov16 	& 			&			& 		&   			\\
\hline
\end{tabular}
\end{adjustbox}
\label{Tab:Observations}
\end{table*}

\subsection{Exoplanets}
Hot-Jupiters are defined as gaseous giant exoplanets with mass and radius similar to Jupiter’s, orbiting close to their host star.
Although they all respect this basic definition, they span a large range of properties arising from different planetary (radius, mass, distance to host) and stellar (stellar type) properties.
The combined effect of these parameters results in different atmospheric properties and particularly a large range of atmospheric temperatures.

Our study focuses on the ten hot-Jupiters (\cref{Tab:properties}) reported by \cite{Sing16}.
The observed transit spectra for each planet were based on measurements through the Hubble Space Telescope’s (HST) Imaging Spectrograph (STIS) and Wide Field Camera 3 (WFC3), as well as, the Infrared Array Camera (IRAC) on board the Spitzer Space Telescope (SST).
As an overall picture, the observations reveal the presence of a varying degree slope in the UV-Visible range followed by the detection (in most but not all cases) of the Na and K lines in the visible and the signature of $\rm H_2O$ absorption in the 1.4 $\mu m$ band probed by WFC3.
The UV slope, the visibility of the Na and K wings, and the degree of muting of the $\rm H_2O$ band in the near IR are strong indicators for the contribution of clouds and/or hazes.
While Na and K are expected to behave similarly, since they both are alkali metals, and to be both present in most hot-Jupiters atmosphere, the observations reveal a more complex behavior with sometimes only Na or K detected.
As temperature decreases, the sodium condenses first, it is therefore possible to have the sodium cold trapped and the potassium remaining in the gas phase, though this requires quite precise pressure/temperature conditions.
Since potassium is less abundant than sodium (log [Na/H] = -5.7 against log [K/H] = -6.9 for a solar metallicity), its signature in the transit spectrum can be hidden, by haze opacity for instance, letting the sodium feature only to appear in the transit spectrum.
However, this would require very precise metallicity conditions and would be followed by a strong weakening of the Na feature.
An explanation would arise from major differences in their elemental abundances, though these are not expected \citep{Lavvas14}.
For very hot exoplanets, the detection of other atomic and molecular lines in the visible provide further constraints for the atmospheric temperature and composition \citep{Yan22}.
The Spitzer observations provide additional constraints, albeit of lower precision, at longer wavelengths.
For some exoplanets, however, Spitzer data are not consistent with WFC3 observations and, therefore, are not accounted in our $\chi^2$ calculation.
These inconsistencies may arise from the use of different instruments or observation periods.
The main features detected in the \cite{Sing16} survey, as well as subsequent space-borne analyses/observations along with additional ground-based observations are discussed hereafter for each planet case ordered by decreasing stellar type (see \cref{Tab:Observations} for an overview of observations for each planet).

WASP-17b and WASP-31b orbit F spectral type stars, the hottest among this study, with effective temperatures above 6,000 K.
However, due to their relatively large semi-major axis (near 0.05 AU), they are not the hottest of this study, with $T_{eff}$ of 1,740 and 1,580 K, respectively.
The \cite{Sing16} analysis suggested the presence of Na in the atmosphere of WASP-17b, but the detection of K was not as clear.
Following ground-based observations with the FOcal Reducer and Spectrograph (FORS2) on the VLT in the visible implied the presence of the K wing in the transit spectrum \citep{Sedaghati16}, indicating a rather weak haze/cloud opacity contribution, which is consistent with the weak muting seen in the 1.4 $\mu m$ H$_2$O band based on the HST/WFC3 observations \citep{Sing16}.
\cite{Saba21} found potential implications for stellar activity and weak contribution of haze particles.
For WASP-31b the \cite{Sing16} survey revealed a Rayleigh slope towards the UV followed by a lack of Na but a strong signal of K and a significantly muted water band.
However, subsequent ground based observations with VLT \citep{Gibson17,Gibson19} suggested the lack of absorption at both alkali lines for this planet.

The hottest planet we are studying, WASP-12b, orbits a G0V star at a close distance of 0.0234 AU, leading to an equilibrium temperature of 2,510 K.
Currently, observations from \cite{Sing16} provide the best spectral coverage for the transit spectrum of this target, showing a rather flat spectrum with a highly muted H$_2$O band and no signatures of Na and K lines.
Such ultra-hot-Jupiters are not expected to contain photochemical haze \citep{Lavvas17} and the presence of atomic lines of heavy metals is expected to affect significantly their UV transit slope \citep{Lothringer20}.
Here we explore the validity of this anticipation.

HAT-P-1b and HD-209458b also orbit G0V type stars with larger semi-major axis (near 0.05 AU), leading to smaller equilibrium temperatures of 1,320 and 1,450 K, respectively.
Observations for HAT-P-1b are based on \cite{Sing16}, which are consistent with ground based observations \citep{Montalto15,Turner16,Todorov19} with a clear sodium detection but no or a weak potassium line and a strong water band.
Similarly, observations for HD209458b are based on the \cite{Sing16} analysis and presents both Na and K lines with a slightly muted water band.

WASP-19b, the second hottest planet of our study, orbits a G8-type star at a very short distance of 0.01616 AU, explaining the large equilibrium temperature of 2,050 K.
The WASP-19b space-borne observations reveal a rather flat spectrum, which is consistent with more recent \citep{Espinoza19} ground-based observations at visible wavelengths with the Inamori-Magellan Areal Camera and Spectrograph (IMACS).
Moreover the ground based observations revealed the major implications that could arise from the strong activity of this star due to the presence of occulted or unocculted stellar spots.
However, \cite{Sedaghati21} used high resolution spectroscopy to separate planet from stellar features and found that stellar activity alone can not explain the observed UV-slope.

WASP-6b and WASP-39b also orbit G8 type stars but with larger semi-major axes, thus leading to lower temperatures around 1,150 K.
The latest analysis of WASP-6b observations \citep{Carter20} combined HST and SST with ground-based Very Large Telescope (VLT) spectroscopic measurements, as well as, photometric measurements by the Transiting Exoplanet Survey Satellite (TESS), the latter allowing for the correction of stellar heterogeneity effects.
The results reveal the signature of Na, K and H$_2$O absorption features as well as the presence of haze, but also the sensitivity of the results on the degree of stellar variability considered.
Similarly for WASP-39b, \cite{Fischer16} reanalyzed the HST/STIS and SST/IRAC observations previously used in \cite{Sing16} and found similar results regarding the presence of Na and K and of a Rayleigh slope towards the UV, while photometric measurements suggested that the activity of WASP 39 was not significant to affect the transit spectrum.
\cite{Nikolov16} obtained new VLT observations for the visible spectrum of WASP-39b that provided consistent results with the previous studies.
Here we retain the \cite{Sing16} transit spectrum for our study since it includes also the HST/WFC observations for the 1.4$\mu m$ H$_2$O band from \cite{Wakeford18}.

HD-189733b orbiting a K-type star has an equilibrium temperature of 1,200 K.
This planet is a particular case due to its relatively small radius and large mass.
Indeed, while most of the planets we study present super-Jupiter radii (from 1.22 $R_J$ for WASP-6b to 1.89 $R_J$ for WASP-17b) with sub-Jupiter masses (from 0.28 $M_J$ for WASP-39b to 0.69 $M_J$ for HD-209458b), leading to small surface gravities (from 3.6 $ m.s^{-2} $ for WASP-17b to 9.4 $ m.s^{-2} $ for HD-209458b), HD-189733b has a relatively small radius of 1.14 $R_J$ and a large mass of 1.14 $M_J$ providing the largest surface gravity (21.4 $ m.s^{-2} $) within this study.
Although WASP-12b and WASP-19b also have super-Jupiter masses, their large radii (1.73 and 1.41 $R_J$, respectively) lead to much smaller surface gravities with 11.6 and 14.2 $ m.s^{-2} $, respectively.
Observations for HD-189733b are based on the \cite{Sing16} analysis and demonstrate the steepest slope among the studied exoplanets.
They found strong evidences for the presence of sodium but no potassium with a highly muted water band consistent with the presence of haze.
We note that the activity of HD 189733 has also been considered as a reason for the apparent strong transit depth at short wavelengths \citep{McCullough14}, which would therefore decrease the need for a high haze opacity in the atmosphere.
Nevertheless, \cite{Angerhausen15} conducted air-borne measurements with the Stratospheric Observatory For Infrared Astronomy (SOFIA) and found a steep UV slope in agreement with \cite{Sing16} observations though slightly offseted to lower transit depths due to lower stellar activity, thus suggesting that haze is present.

Finally, HAT-P-12b, in addition to being the smallest planet (0.96 $R_J$) and the lightest (0.21 $M_J$) within our study, it is also the coldest ({T$_{eq}$ = }960 K) due to its cold K5-type host star.
This target has received much attention due to the divergent conclusions derived for the presence of a UV slope from the analysis of space borne \citep{Sing16} or ground based \citep{Mallonn15} observations.
\cite{Alexoudi18} demonstrated that the orbital parameters assumed in the analysis have a strong impact on the presence of the slope and was able to reconcile the two type of measurements, demonstrating that the UV slope was a persistent feature.
A recent analysis by \cite{Wong20}, including additional HST observations of the muted water band, provides the latest transit spectrum we use in our analysis, while we also compare with the \cite{Yan20} transit spectrum based on the Multi-Object Double Spectrograph (MODS) of the Large Binocular Telescope (LBT).
These three datasets are in agreement about the absence of sodium while they disagree about the presence of potassium.
\cite{Yan20} found no clear evidences for the presence of K, while \cite{Alexoudi18} and \cite{Wong20} were able to detect the contribution of this element.

\subsection{Approach}

Our goal is to evaluate the potential contribution of photochemical hazes in the observed transit spectra of the above exoplanets.
We note that the contribution of clouds could also be significant for some of these planets.
Thus, our investigation under the assumption of photochemical hazes attempts to identify which among the hazy planets is best characterized by a photochemical haze component, and how the mass flux of this component varies among the different planets.
The mass flux is the postulated flux of mass that is transformed from the gas to the solid phase at a prescribed high altitude and that describes the nascent population of haze particles that evolve as they settle in the atmosphere.
This value is treated as a free parameter in our calculation and is limited by photochemistry results as discussed later on.
Our approach is divided in three steps of increasing complexity discussed hereafter.

\subsubsection{Step 1}

In the first step of our approach, we use a haze microphysics model \citep{Lavvas17} to evaluate the range of atmospheric and haze parameters required to fit the observed spectra.
The model solves the continuity equation for each particle size taking into account their transport, their growth through coagulation, as well as, their possible loss through sublimation.
The growth of aerosols, driven by their random collisions, is computed through a Brownian coagulation kernel, which is the main contribution for the expected (sub $\mu$m) particle size.
Therefore, increasing the temperature will enhance the particle collision probability and then will lead to larger particles.
Charge effects on the particle growth may also be taken into account.
Indeed, charges will impede particle coagulation through electrostatic repulsion, thus result in smaller particles.
However, the number of charges carried by a particle is proportional to its radius, therefore, considering the small particle radii expected in hot-Jupiter atmospheres at the regions probed by the transit (a few nm), we decided not to take this effect into account.
The model assumes a production profile in the upper atmosphere and follows the evolution of the particles down to the deep atmosphere (10$^3$ bar).
The particle production considers a gaussian profile nominally centered at the 1 $\mu bar$ level, above which most of the high-energy photons, responsible for the photochemistry, are deposited, with a profile width of 1/3$^{rd}$ of the scale height.
The integrated surface of this profile is defined by the haze mass flux.

\begin{figure}
\includegraphics[width=0.5\textwidth]{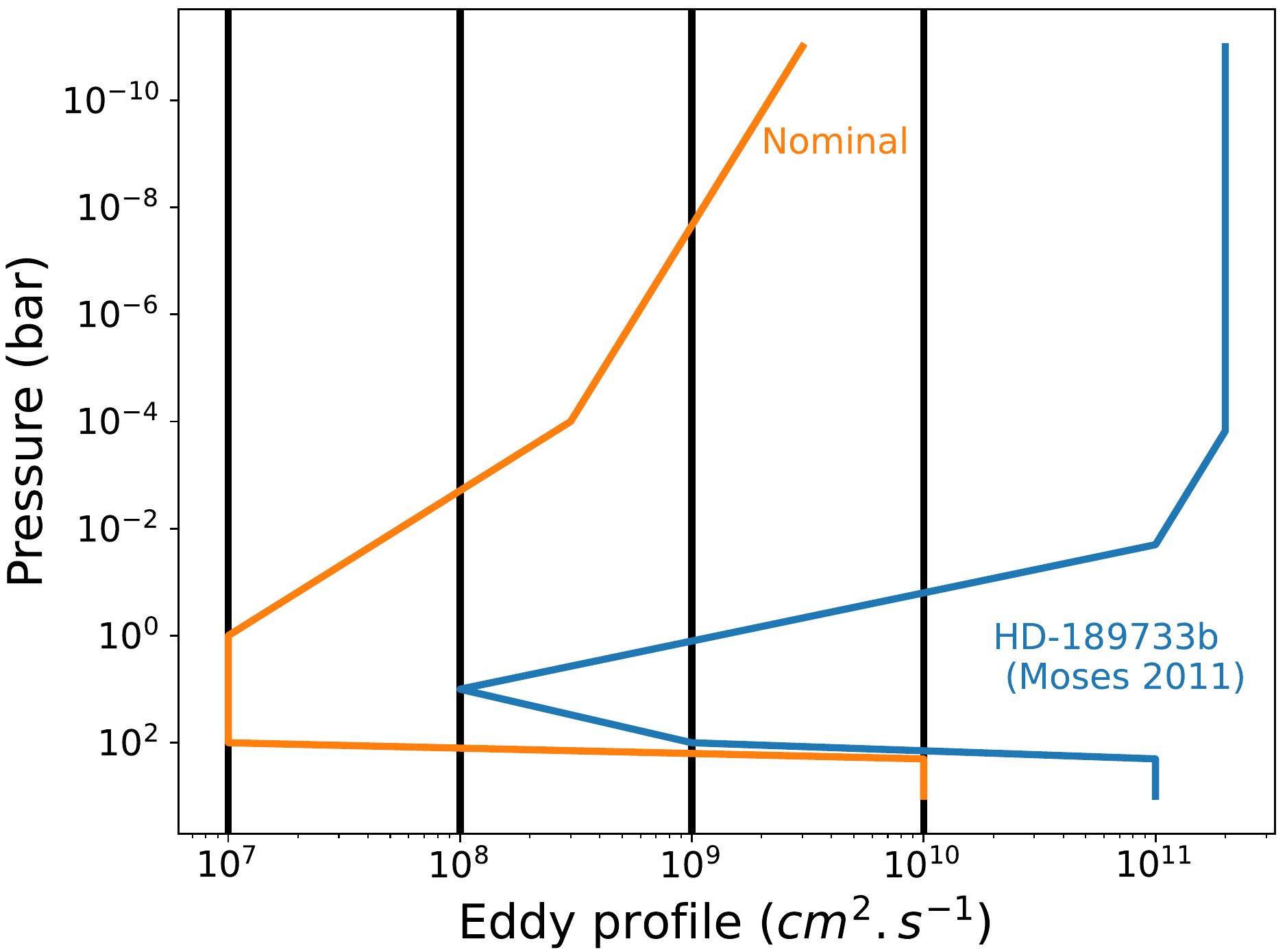}
\caption{Eddy mixing profiles used in this study. The nominal and Moses 2011 profiles come from GCMs calculations while the four black curves are simple trials, the 10$^{10}cm^2.s^{-1}$ being tried for HD-189733b only.}
\label{Fig:EddyProfiles}
\end{figure}

In this preliminary step we assume that the chemical composition follows a thermochemical equilibrium.
We evaluate the abundances of gases with the CEA model (Chemical Equilibrium and Applications) from NASA \citep{CEA}.
This model takes into account hundreds of species and condensation effects.
The atmospheric metallicity is considered to be solar with elemental abundances taken from \cite{Lodders10}, but we also tested sub-solar metallicities for some planets.
We use atmospheric temperature profiles from \cite{Sing16}, evaluated assuming equilibrium conditions and no haze, as input to the microphysics and thermochemical equilibrium models.
These initial models are used to generate theoretical transit spectra (see \cref{SSec:TransitSim}) which can be confronted to the observations to determine the range of haze mass fluxes and eddy mixing profiles that are consistent with the observations.

The haze mass flux is the main parameter of interest being a measurement of the amount of haze present in the atmospheres.
We first tested three possibilities for the haze mass flux with: 10$^{-12}$, 10$^{-13}$ and 10$^{-14}$ $g.cm^{-2}.s^{-1}$ as well as a haze-free model.
This range of values is based on photochemical computations for HD-189733b \citep{Lavvas17} and their applicability to other planets will be evaluated here.
For the planets best fitted by the haze-free theoretical model, additional mass flux cases were tested with values down to 10$^{-16}$ $g.cm^{-2}.s^{-1}$, while larger mass fluxes have been tried for HD-189733b.
We do not attempt to provide definitive values for the haze mass fluxes required, but to constrain the range of haziness that can be expected in hot-Jupiters based on the available observations.

For the eddy profiles we consider different constant profiles (\cref{Fig:EddyProfiles}) as well as a nominal profile based on GCM simulations. Evaluations based on the rms vertical velocity \citep{Moses11} were shown to overestimate the Kzz magnitude by up to two orders of magnitude compared to the tracer-distribution-based retrieval method \citep{Parmentier13}. Although the latter does not capture sub-scale motions that can have an important contribution \citep{Steinrueck21}, we chose to scale the \cite{Moses11} suggested Kzz profile by two orders of magnitude and consider this as our nominal profile.
The \cite{Moses11} profile was derived from GCM computations of HD-189733b, therefore we also test it for this planet without down-scaling it. The constant eddy profiles are not physically anticipated and we use them in this step to evaluate the sensitivity of our result on the magnitude of this parameter.

\begin{figure}
\begin{subfigure}{0.5\textwidth}
\includegraphics[width=0.595\textwidth]{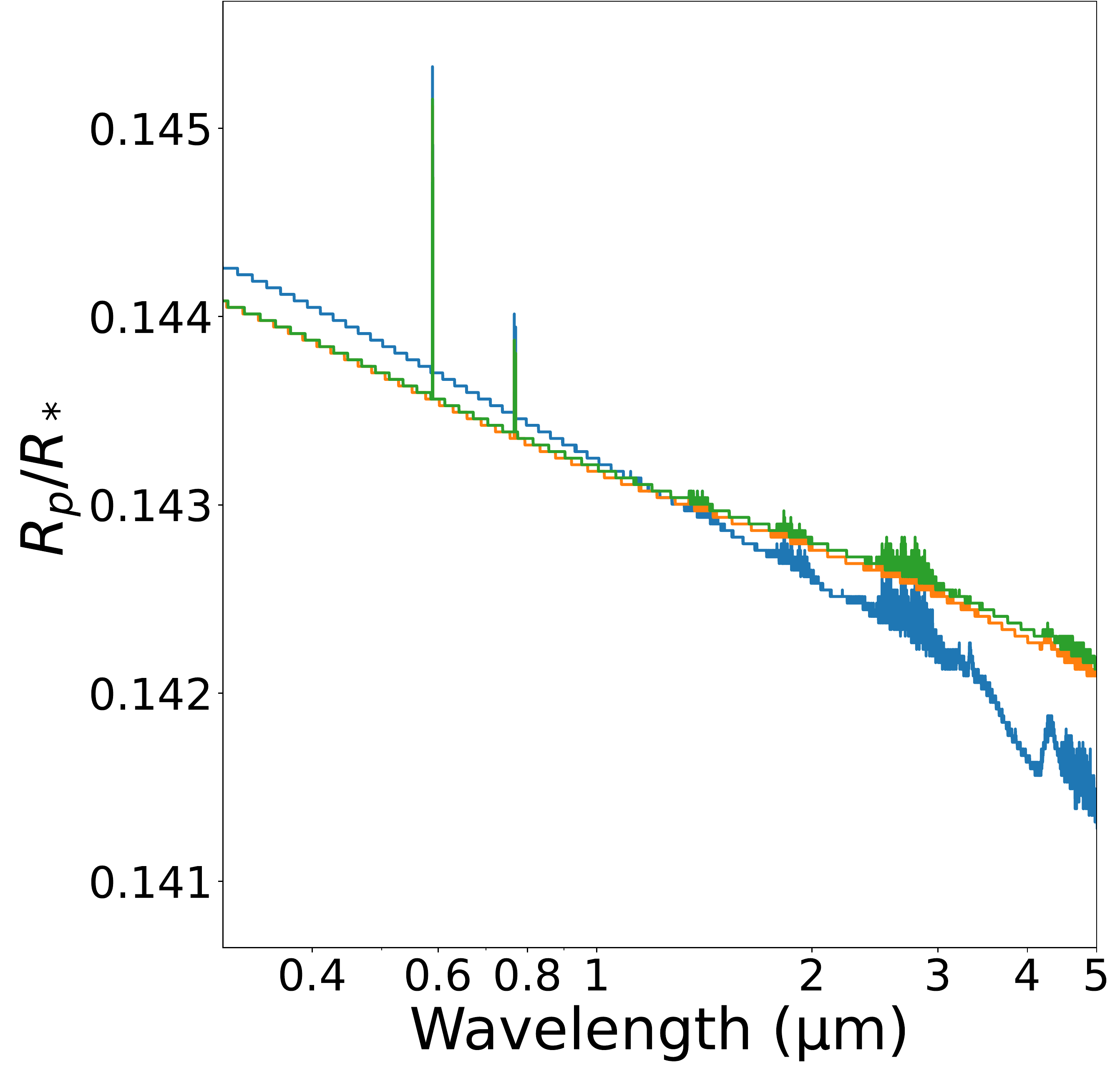}
\includegraphics[width=0.395\textwidth]{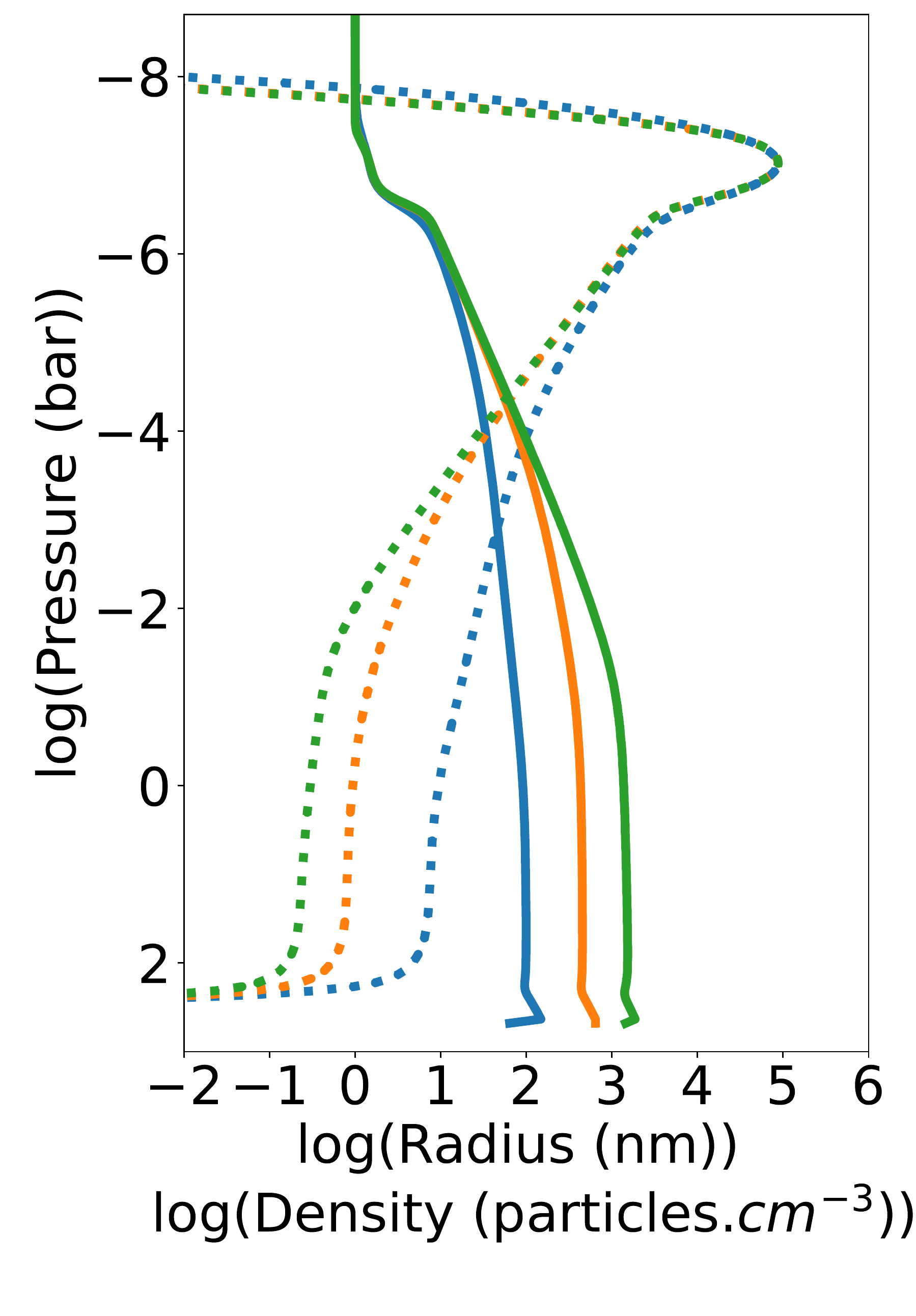}
\caption{}
\label{Fig:eddy1}
\end{subfigure}
\begin{subfigure}{0.5\textwidth}
\includegraphics[width=0.59\textwidth]{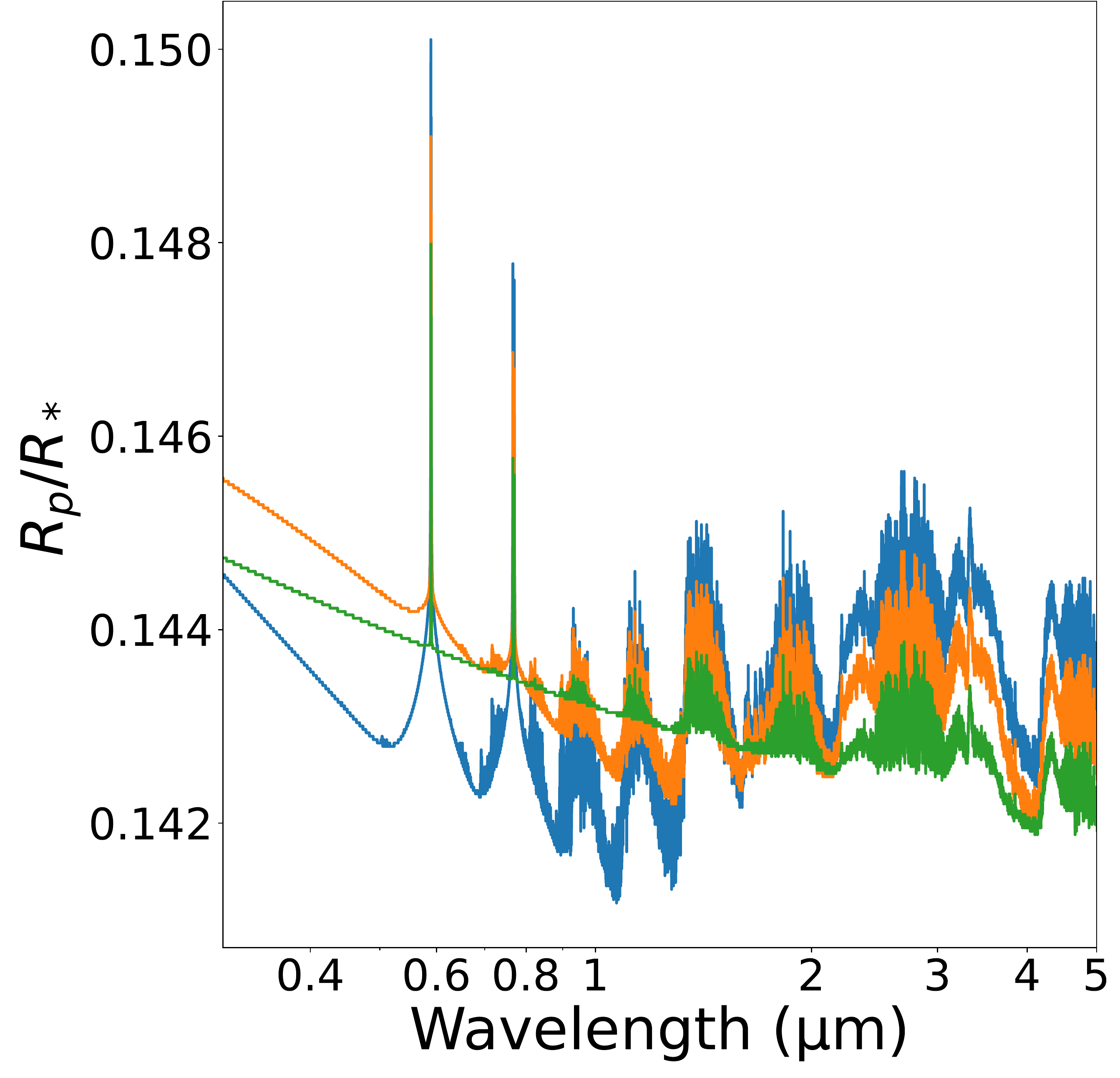}
\includegraphics[width=0.395\textwidth]{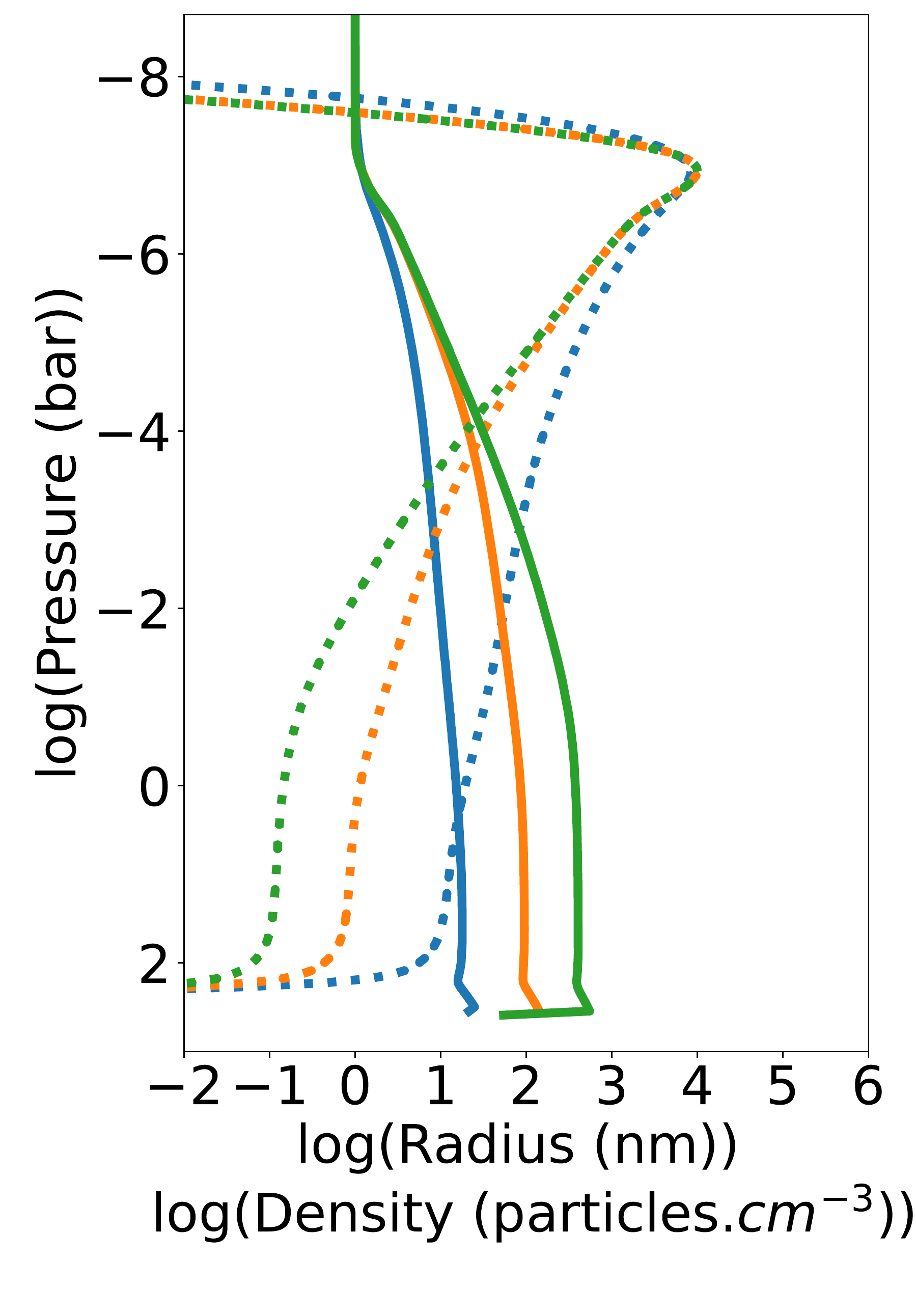}
\caption{}
\label{Fig:eddy2}
\end{subfigure}
\caption{WASP-6b for different eddy profiles with mass fluxes of 10$^{-12}$ $g.cm^{-2}.s^{-1}$ (\protect\cref{Fig:eddy1}) and 10$^{-14}$ $g.cm^{-2}.s^{-1}$ (\protect\cref{Fig:eddy2}).
The blue lines correspond to 10$^{9}$ $cm^{2}.s^{-1}$, the oranges to 10$^{8}$ $cm^{2}.s^{-1}$ and the greens to 10$^{7}$ $cm^{2}.s^{-1}$.
Within each panel, the left figure shows the spectra and the right, the haze distribution; the dashed line is the number density and the solid line the average radius.}
\end{figure}

Hazes tend to increase the UV slope but not under all conditions. A too strong haze abundance results in large particles that tend to act as grey absorbers rather than Rayleigh scattering particles, thus lead to a flat spectrum.
The magnitude of the atmospheric mixing has an important role in this interplay of haze abundance and UV slope. Indeed, a powerful vertical mixing (or equivalently a strong gravity field) will impede particle coagulation and spread them over altitudes. Thus, haze particles will have a small size that demonstrates a Rayleigh type behaviour and will lead to an increase of the UV slope \citep{Lavvas17,Ohno20}. On the opposite limit, a weak eddy mixing will allow for the faster coagulation of particles, resulting in large particles having a grey absorber behaviour, i.e. a rather flat transit signature.
We also note that the eddy magnitude needs to increase with increasing mass flux, for the UV slope to remain steep.
As shown by \cref{Fig:eddy1}, with a strong mass flux of 10$^{-12}$ $g.cm^{-2}.s^{-1}$, there is nearly no differences between the 10$^{8}$ $cm^{2}.s^{-1}$ (orange line) and 10$^{7}$ $cm^{2}.s^{-1}$ (green line)  Kzz cases, and a mixing profile of 10$^{9}$ $cm^{2}.s^{-1}$ (blue line) shows a relatively small influence.
However, for a much smaller mass flux of 10$^{-14}$ $g.cm^{-2}.s^{-1}$ (\cref{Fig:eddy2} with the blue line corresponding to 10$^{9}$ $cm^{2}.s^{-1}$, the orange to 10$^{8}$ $cm^{2}.s^{-1}$ and the green to 10$^{7}$ $cm^{2}.s^{-1}$), the eddy profile plays an important role leading to a steeper slope as it increases.

\subsubsection{Step 2}

In the second step we advance over the assumption of thermochemical equilibrium and evaluate the disequilibrium chemistry composition in order to provide a more detailed picture of the studied planet atmospheres.
This calculation allows to account for transport of species and photolysis reactions \citep{Lavvas14}.
The dis-equilibrium model calculates the chemical composition by solving for the continuity equation:
\begin{equation}
\frac{\partial n_i}{\partial t} = - \nabla \phi_i + P_i - L_i n_i
\end{equation}
The time variation of the density $n_i$ of species \it i \rm depends on the flux divergence $- \nabla \phi_i $, the gain $P_i $ and the losses $- L_i n_i$.
The flux is driven by molecular diffusion and atmospheric eddy mixing, and the sources and losses involve only chemical and photochemical processes related to the rates of the different reactions involved for species \it i\rm.
We use the nominal eddy profile for the evaluation of the disequilibrium chemistry, even if a constant profile provides a better fit in step 1, since a constant profile is not physically anticipated.
For HD-189458 b we consider the eddy profile from \cite{Moses11}.
The boundary conditions at the bottom boundary (10$^3$ bar) are provided by the thermochemical equilibrium model.

The model is coupled to the microphysics model and uses the best-fitting haze parameters retrieved from step 1.
Photolysis reactions are the initial step of photochemical haze formation.
We assume a soot type composition for the haze, therefore haze precursors are anticipated to come from carbon chemistry.
Molecules with significant photolysis rates and/or large upper atmosphere abundances, are expected to be the main precursors.
We thus consider CH$_4$, HCN, CO and C$_2$H$_2$ as the most likely precursors for soot type particles.
Our model is therefore able to provide the haze precursors photolysis mass flux, which corresponds to an upper limit for the haze mass flux potentially available.
The photochemical losses of haze precursors are integrated with altitude over the region above 10$^{-5}$ bar and summed up.
A large part of these products will react to form other molecules and only a fraction, i.e. the formation yield, will actually form hazes.
The haze formation yield can only be constraint by lab experiment or observations (see \cite{Lavvas17,Lavvas19}).
Here we evaluate its range of values for the studied exoplanets.

\subsubsection{Step 3}

In the last step of our approach, we explore the impact of simulated hazes and disequilibrium chemistry on the atmospheric structure. For this purpose, we use a radiative-convective model coupled to the disequilibrium chemistry/microphysics model from step 2. This way, we account for haze and disequilibrium chemistry radiative feedbacks on the temperature profile, as well as the feedbacks of this modified temperature profile on the chemical composition and haze distribution.
Although hot-Jupiters are expected to be tidally locked, owing to their short distance to the host star, we found that a full heat redistribution produces temperature profiles in better agreement with 3D circulation models for the terminators probed in transit \citep{Lavvas21}, thus consider this condition for the energy re-distribution.

Multiple phenomena providing additional heat sources in planet deep interiors, as radioactivity, ohmic dissipation or tidal damping for instance, are required to match the inflated radii of hot-Jupiters \citep{Guillot02,Thorngren18}.
The energy released by those processes propagates to the atmosphere, and is accounted in radiative-convective calculations via an intrinsic flux assumed equivalent to a black-body at a temperature T$_{int}$. We use the typical 100 K value assumed for Jupiter like planets \citep{Iro05}. \cite{Thorngren19b} derived a parameterization to evaluate this intrinsic temperature as a function of the planet effective temperature and found values up to 700 K. The implications of this large intrinsic temperature will be explored in a future study.

Haze radiative feedback can have major consequences on the temperature profile.
Indeed, particles absorb light from the planet host star and transfer the heat to their surrounding environment through conduction, with this contribution leading to larger upper atmosphere temperatures \citep{Lavvas21}.
Since a part of the star light is absorbed by haze in the upper atmosphere, it is no more available for the deep atmosphere resulting in colder temperatures in this region. Moreover, the atmospheric scale height will increase in the region heated by the particles. Therefore, the planet will look larger when observations are probing at or above the heated region. As UV observations probe high in the atmosphere, particle heating thus results in an enhanced UV-slope.

Disequilibrium chemistry has an important impact on the thermal structure of the atmosphere.
In addition to variations in composition that can locally impact the temperature profile, disequilibrium chemistry reveals the presence of the thermosphere. When molecular diffusion dominates over eddy diffusion, gravity separates the molecules with the heavier remaining deeper in the atmosphere and the above layers dominated by H, H$_2$ and He.
The absence of cooling species in the region above the homopause results in a rapid increase of the temperature.
The presence of the thermosphere has major implications in our model that are discussed below.

\subsection{Transit simulations}
\label{SSec:TransitSim}

We use a spectral model that computes the theoretical transit spectrum of a planet taking into account Rayleigh scattering and absorption by the atmospheric main gaseous components and Mie extinction by photochemical aerosols.
We consider contributions from atomic and molecular hydrogen, $\rm H_2O$, $\rm CH_4$, $\rm NH_3$, $\rm HCN$, $\rm H_2S$, $\rm CO_2$, $\rm Na$ and $\rm K$, as well as collision-induced absorptions by $\rm H_2-H_2$ and $\rm He-H_2$ pairs.
We also consider absorption in atomic lines of AlI, CaI, FeI, FeII, MgI, MgII, TiI, TiII and SiI as well as $\rm SiO$, $\rm TiO$ and $\rm VO$ absorption bands for the hottest planets.
Opacity sources are described in \cite{Lavvas21}.
For the evaluation of the particle optical properties we consider a soot type composition.
Soot particles demonstrate a strong resistance to sublimation at high temperatures, therefore are capable to survive at the atmospheric conditions anticipated in hot-Jupiters \citep{Lavvas17}.
The probed pressure is the isobar level below which negligible amount of light cross through the atmosphere, it therefore corresponds to the planet apparent radius at the given wavelength.
To determine the probed pressure, we therefore need to map our pressure grid to the planet radius.
This calculation is achieved assuming an hydrostatic equilibrium and assuming a reference pressure for which the corresponding altitude is known.
This reference point is determined in the water band and the radius is obtained by integrating the planet to star radius ratio observed with WFC3.

\subsection{Stellar spectra}

\begin{figure}
\includegraphics[width=0.5\textwidth]{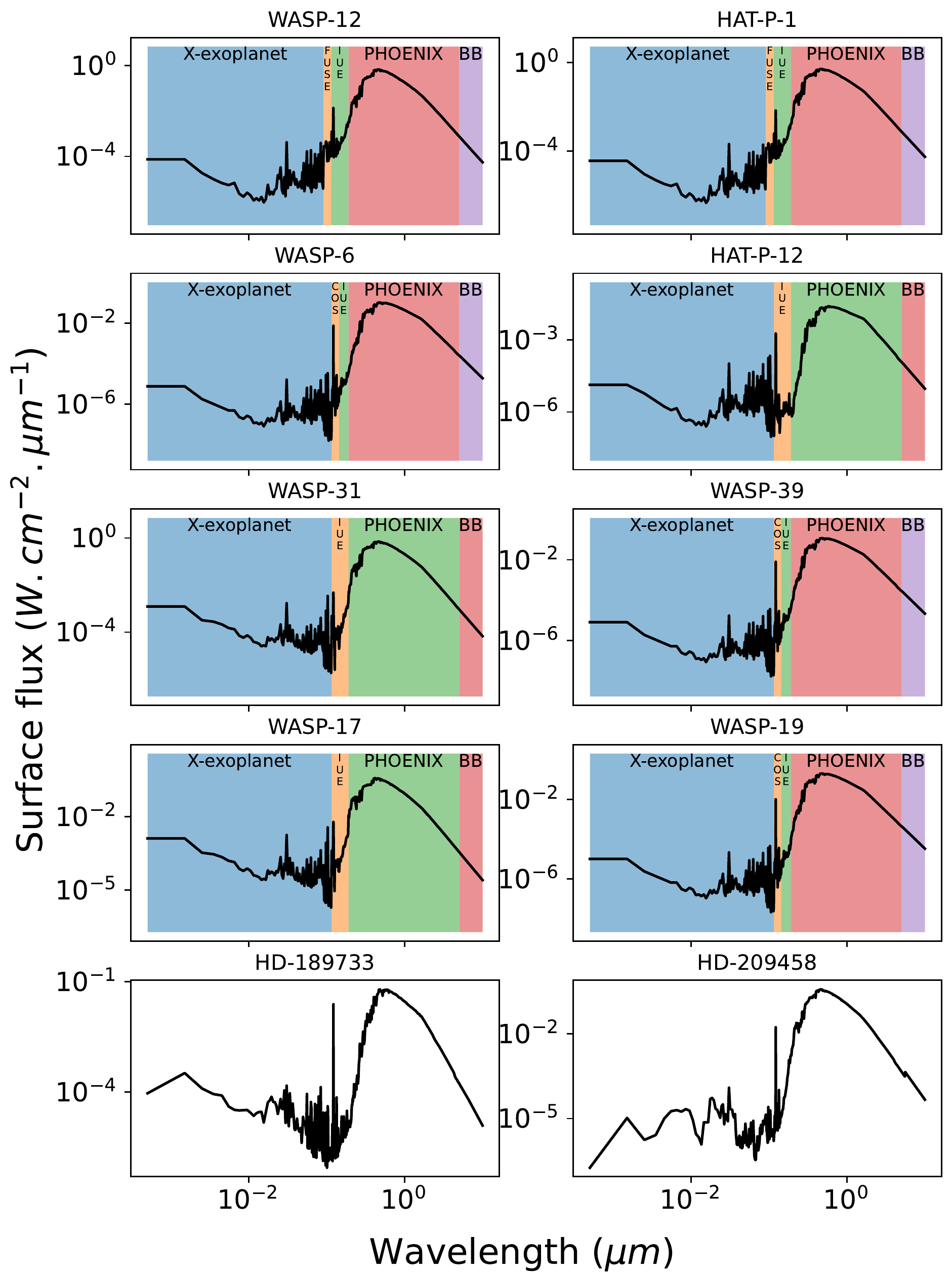}
\caption{Stellar spectra of the studied exoplanet host stars at 1 AU.
The colored areas present the wavelength range within which we use the corresponding model/instrument.}
\label{Fig:StellarSpectra}
\end{figure}

\begin{table}
\centering
\caption{Studied stars in the study (left) and proxy stars (right) used for the evaluation of the stellar fluxes. Stellar radii are provided in $R_{Sun}$ and ages in Gyr. The wavelengths correspond to the range where each proxy star spectrum is used.
}
\begin{tabular}{cc|cc}
\hline
\multicolumn{2}{c}{Studied stars} 	& 	\multicolumn{2}{c}{Proxy stars}					\\
\hline
\multirow{5}{*}{WASP-17}	&		& $\tau$ Bootis A 		&			Name		\\
					&		&		5 to 1900 \AA{}	& 		Wavelength		\\
					& F4V	&		F7V			&	Type					\\
					& 1.38	&		1.331		&	Radius				\\
					& 3.0		&		2.52			&	Age					\\
\hline
\multirow{5}{*}{WASP-31}	&		& $\tau$ Bootis A 		&						\\
					&		&		5 to 1900 \AA{}	& 						\\
					& late-F	&	F7V				&						\\
					& 1.24	&	1.331			&						\\
					& 1.0		&	2.52				&						\\
\hline
\multirow{5}{*}{WASP-12}	&		& HD-216435			& $\chi^1$ Orionis			\\
					&		&		5 to 910 \AA{}	& 	910 to 1900 \AA{}		\\
					& G0V	&	G0V				&	G0V					\\
					& 1.657	&	2.0				&	0.979				\\
					& 1.7		&	5.27				&	0.4					\\
\hline
\multirow{5}{*}{HAT-P-1}	&		& HD-216435			& $\chi^1$ Orionis			\\
					&		&		5 to 910 \AA{}	& 	910 to 1900 \AA{}		\\
					& G0V	&	G0V					&	G0V				\\
					& 1.174	&	2.0				&	0.979				\\
					& 3.6		&	5.27				&	0.4					\\
\hline
\multirow{5}{*}{WASP-19}	&		& 55 Cancri			& 61 Ursae Maj.			\\
					&		&		5 to 1440 \AA{}	& 	1440 to 1900 \AA{}		\\
					& G8V	&	K0V				&		G8V				\\
					& 1.004	&	0.98				&		0.86				\\
					& 11.5	&	10.2				&		3.5				\\
\hline
\multirow{5}{*}{WASP-6}	&		& 55 Cancri			& 61 Ursae Maj.			\\
					&		&		5 to 1440 \AA{}	& 	1440 to 1900 \AA{}		\\
					& G8V	&	K0V				&		G8V				\\
					& 0.87	&	0.98				&		0.86				\\
					& 11.0	&	10.2				&		3.5				\\
\hline
\multirow{5}{*}{WASP-39}	&		& 55 Cancri			& 61 Ursae Maj.			\\
					&		&		5 to 1440 \AA{}	& 	1440 to 1900 \AA{}		\\
					& G8V	&	K0V				&		G8V				\\
					& 0.895	&	0.98				&		0.86				\\
					& 9.0		&	10.2				&		3.5				\\
\hline
\multirow{5}{*}{HAT-P-12}	&		& HD-218566			& 	61 Cygnus A			\\
					&		&		5 to 1150 \AA{}	& 	1150 to 1900 \AA{}		\\
					& K5V	&	K3V				&	K5V					\\
					& 0.7		&	0.86				&	0.665				\\
					& 2.5		&	8.5				&	6.1					\\
\hline
\end{tabular}
\label{Tab:ProxyStars}
\label{Tab:StarData}
\end{table}

\begin{table*}
\caption{Retrieved haze properties from step 1. The average particle sizes and the number densities are taken in the region probed by observations in the FUV part of the spectra (around 0.35 $\mu m$).}
\begin{tabular}{c|cccccc}
\hline
Planet					& $\chi^2$ 	& Mass flux  			& Eddy profile 		& Metallicity 		& Average radius	& Density 				\\
						& 			& ($g.cm^{-2}.s^{-1}$)	& ($cm^{2}.s^{-1}$) 	& ($\times$solar)	& (nm) 			& ($particles.cm^{-3}$)	\\
\hline
HAT-P-1b					&	1.474	& $ 10^{-16}$ 			& nominal			& 0.1				& 10				& 6					\\
						&	1.314	& $ 10^{-12}$ 			& nominal			& 1				& 150			& 8					\\
HAT-P-12b				&  	0.987	& $ 10^{-14}$ 			& 1e7 			& 1				& 15 				&  60					\\
HD-189733b				&  	9.364	& $ 5.10^{-12}$ 		& 1e10		 	& 1				& 2	 			& 2x10$^{5}$			\\
HD-209458b				&  	2.207	& $ 10^{-14}$			& nominal  		& 1				& 10				& 80		 			\\
WASP-6b					&  	1.616	& $ 10^{-14}$ 			& 1e8	  		& 0.1				& 20				& 30		 			\\
						&  	2.455	& $ 10^{-14}$ 			& 1e8	  		& 1				& 20				& 30		 			\\
WASP-12b				&  	0.844	& $ 10^{-13}$ 			& nominal  		& 1				& 6 				& 750				\\
WASP-17b				&  	1.523	& $ <10^{-16}$ 			& 		 		& 1				& 				&					\\
WASP-19b				&  	2.284	& $ 10^{-11}$ 			& nominal  		& 1				& 1.5				& 10$^5 $ 			\\
WASP-31b				&	1.575	& $ 10^{-12}$			& nominal			& 1				& 1.5				& 5x10$^{4}$ 			\\
						&	1.659	& $ <10^{-16}$			& 				& 0.1				& 				& 					\\
WASP-39b				&  	4.706	& $ <10^{-16}$ 			& 		 		& 1				& 				& 		 			\\
\hline
\end{tabular}
\label{Tab:Results}
\end{table*}

There are currently no available observations for the stars within this study that cover the whole spectrum, therefore, we need to create composite spectra based on theoretical models and proxy star observations.
We use two different models in our composite spectra.
First, for optical and NIR spectrum, we used the PHOENIX database \citep{Husser13} that takes as input the star temperature, surface gravity and metallicity.
Then, we need information on the region from X-rays to EUV.
EUV radiation is strongly absorbed by the interstellar medium, therefore there are limited observational constraints for this part of the stellar spectrum.
We thus use the X-Exoplanets database \citep{Sanz-Forcada11}, which uses a coronal model fitted on X-ray observations to extrapolate the EUV part of the spectrum to cover this gap.
This database provides spectra for a few stars spanning a large range of spectral types.
Additional information is required in the FUV part of the spectra to connect the X-Exoplanet data to the PHOENIX model.
Space-borne observations with the International UV Explorer (IUE) and the Far-UV Space Explorer (FUSE) facilities (MAST database), as well as, the Cosmic Origin Spectrograph (COS) on board the HST (ESA database), provide such data.
The spectra, along with the corresponding datasets, are shown in \cref{Fig:StellarSpectra}.
\cref{Tab:ProxyStars} presents the proxy star properties for each studied star.
Every stellar spectrum is composed of an X-Exoplanet \citep{Sanz-Forcada11} model from 5 to 1150 \AA{} except the GOV-type for which the region 910 to 1150 \AA{} is replaced by FUSE observations of $\chi^1$ Orionis \citep{Guinan02}.
For the G8V-type stars we use as proxy the 55 Cancri COS observations \citep{Bourrier18} from 1150 to 1440 \AA{} followed by IUE observations of 61 Ursae Majoris \citep{Landsman93} from 1440 to 1900 \AA{}.
For every other star, in the range 1150 to 1900 \AA{}, we use as proxy IUE observations of $\chi^1$ Orionis \citep{Cappelli89} for the G0V, 61 Cygnus A (\url{https://archive.stsci.edu/proposal_search.php?id=QC008&mission=iue}) for K5V and $\tau$ Bootis A \citep{Bohm-Vitense83} for F-type stars.
The measured fluxes of the proxy stars are scaled to the radius of the host star of each planet.
Finally, we use the PHOENIX model from 0.19 to 5 $\mu m$ and we expand it to 10 $\mu m$ assuming a black body emission.
For HD-189733 and HD-209459 we use stellar spectra from \cite{Lavvas17}.

\section{GENERAL RESULTS}
\label{Sec:GenResults}

In this section, we discuss the general results obtained from the three steps of our work.

\subsection{Step 1: Thermochemical equilibrium}

\begin{figure*}
	\includegraphics[width=0.9\textwidth]{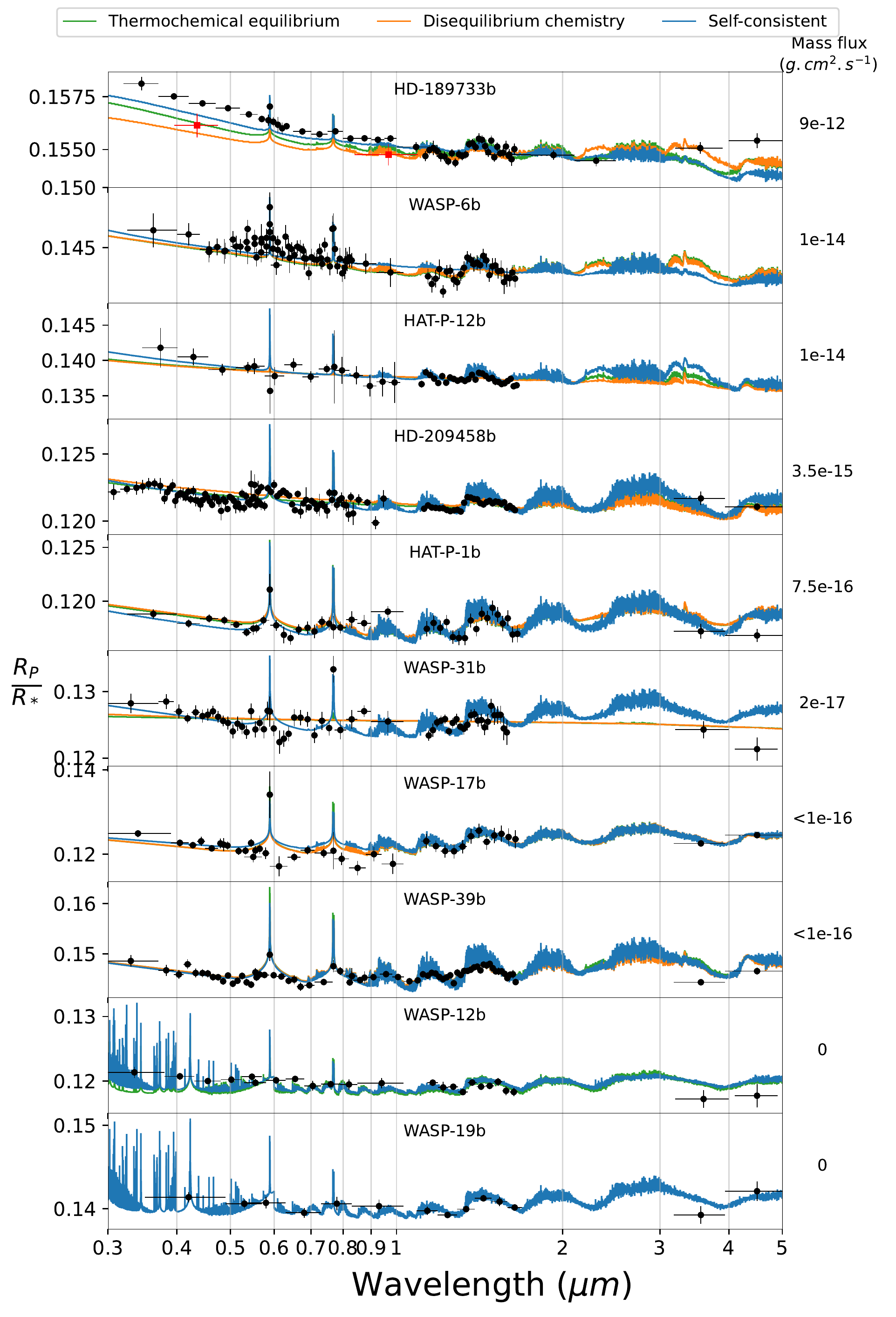}
	\caption{Transit spectra from step 1 (green), 2 (orange) and 3 (blue) sorted by mass flux.}
	\label{Fig:FinalSpectra}
\end{figure*}

In this first step, multiple parameters were tested for the different planets in order to fit the observed transit spectra. These best-fitting parameters are gathered in \cref{Tab:Results}. The tested models are assessed via a $\chi^2$ calculation. However, some planets present degeneracies, with sometimes all evaluated models having residuals within the 3$\sigma$ of the observations (e.g. HAT-P-1b and WASP-31b). In such cases the minimum $\chi^2$ case is typically supporting a hazy atmosphere although the spectral features in the observations correlate favorably with a low or haze-free atmosphere. We preserve the latter case in our results from the step 1 evaluation for these two planets. Such cases demonstrate that care must be taken when arguing on optimal fits based upon $\chi^2$ values, and highlight the need for more precise observations. These issues are further discussed in \cref{Sec:PlanResults} for each planet.
We find that haze mass fluxes range from clear atmosphere values, which is either no haze or a haze mass flux producing spectra that do not resolve the presence of haze, to values up to 5x10$^{-12} g.cm^{-2}.s^{-1}$, which is larger than what is observed in the solar system. The best-fits of the transit spectra obtained in this first step are the green curves in \cref{Fig:FinalSpectra}.
The nominal eddy diffusion profile is the best case for most planets but for specific cases (HAT-P-12b, WASP-6b) the constant profiles provide better $\chi^2$ values.

\begin{figure*}
{\includegraphics[width=\textwidth]{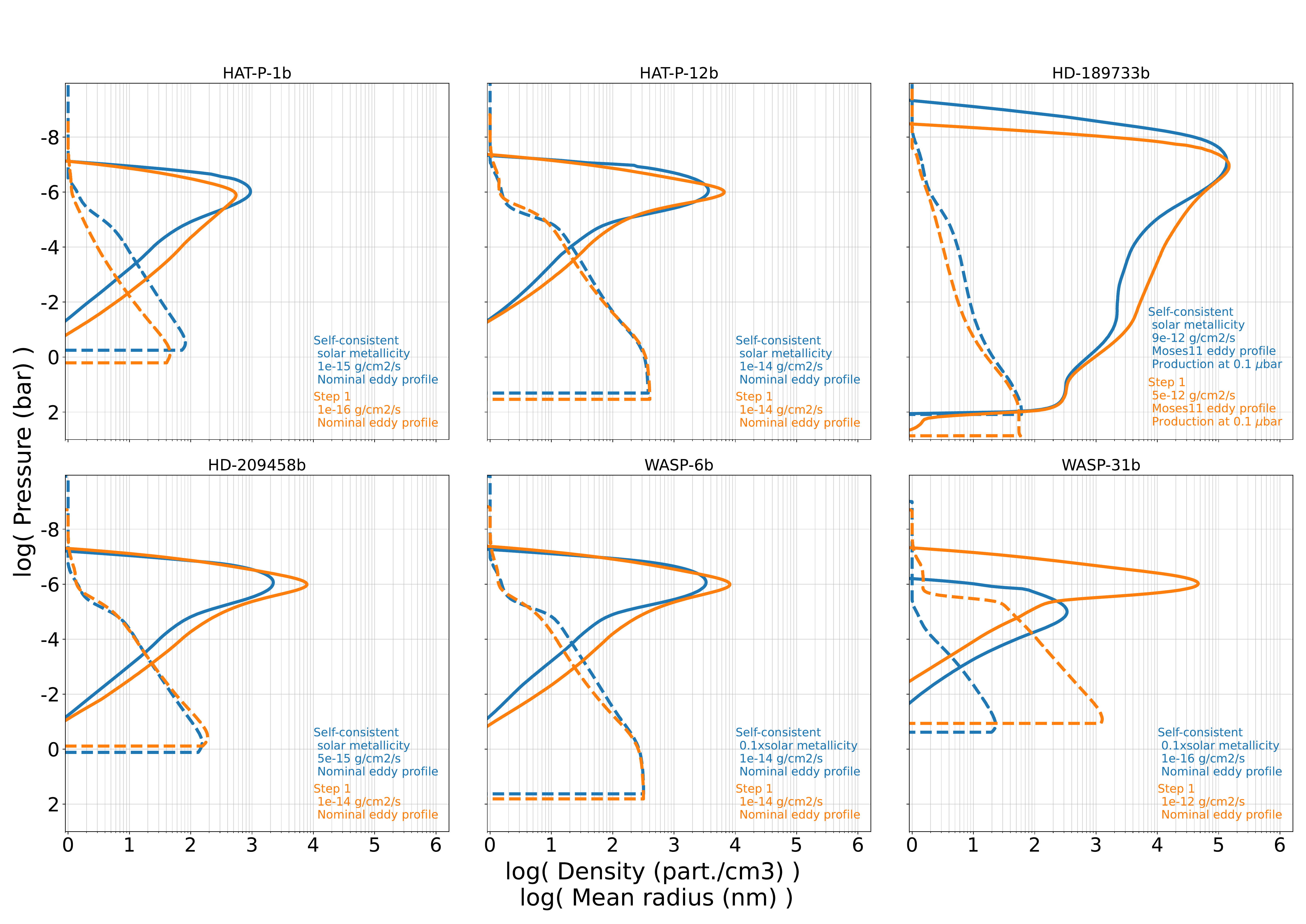}}
\caption{Self-consistent (blue lines) and step 1 (orange lines) haze distributions. The solid lines correspond to the particle number densities and the dashed lines are the mean particle radii.
For HAT-P-12b and WASP-6b, the step 1 profiles are provided for a nominal eddy (even when a constant eddy is providing a better fit) to allow the comparison between the two steps.
The other planets assume different haze mass fluxes between step 1 and 3.
For HD-189733b, we use a higher haze production altitude centered at 0.1 $\mu$bar instead of the nominally 1 $\mu$bar used for the other planets.}
\label{Fig:HazeDistributions}
\end{figure*}

The strength of the diffusion has important ramifications for the haze distribution, especially, a strong eddy prevents particle coagulation.
However, for a similar eddy profile, particle radius and density grow with increasing haze production (\cref{Fig:HazeDistributions}).
Indeed, in the probed regions, the lowest particle density is obtained for HAT-P-1b with 6 $particles.cm^{-3}$ for a mass flux of 10$^{-16} g.cm^{-2}.s^{-1}$, while HAT-P-12b, WASP-6b and HD-209458b, fitted with a 10$^{-14} g.cm^{-2}.s^{-1}$ mass flux, present particle densities ranging from 60 to 80 $particles.cm^{-3}$ (\cref{Tab:Results}).
HD-189733b, with the strongest eddy profile, obtains the largest particle density of 2x10$^{5}$ particles.cm$^{-3}$ for a mass flux of 5x10$^{-12} g.cm^{-2}.s^{-1}$.
The resulting particle densities are further modulated by thermal conditions of each planet that affect their coagulation, with higher temperatures leading to larger particles therefore to smaller particle densities.

As anticipated, we find small particles in the region probed by UV observations with a maximum average radius of 20 nm (\cref{Tab:Results}). The particles keep growing below the probed region, as they are transported by diffusion and gravitational settling, and reach radii of a few hundred nm (\cref{Fig:HazeDistributions}).
HAT-P-12b, WASP-6b and HD-209458b, fitted with a 10$^{-14} g.cm^{-2}.s^{-1}$ mass flux, have particles with mean radii smaller than 400 nm.
HAT-P-1b, fitted with a 10$^{-16} g.cm^{-2}.s^{-1}$ haze mass flux, presents particle size up to 50 nm.
HD-189733b demonstrates a maximum particle radius smaller than 20 nm while it assumes a large mass flux of 5x10$^{-12} g.cm^{-2}.s^{-1}$.

Most of the studied planets are well fitted with a solar metallicity. For specific cases we evaluated results for sub-solar metallicities, a decision driven by differences in the observed and simulated shapes of the Na and K lines. Indeed, a lower metallicity results in weaker absorption features, and Rayleigh scattering by H$_2$ becomes relatively more important because of the referencing of the simulated spectra to the H$_2$O band. All  these factors lead to an enhanced UV slope, thus, the resulting spectrum mimics the presence of haze, reducing the need for a strong haze mass-flux to fit the observations. 
We note that changes in metallicity may modify the thermal structure of the planet, while the temperature profiles used in step 1 were calculated based on solar metallicity \citep{Sing16}.
The implications of this inconsistency are explored in the step 3 of our analysis where we calculate self-consistent temperature profiles.
Changing the metallicity alone, keeping the other parameters constant, has major consequences in the spectra and can strongly impact the fit to the observations.
For instance, the best-fitting sub-solar metallicity case for HAT-P-1b and WASP-6b present $\chi^2$ of 1.474 and 1.616, respectively, against 1.820 and 2.455 for the corresponding solar metallicity case assuming the same haze and eddy parameters.
However, when looking at the best fits in both metallicity cases independently, we find that the changes are not sufficiently significant to draw out any definitive conclusions for HAT-P-1b and WASP-31b with the precision of the currently available observations.
Indeed, as shown in \cref{Tab:Results}, the best-fit sub-solar metallicity case for HAT-P-1b and WASP-31b present $\chi^2$ of 1.474 and 1.575, respectively, against 1.314 and 1.659 for the best-fitting solar metallicity cases.
On the other hand, WASP-6b presents much larger differences between solar and sub-solar best-fits with a $\chi^2$ of 1.616 for a sub-solar metallicity against 2.455 for the best solar case.
High resolution measurements may help to constrain the relative abundances of the different elements and resolve these degeneracies.

\subsection{Step 2: Disequilibrium chemistry}

\begin{table*}
\caption{Photochemical mass fluxes of each haze precursor ($g.cm^{-2}.s^{-1}$) integrated above 10$^{-5}$ bar altitude with self-consistent calculation and the total mass flux summing all the individual contributions.
The last columns provide the mass flux retrieved from step 1 and 3 (final value) and the yield corresponding to the latter case.
The last column indicates the metallicity assumed for the planet.
}
\begin{tabular}{c|cccc|c|c|cc|c}
\hline
Planets			&	$CH_4$	&	$CO$	&	$HCN$	&	$C_2H_2$&	Total		&	Step 1 value		& Final value 		& Yield (\%)			& Metallicity	\\
\hline
HAT-P-1b			&	6.8(-22)	&	2.6(-12)	&	2.5(-13)	&	2.4(-17)	&	2.9(-12)	&	1.0(-15)		&	7.5(-16)			&		0.03 			&		1		\\
HAT-P-12b		&	5.8(-18)	&	1.4(-12)	&	2.3(-11)	&	1.7(-14)	&	2.4(-11)	&	1.0(-14)		&	1.0(-14)			&		0.04			&		1		\\
HD-189733b		&	2.2(-15)	&	1.4(-13)	&	1.4(-10)	&	6.0(-12)	&	1.4(-10)	&	5.0(-12)		&	9.0(-12)			&	 	6.4			&		1		\\
HD-209458b		&	2.0(-23)	&	1.6(-13)	&	3.4(-15)	&	2.3(-18)	&	1.6(-13)	&	1.0(-14)		&	3.5(-15)			&		2.2			&		1		\\
WASP-6b			&	1.0(-19)	&	1.1(-15)	&	2.6(-13)	&	1.7(-18)	&	2.6(-13)	&	1.0(-14)		&	1.0(-14)			&		3.8			&		0.1		\\
WASP-12b		&	7.8(-23)	&	9.5(-11)	&	2.6(-14)	&	5.7(-18)	&	9.5(-11)	&	1.0(-13)		&	0.0				&		0.0			&		1		\\
WASP-17b		&	2.6(-22)	&	4.4(-13)	&	6.8(-14)	&	3.2(-19)	&	5.1(-13)	& $<$1.0(-16)		&	$<$1.0(-16) 		&	$<$0.02			&		1		 \\
WASP-19b		&	2.0(-23)	&	5.1(-11)	&	9.3(-17)	&	1.2(-20)	&	5.1(-11)	&	1.0(-11)		&	0.0						&		0.0	&		1		\\
WASP-31b		&	3.3(-20)	&	1.2(-13)	&	6.5(-14)	&	3.7(-19)	&	1.9(-13)	&	1.0(-12)		&	2.0(-17)			&		0.01			&		0.1		\\
WASP-39b		&	2.9(-23)	&	2.3(-14)	&	2.1(-14)	&	8.0(-20)	&	4.4(-14)	&	$<$1.0(-16)	&	$<$1.0(-16)		&		$<$0.2		&		1		\\
\hline
\end{tabular}
\label{Tab:PhotoFlux}
\end{table*}

A benefit of using a disequilibrium chemistry model is the ability to calculate the mass flux arising from haze precursors photolysis.
These mass fluxes provide an upper limit value for the haze mass flux. In \cref{Tab:PhotoFlux}, the first four columns provide the photolysis mass fluxes of the main haze precursors while the 5$^{th}$ column shows the total mass flux. As discussed later on, the temperature variations brought by the step 3 of our analysis impact the haze precursors profiles and then their photolysis mass fluxes. Therefore, the values provided in \cref{Tab:PhotoFlux} are from the self-consistent dis-equilibrium calculations. Although we do find differences between the precursor profiles evaluated with the \cite{Sing16} temperature profiles and those with our self-consistent profiles, these do not modify the general picture provided below regarding the importance of each precursor. 
We note that the haze precursors mass fluxes calculated for the planets are consistent with (i.e. they are larger than) the haze mass flux retrieved from the first part of our investigation. This result indicates that a detectable amount of photochemical haze can actually form in the atmospheres of these planets.

\begin{figure*}
	\includegraphics[width=\textwidth]{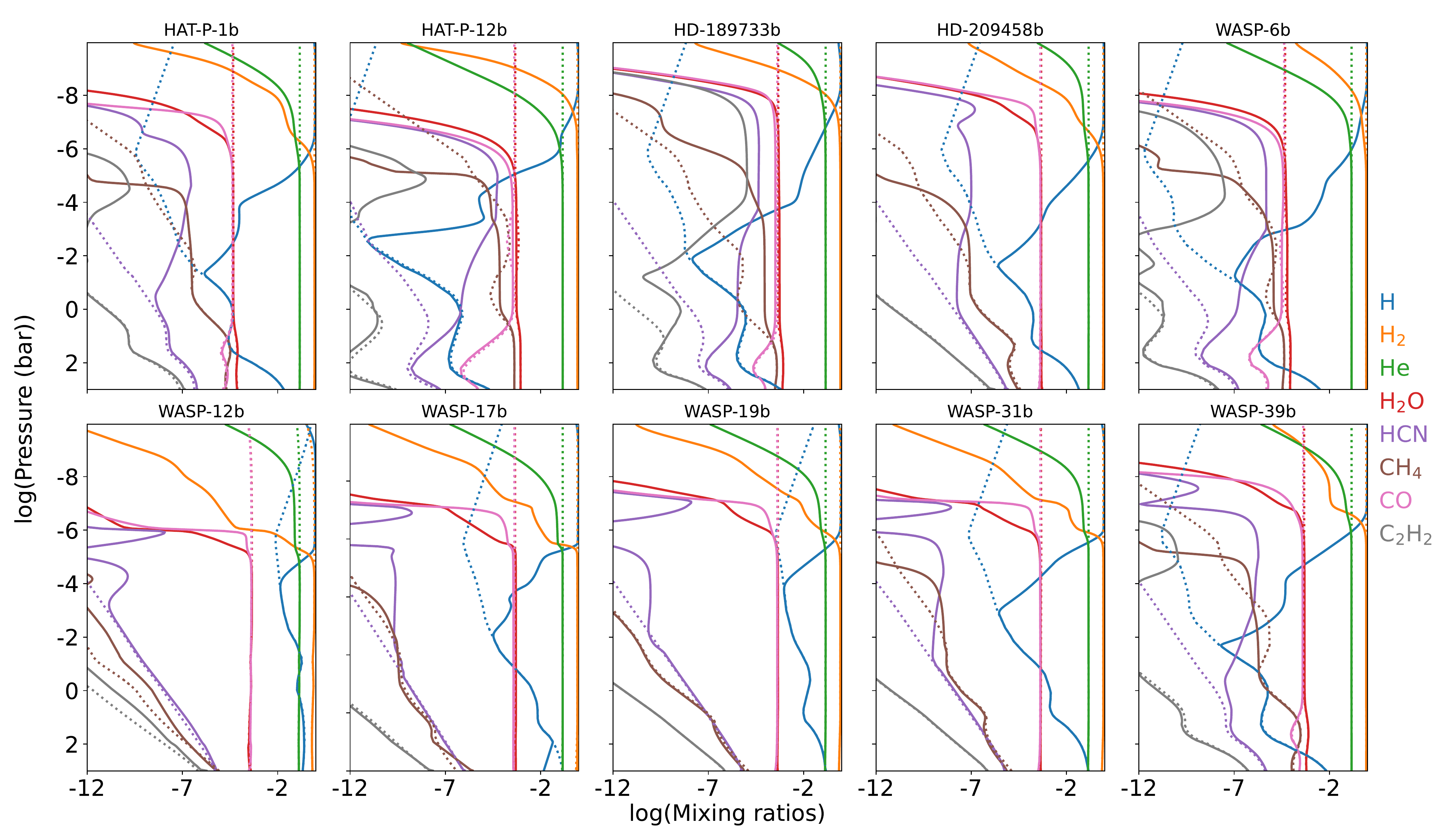}
	\caption{Disequilibrium (solid line) and thermochemical equilibrium (dotted lines) chemical composition with \protect\cite{Sing16} temperature profiles for the major consitituents H, H$_2$, He, H$_2$O and for the haze precursors CH$_4$, HCN, C$_2$H$_2$ and CO.}
	\label{Fig:ChemCompTherm}
\end{figure*}

Disequilibrium effects have major implications for the atmospheric chemical composition in the upper atmosphere \citep{Liang03,Moses11,Lavvas14}.
The quenching of chemical species related to transport can impact their upper atmosphere abundances (\cref{Fig:ChemCompTherm}).
The location of the quench level for each species will impact its whole atmosphere profile.
A relatively cold atmosphere with large eddy will present a transition at low altitudes as HD-189733b with a HCN transition altitude at 100 bar, while a hot exoplanet with smaller $K_{ZZ}$ shows a transition at much higher altitude as WASP-12b with an altitude of 10$^{-2}$ bar for HCN (\cref{Fig:ChemCompTherm}).
This can have an important impact on the haze formation.
Indeed, most of the haze precursors (CH$_4$, HCN and C$_2$H$_2$) show decreasing mixing ratios with increasing temperature in thermochemical equilibrium.
Since temperature decreases with altitude in the deep atmosphere, CH$_4$, HCN and C$_2$H$_2$ show decreasing mixing ratios with increasing altitude.
Then, a quench level at high altitude and so at low mixing ratio, leads to smaller abundances in the upper atmosphere where photolysis occurs.
A relatively cold exoplanet with a strong eddy profile would then be expected to present an enhanced haze mass flux as demonstrated by HD-189733b.
However, we note that CO, one of the considered haze precursors, has the opposite behavior with abundances increasing with the temperature reaching an almost constant profile relatively deep down in the atmosphere.
This species is less impacted by the location of the quench level, and is always available for photolysis.
As a result, hot planets may present haze mass fluxes owing to CO photodissociation, even if they present quench levels high up in the atmosphere, as demonstrated by WASP-17b in \cref{Tab:PhotoFlux}.
CO is then the dominant haze precursor for this kind of planets.

On the other hand, photochemistry also has an important impact on the haze precursors abundances.
For HCN, we indeed note the large upper atmosphere abundances that cannot be explained by the quenching since the latter should lead to a constant mixing ratio.
Instead, the observed behavior is explained by the photolysis of NH$_3$ and CH$_4$ as well as their reactions with atomic hydrogen, which also is a product from photochemistry \citep{Liang03,Moses11}.
The nitrogen and CH$_3$ produced from dissociation of NH$_3$ and CH$_4$, respectively, reacts via different channels to produce HCN, either directly (CH$_3$ + N $\rightleftarrows$ HCN + H$_2$) or via an intermediate species: H$_2$CN (CH$_3$ + N $\rightleftarrows$ H$_2$CN + H), which will produce HCN via thermal dissociation (H$_2$CN + M $\rightleftarrows$ HCN + H + M) or via reaction with H (H$_2$CN + H $\rightleftarrows$ HCN + H$_2$).
This happens around 1 mbar where the HCN mixing ratio reaches its largest value. Transport then propagates the produced HCN to the above atmosphere.
C$_2$H$_2$ also presents upper atmosphere mixing ratios larger than the quench value resulting from the formation of C$_2$H$_6$ by CH$_3$ (2CH$_3$ $\rightleftarrows$ C$_2$H$_6$) followed by its consecutive dissociations to C$_2$H$_5$, C$_2$H$_4$, C$_2$H$_3$ and then C$_2$H$_2$.
C$_2$H$_4$ is also produced by CH$_2$ and CH$_3$ (CH$_2$ + CH$_3$ $\rightleftarrows$ C$_2$H$_4$ + H), which further contributes to the formation of C$_2$H$_2$.
Despite this production of C$_2$H$_2$, its loss by photodissociation prevents this species to reach very large amounts in most cases and its abundance remains smaller than HCN and CO.

CO photodissociation requires high energy radiation ($\lambda$ < 1100 \AA{}) that are quickly absorbed high up in the atmosphere. This species is then expected to be an important haze precursors only for strongly irradiated planets.
CH$_4$ and HCN have smaller photolysis energy thresholds ($\lambda$ $\sim$ 1500 \AA{}) allowing them to efficiently contribute to the haze precursors photolysis mass flux even for less irradiated planets, though the low abundance of CH$_4$ leads to a contribution smaller than for HCN.
C$_2$H$_2$ cross section further extends in the middle UV up to 2300 \AA{} which allows for the photolysis at much larger pressures than the other precursors.
This loss of C$_2$H$_2$ deeper in the atmosphere is partly responsible for the low abundance demonstrated by this species in the upper atmosphere where the precursors photolysis mass flux is calculated.
This species therefore does not strongly contribute to the haze precursors photolysis mass flux.

The weaker abundances of CH$_4$ and C$_2$H$_2$ make these species weaker contributors to the haze production compared to HCN and CO as presented in \cref{Tab:PhotoFlux}.

\subsection{Step 3: Self-consistent calculation}

\begin{figure*}
\includegraphics[width=\textwidth]{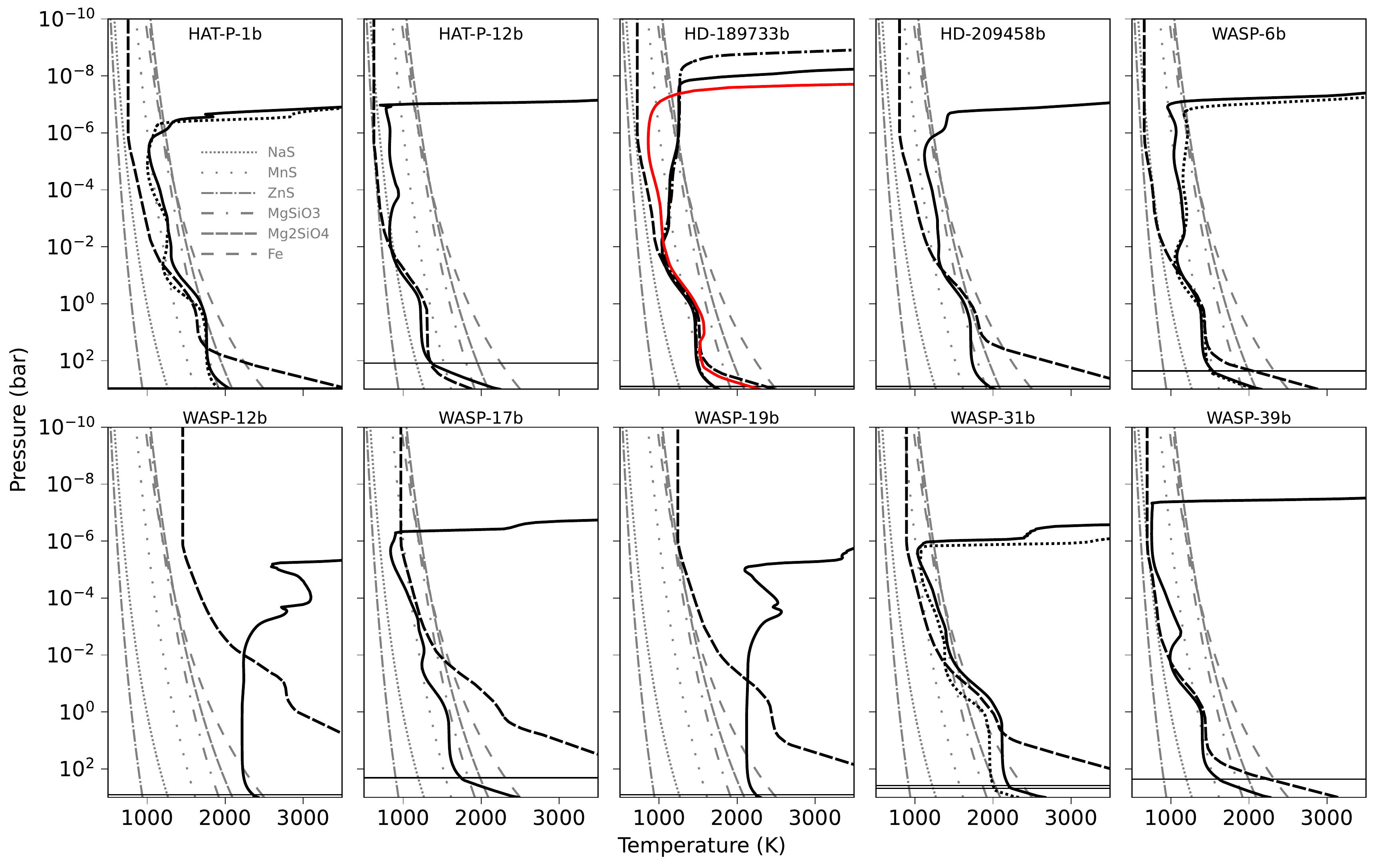}
\caption{Initial temperature profiles from  \protect\cite{Sing16} (dashed lines) and final self-consistent temperature profiles assuming a solar (solid lines) and a sub-solar (dotted lines for relevant cases) metallicity. The red line for HD-189733b corresponds to an additional tested profile from \protect\cite{Moses11}.
The dashed-dotted line for HD-189733b assumes a 0.1 $\mu$bar haze production altitude.
Radiative-convective boundaries (RCB) are also shown as horizontal black lines. Condensation curves of condensate candidates are also provided (grey lines).}
\label{Fig:FinalPTprofiles}
\end{figure*}

In \cref{Fig:FinalPTprofiles} we compare our self-consistent temperature profiles (solid lines) with those derived by \cite{Sing16} (dashed lines) and used in the previous steps of our approach. We also present temperature profiles evaluated for sub-solar (0.1xSolar) conditions (dotted lines) which are rather close to the corresponding solar cases. This result suggests that the approximation in step 1 of using temperature profiles of solar composition for the sub-solar cases is acceptable at that preliminary step. However, we note the differences in the deep atmosphere when comparing the step 3 profiles with those from step 1, related to different assumptions for the intrinsic temperatures in the two calculations. Nevertheless, in most cases, these differences are limited to the deep atmosphere (pressures higher than 1 bar) and should not affect the transit spectra.

In the upper atmosphere, our self-consistent profiles are typically hotter than the \cite{Sing16} profiles due to haze absorption as well as differences in the composition due to the disequilibrium chemistry.
Our simulated profiles also show a sharp temperature increase above the 1 $\mu$bar pressure level related to the drop in water abundance. Indeed, H$_2$O absorbs a large part of UV radiation and emits this excess energy in the infrared. The absence of this coolant leads to a rapid increase of the temperature. This feature does not have a strong impact on the transit spectra as this region is not usually probed in the studied exoplanets (with the exception of HD-189758b), but can be important for the haze formation.
The location of the thermospheric inversion has an impact on the haze precursors mass flux as this parameter is evaluated from the integration of the photolysis rates at altitudes above 10 $\mu$bar. A deep thermospheric inversion will reduce the integrated column density of the species resulting in smaller mass fluxes. Also, particles are expected to sublimate at the thermosphere temperatures, therefore a deep thermosphere may prevent the formation of hazes, as seen for WASP-31b. For this case we assumed a lower location for the haze production profile to evaluate the possible contribution of hazes in the observed spectrum.

While temperature increases due to particle heating, the coagulation of these particles is enhanced allowing them to reach larger radii. This trend is observed in \cref{Fig:HazeDistributions} which shows larger average radii and lower particle densities in the heated part of the atmospheres (for planets with the same haze production parameters between step 1 and 3). We further note that the coagulation enhancement is stronger for large haze mass fluxes as expected. Indeed, more numerous particles result in a larger temperature increase, therefore in a stronger enhancement of the coagulation. A consequence of this larger and less numerous particles would be a slightly flatter UV slope. This is however not observed (blue lines in \cref{Fig:FinalSpectra}) owing to the atmospheric expansion.
This allows for example a better fit of HD-189733b observations, although still not satisfactory. The effect of the atmospheric expansion is more pronounced for some planets. This is related to the magnitude of the haze mass flux, with higher values leading to a stronger temperature increase. However, the final effect on the transit spectrum depends also on the pressure probed by the observations. If the observations are probing below the region where the atmosphere expansion occurs, the effect will be negligible.
\cref{Tab:ProbedPressure} presents the range of pressure probed by observations for the main features observed: the UV slope, the Na and K lines and the water band.
The sodium and potassium lines are probing a wide range of pressure, the wings probing the deep atmosphere and the core the upper atmosphere.
The presence of haze will therefore block information from the deep atmosphere, hiding the wings of the features.
Thus, haze-free planets present Na and K lines probing larger pressures as demonstrated in \cref{Tab:ProbedPressure}.
The same behavior is observed in the water band with observations at larger pressures for haze-free exoplanets.

In the deep atmosphere, as the temperature increases, sublimation becomes important leading to the destruction of all particles below a given level. Based on our assumed vapor pressure for the haze sublimation \citep{Lavvas17}, particles should sublimate at particle temperatures above $\sim$1800 K. The location of this particle destruction level in \cref{Fig:HazeDistributions} depends on the temperature profiles used in each step of our investigation.
Particles are heated by UV radiation in the upper atmosphere, therefore they can present much hotter temperatures than the atmosphere in this region.
This effect is prominent above 10 $\mu$bar while deeper in the atmosphere, particles are at atmospheric temperature.
Our results indeed suggest that particles can be hot enough to sublimate in the atmosphere of HAT-P-1b, HD-209458b and WASP-31b.
For HAT-P-1b and HD-209458b, we obtain effective mass fluxes of 7.5x10$^{-16}$ and 3.5x10$^{-15} g.cm^{-2}.s^{-1}$, respectively, instead of the assumed 10$^{-15}$ and 5x10$^{-15} g.cm^{-2}.s^{-1}$, respectively.
This results in negligible differences in the self-consistent atmosphere temperature profiles and transit spectra with $\chi^2$ of 1.505 and 3.170 instead of 1.314 and 3.092 when using atmospheric temperature for HAT-P-1b and HD-209458b, respectively.
For WASP-31b, the effective mass flux produced by the calculation is much smaller than the value we fix with 2x10$^{-17} g.cm^{-2}.s^{-1}$ instead of 10$^{-16} g.cm^{-2}.s^{-1}$, though this value is already very small and produces a spectrum close to a haze-free atmosphere. Therefore, the spectrum obtained using particle temperature in the sublimation rates is very similar to the spectrum using atmospheric temperature, with $\chi^2$ of 1.879 for the former against 1.818 for the latter.
Moreover, as further discussed in \cref{Sec:PlanResults}, we suspect clouds to play an important role in the transit spectra of HD-209458b and WASP-31b, which would decrease the need for haze opacities.
For HAT-P-1b, a smaller mass flux of 10$^{-16} g.cm^{-2}.s^{-1}$ can avoid this upper atmosphere particle sublimation but requires a sub-solar metallicity to fit the observations.
A value closer to the effective mass flux of 7.5x10$^{-16}g.cm^{-2}.s^{-1}$ can also work out the problem.

\begin{table}
	\caption{Pressures probed by the observations for the different features observed in the spectra. The pressures are provided in bar.}
	\label{Tab:ProbedPressure}
	\begin{adjustbox}{max width=0.5\textwidth}
		\begin{tabular}{c|c|c|c|c}
			\hline
			Wavelength		&	UV-slope			&	Na line 			& K line 				& Water band 			\\
			range ($\mu$m)	&	0.3 -  0.5			&	0.5 -  0.7			&	0.7 -  0.8			&	0.8 -  3			\\
			\hline
			HD-189733b		&	3(-6) - 6(-8)		&	4(-6) -  5(-8)		&	9(-6) -  4(-7)		&	9(-5) -  6(-6)		\\
			HAT-P-12b		&	3(-4) - 4(-5)		&	5(-4) -  2(-7)		&	7(-4) -  4(-6)		&	2(-3) -  1(-4)		\\
			WASP-6b			&	5(-4) - 4(-5)		&	6(-4) -  2(-6)		&	1(-3) -  2(-5)		&	3(-3) -  4(-4)		\\
			HD-209458b		&	2(-3) - 2(-4)		&	4(-3) -  9(-7)		&	4(-3) -  7(-6)		&	1(-2) -  2(-4)		\\
			HAT-P-1b			&	1(-2) - 1(-3)		&	2(-2) -  8(-7)		&	2(-2) -  6(-6)		&	5(-2) -  2(-4)		\\
			WASP-31b		&	4(-2) - 4(-3)		&	1(-1) -  4(-6)		&	1(-1) -  6(-5)		&	3(-1) -  1(-3)		\\
			WASP-17b		&	4(-2) - 3(-3)		&	6(-2) -  1(-6)		&	6(-2) -  7(-6)		&	2(-1) -  2(-4)		\\
			WASP-39b		&	4(-2) - 3(-3)		&	7(-2) -  7(-7)		&	7(-2) -  1(-5)		&	3(-1) -  2(-4)		\\
			WASP-12b		&	2(-2) - 3(-6)		&	1(-3) -  2(-5)		&	2(-2) -  2(-4)		&	3(-2) -  1(-3)		\\
			WASP-19b		&	3(-2) - 2(-6)		&	1(-3) -  6(-6)		&	2(-2) -  1(-4)		&	4(-2) -  5(-4)		\\
			\hline
		\end{tabular}
	\end{adjustbox}
\end{table}

\begin{figure*}
	\includegraphics[width=\textwidth]{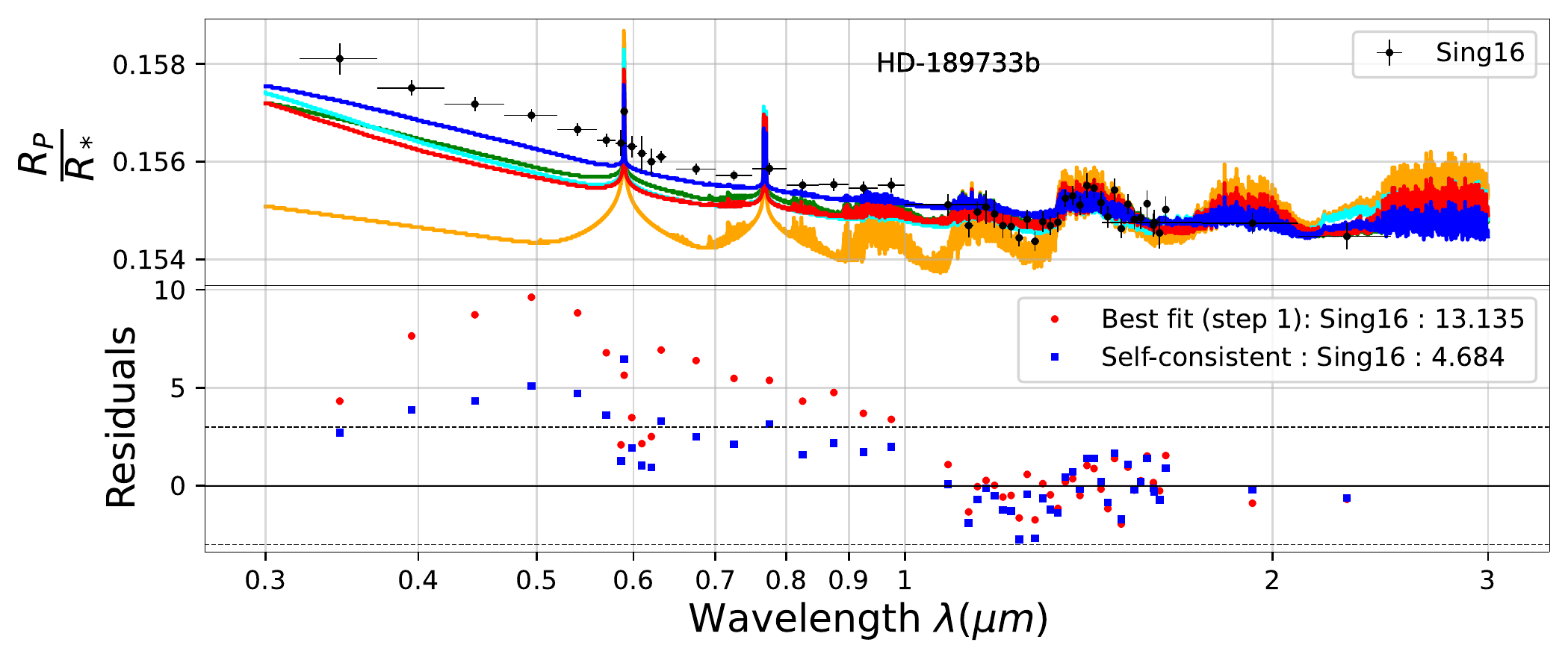}
	\caption{Comparison of observed and simulated transit spectra of HD-189733b along with the residuals of the best-fitting models. Crosses correspond to observations. The red and cyan lines are the 5x10$^{-12}g.cm^{-2}.s^{-1}$ mass flux case with \protect\cite{Moses11} eddy from the first and third step, respectively. The green line assumes a constant 10$^{10}cm^2.s^{-1}$ eddy profile with a 5x10$^{-12}g.cm^{-2}.s^{-1}$ mass flux, from step 1. The blue line is the self-consistent model with a higher haze production at 0.1 $\mu bar$ and a larger haze mass flux of 9x10$^{-12} g.cm^{-2}.s^{-1}$. }
	\label{Fig:189}
\end{figure*}

The temperature variations presented in this section have important ramifications for the haze precursors abundances, therefore, the photolysis mass fluxes presented in \cref{Tab:PhotoFlux} are calculated at the end of this third step of our analysis.
We found photolysis mass fluxes ranging from 4.4x10$^{-14}$ to 1.4x10$^{-10} g.cm^{-2}.s^{-1}$ that mainly reflect changes in stellar insolation and in haze precursor abundances among the studied planets. 
The corresponding yields remain rather small with a maximum 6.4\% obtained for HD-189733b. As the haze mass flux values we retrieve from the observations are supported by the photochemistry, it is clear that the observed features can be attributed to photochemical hazes, although clouds may contribute in specific cases as discussed below.
Among the hazy planets, the yields are spanning nearly four orders of magnitude from 0.01 to 6.4\%, suggesting that the processes between the haze precursors photolysis and the particle formation are quite sensitive to the atmospheric conditions.
We further investigate the behavior of the haze formation yield with the atmospheric and planetary parameters in \cref{Sec:Correlations}.
Finally, we note the particular case of HD-189733b among the hazy planets, as it requires a mass flux three orders of magnitude larger than any of the others.
The haze formation yield we retrieve for this planet is consistent with the values obtain for the other planets, though this is related to the strong eddy profile required for this planet (two orders of magnitude stronger than the nominal) that enhances the haze precursors abundances.
Stellar contamination is expected to, at least partially, affect the steep UV-slope observed, weakening the need for such a strong mass flux \citep{McCullough14}. Future observations of this target should resolve the relative contributions of haze and stellar effects in this atmosphere.

\begin{figure}
	{\includegraphics[width=0.5\textwidth]{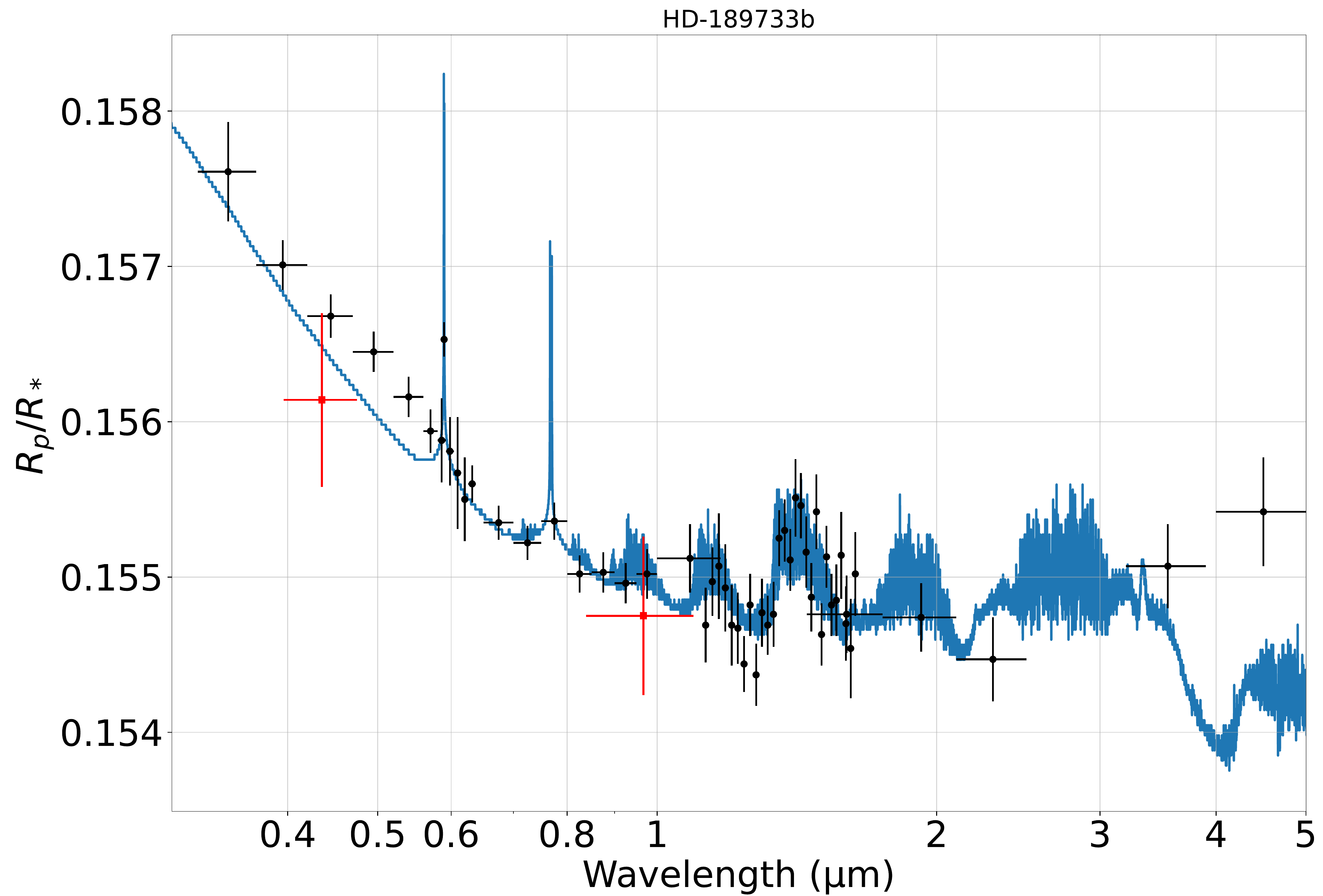}}
	\caption{Self-consistent HD-189733b spectrum with a 0.1 $\mu$bar haze production altitude and a mass flux of 5.10$^{-12}g.cm^{-2}.s^{-1}$.
	The black and red points are observations from \protect\cite{Sing16}  and \protect\cite{Angerhausen15}, respectively.
	The STIS \protect\cite{Sing16} data are shifted downward by $|\Delta R_P/R_*| = 0.0005$.}
	\label{Fig:189Offset}
	\includegraphics[width=0.5\textwidth]{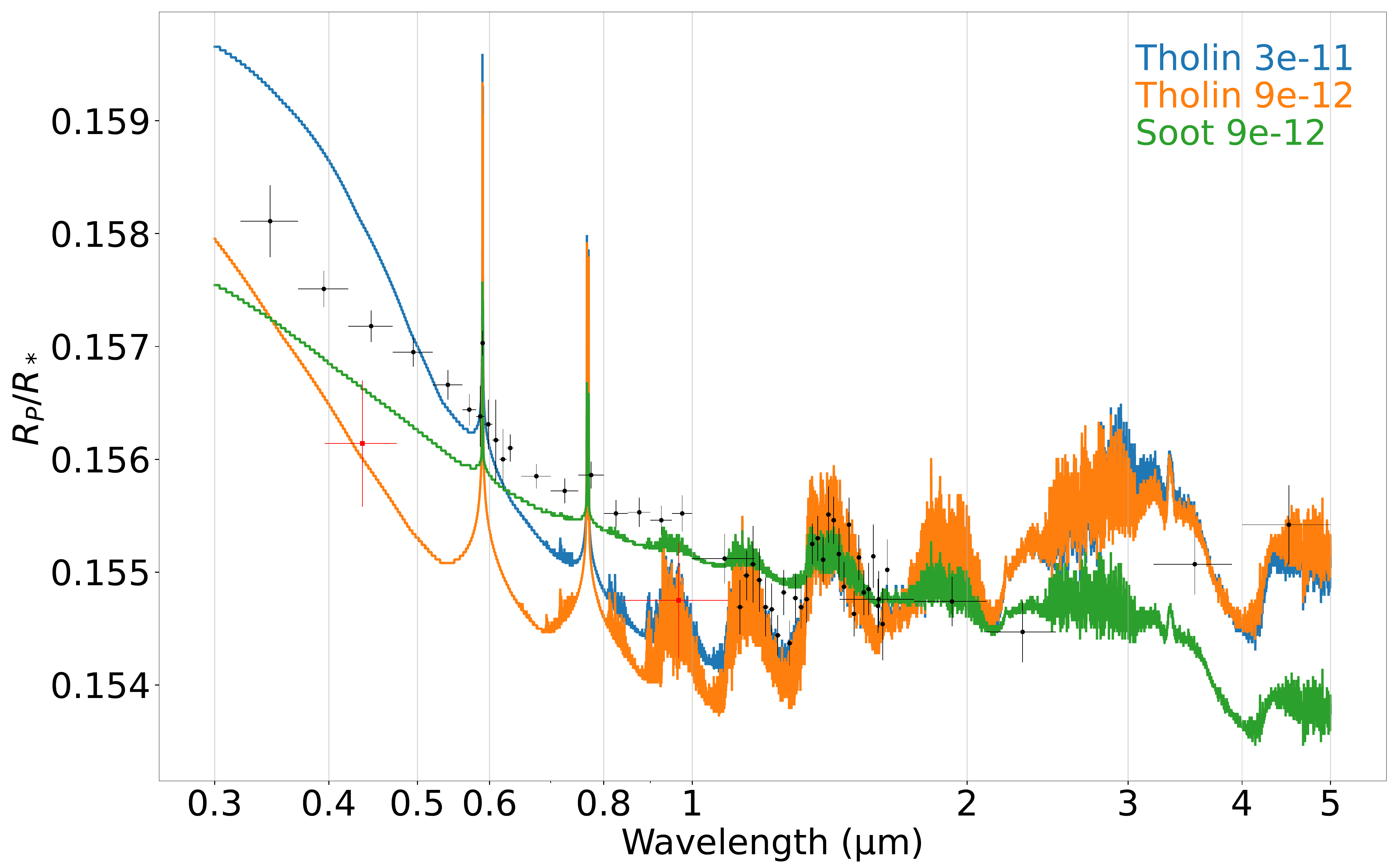}
	\caption{Haze index impact for HD-189733b: the spectra are calculated from step 3, the green and orange assumes a 9.10$^{-12}g.cm^{-2}.s^{-1}$ mass flux with a soot and tholin type refractive indices, respectively, and the blue assumes a larger 3x10$^{-11}g.cm^{-2}.s^{-1}$ with a tholin refractive index.}
	\label{Fig:HazeIndex}
\end{figure}

\section{PLANET BY PLANET RESULTS}
\label{Sec:PlanResults}

In this section we provide a detailed description of the results for each planet. The temperature variations in the self-consistent calculations modify the disequilibrium chemistry results from step 2. Thus, we focus our attention on the results from initial evaluation of the haze mass flux and eddy (step 1) and on how these results are modified through the self-consistent evaluation (step 3).

\subsection{HD-189733b}

\bf Step 1: \rm
For HD-189733b, the best fit in our preliminary minimization is obtained with a large mass flux of 5x10$^{-12}g.cm^{-2}.s^{-1}$.
This best fit assumes the \cite{Moses11} temperature profile and a constant eddy of 10$^{10}cm^2.s^{-1}$ ($\chi^2$=9.364, green line in \cref{Fig:189}).
The combination of a high mass flux and strong atmospheric mixing leads to small and numerous particles in the probed region, thus producing a steep UV slope. However, this combination is not sufficient to reproduce the observed slope. We also tested the \cite{Moses11} eddy profile, which resulted in a slightly worse fit ($\chi^2$=13.13, red line in \cref{Fig:189}). Assuming a larger mass flux of 10$^{-11}g.cm^{-2}.s^{-1}$ leads to a more shallow UV slope as both particle radii and number density increase (not shown).

\bf Step 3: \rm
As the step 1 evaluation demonstrated the need for a strong atmospheric mixing for this planet we decided to use the \cite{Moses11} eddy profile for our self-consistent simulation.
The inclusion of photochemical haze feedback in the radiative transfer leads to a hotter upper atmosphere (dash-dotted line in \cref{Fig:FinalPTprofiles}), yielding a more extended atmosphere and therefore, a larger planet apparent radius. Such an expansion causes an increase of the UV slope and brings the model closer to the observations ($\chi^2$=11.91, cyan line in \cref{Fig:189}), albeit it is still not sufficient to fit the steep UV slope observed.

On the other hand, the observed UV slope probes high up in the atmosphere where particles are produced. We can anticipate that a higher production level would enhance the steepness of the UV slope and bring the model closer to the observations. We thus tried a haze production at 0.1 $\mu bar$, which provides a better fit with a $\chi^2$ of 6.981, but is still not sufficient to bring the model within the 3$\sigma$ of the observations. For this higher altitude production we further explored if increasing the haze mass flux would further improve the fit by testing values up to 10$^{-11}g.cm^{-2}.s^{-1}$. We found that the best fit is obtained with a mass flux of 9x10$^{-12}g.cm^{-2}.s^{-1}$ producing the smallest $\chi^2$ of 4.684 (blue curve in \cref{Fig:189}). Our calculated haze precursor mass flux is of the order of 10$^{-10} g.cm^{-2}.s^{-1}$ for the self consistent calculations with the best fit case providing a value of 1.4x10$^{-10} g.cm^{-2}.s^{-1}$ corresponding to a formation yield of 6.4\%.

\begin{figure*}
\includegraphics[width=\textwidth]{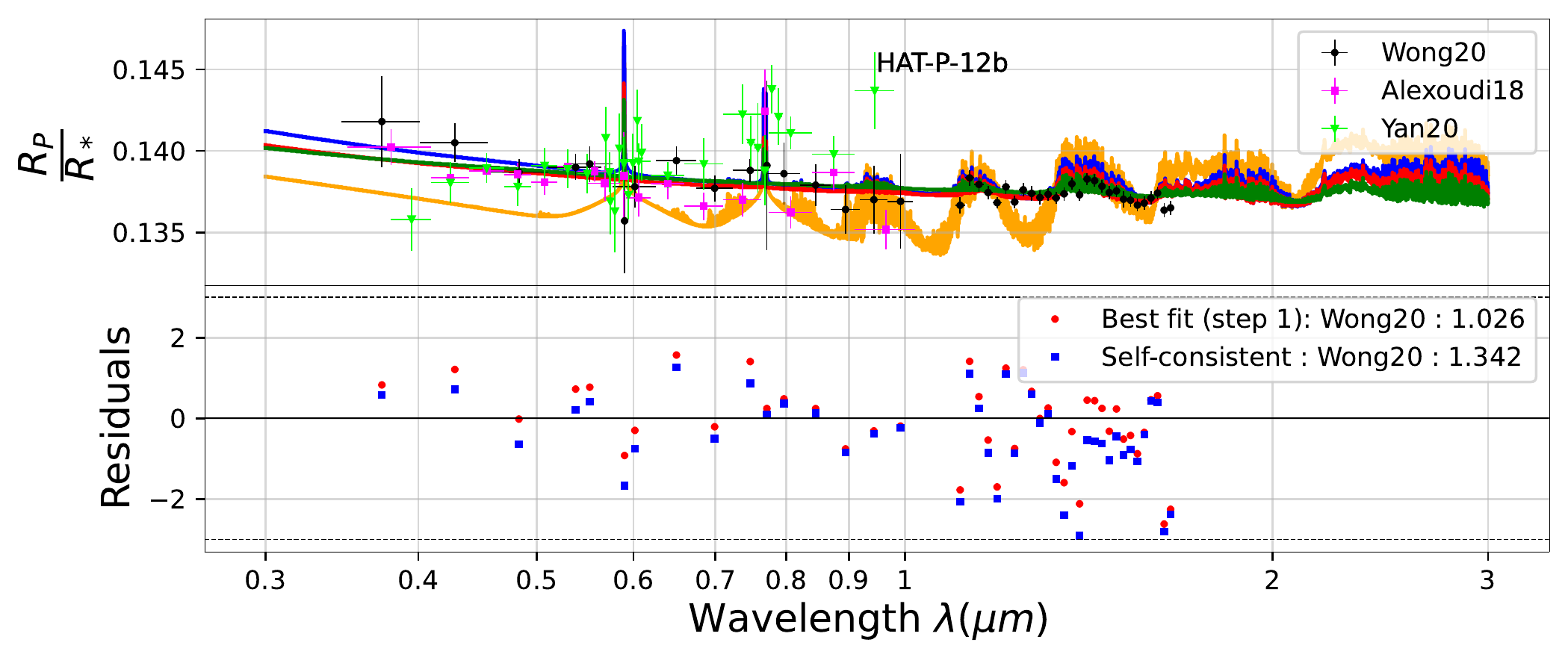}
\caption{HAT-P-12b: The green and red lines correspond to a 10$^{-14}g.cm^{-2}.s^{-1}$ mass flux with a constant and the nominal eddy profile, respectively.
The blue line is the self-consistent model with a 10$^{-14}g.cm^{-2}.s^{-1}$ mass flux and the nominal eddy.
The orange line is the haze-free model from step 1.}
\label{Fig:H12}
\end{figure*}

The water band observed is in agreement with the large amount of haze we retrieve. Although the observations and our model agree about the presence of sodium, our simulations are off the 3$\sigma$ limit of most observations near the Na core (\cref{Fig:189}). The large potassium line anticipated in our calculation does not match the observations either with residuals off the 3$\sigma$, albeit they actually agree with the presence of this specie. We note that the K observation has a low resolution and is mainly affected by the continuum, thus a higher resolution observation could improve the fit to the K line. A better fit of the UV slope would clearly help fitting these features, bringing the residuals within the 3$\sigma$ of the observations. 

We note that an offset of $\Delta R_P/R_* \sim - 0.0005$ in the data below 1 $\mu$m would bring the observations in a very good agreement with our self-consistent model that assumes a 0.1 $\mu$bar haze production altitude and a mass flux of 5.10$^{-12}g.cm^{-2}.s^{-1}$ ($\chi^2$ of 1.062; \cref{Fig:189Offset}).
This offset can be interpreted as a possible overestimation of the UV slope as demonstrated by \cite{McCullough14}, who discuss the potential implication of stellar activity in the observed spectrum.
This is consistent with \cite{Angerhausen15} observations with SOFIA (\cref{Fig:189Offset}) that presents slightly smaller transit depths in the visible due to a lower stellar activity.
This offset allows a good fit of the Na and K lines and keeps the water band in agreement with the WFC3 observations, though the UV slope still requires a large haze mass flux. 
The addition of an offset has to be implied by physical arguments and not to fit a model. Apart from stellar activity, the need for an offset may be related to  differences between the two instruments and the times of the observations. As the impact of these effects is not clear we still consider the steep UV-slope of HD-189733b unfitted, but we point out this suspicious fact.
 Given that HD-189733b is consider as an archetype of a hazy hot-Jupiter, new observations are certainly required to clarify the haze conditions of this atmosphere. 

A different assumption on the haze optical properties may also help to improve the fit of the HD-189733b observations.
The haze refractive index is poorly known and we use a soot refractive index for our study. Although a tholin type composition is not anticipated for the hot-Jupiter conditions \citep{Lavvas17}, we tested how the use of such a refractive index would affect our conclusions for this planet.
Using a tholin refractive index produces smaller transit depth than with the soot index for the same haze production (respectively orange and green line in \cref{Fig:HazeIndex}).
A three time larger value brings the model closer to the observations, though we note that the slope obtained is steeper than the observed one. Indeed, with a 3x10$^{-11}g.cm^{-2}.s^{-1}$ mass flux, our theoretical spectrum underestimates the transit depth between 0.6 and 1 $\mu$m and overestimates the transit depth below 0.6 $\mu$m.
Moreover, the haze precursors photolysis provides a mass flux of 7.2x10$^{-11}g.cm^{-2}.s^{-1}$, corresponding to a 43\% yield.
This yield is significantly larger from the evaluated yields of all other planets in the study, making this scenario suspicious.
Thus, we do not consider this scenario feasible. 
We highlight however the need for laboratory constraints for the refractive index at high temperature exoplanet conditions.

Silicate and iron clouds are expected to form in the deep atmosphere of HD-189733b according to the \cite{Sing16,Moses11} and our temperature profiles (\cref{Fig:FinalPTprofiles}). Similarly, MnS clouds may form between 100 and 1 mbar. However, they are not anticipated to impact the spectrum considering that the pressures probed by the observations are above the 0.1 mbar level (\cref{Tab:ProbedPressure}).

\subsection{HAT-P-12b}

\bf Step 1: \rm
For HAT-P-12b, the \cite{Sing16} data show a steep UV-visible slope that suggests a hazy atmosphere (not shown).
However, \cite{Alexoudi18} reanalyzed these data and found a weaker slope.
Recently, \cite{Wong20} and \cite{Yan20} have conducted HST and VLT observations, respectively, and have found similar results {to \cite{Alexoudi18}}.
We thus decided to use the \cite{Wong20} observations for the residuals and $\chi^2$ computation as they cover the largest wavelength range.
\cite{Jiang21} found strong evidence for stellar spots and faculae contamination of the spectra against the presence of hazes.
However, the muted water band found by \cite{Wong20} is in disagreement with a haze-free model (orange line in \cref{Fig:H12}) suggesting the presence of haze or clouds.
A small haze mass flux of 10$^{-14}g.cm^{-2}.s^{-1}$ provides a good fit of the water band according to our step 1 simulations.
The sensitivity analysis on the eddy profile provides as best fit a low value of 10$^7 cm^2.s^{-1}$.
However, this results in very minor changes on the transit spectra compared to the nominal eddy case with a $\chi^2$ of 0.987 for the constant eddy against 1.026 for the nominal (respectively green and red lines in \cref{Fig:H12}).
Such a low vertical mixing is expected to allow for relatively large particles at the probed pressure, albeit the average particle radius we calculate for this planet lies around 20 nm (\cref{Tab:Results}) as a result of the small haze mass flux required to fit the transit depth of HAT-P-12b.

\bf Step 3: \rm
The larger upper atmosphere temperatures resulting from particle heating and disequilibrium chemistry lead to a slightly enhanced UV slope (blue line in \cref{Fig:H12}) thus providing an even better fit of \cite{Wong20} data in the UV-visible range. However, the larger residuals obtained in the water band, although still within the 3$\sigma$ uncertainties of the observations, lead to a slightly larger $\chi^2$ of 1.342 against 1.026 for step 1. The large haze precursors mass flux of 2.4x10$^{-11}g.cm^{-2}.s^{-1}$ calculated is consistent with the haze mass flux we retrieve corresponding to a formation yield of 0.04\%.

On the other hand, if \cite{Alexoudi18} and \cite{Wong20} observations seem to fall in agreement about the presence of potassium, they provide no clear evidence for the sodium feature. \cite{Yan20} concludes to the absence of Na and K. However, the large error bars of these observations prevent to definitely rule out the presence of this element.
Indeed, our model, although presents a sodium feature, lies within the 3$\sigma$ uncertainties of the observations in this range.
Therefore, higher precision observations are required to clarify the presence of these elements.

Below 10 mbar, \cite{Sing16} and our temperature profiles are very similar and in agreement with the presence of silicate and MnS clouds above 1000 bar, while iron clouds are  expected even deeper in the atmosphere (\cref{Fig:FinalPTprofiles}). Considering that the transit observations are probing above the 2 mbar level (\cref{Tab:ProbedPressure}), we do not expect these clouds to extend to the probed region. Therefore we do not anticipate substantial impact of clouds on the observed spectrum.

\begin{figure*}
\includegraphics[width=\textwidth]{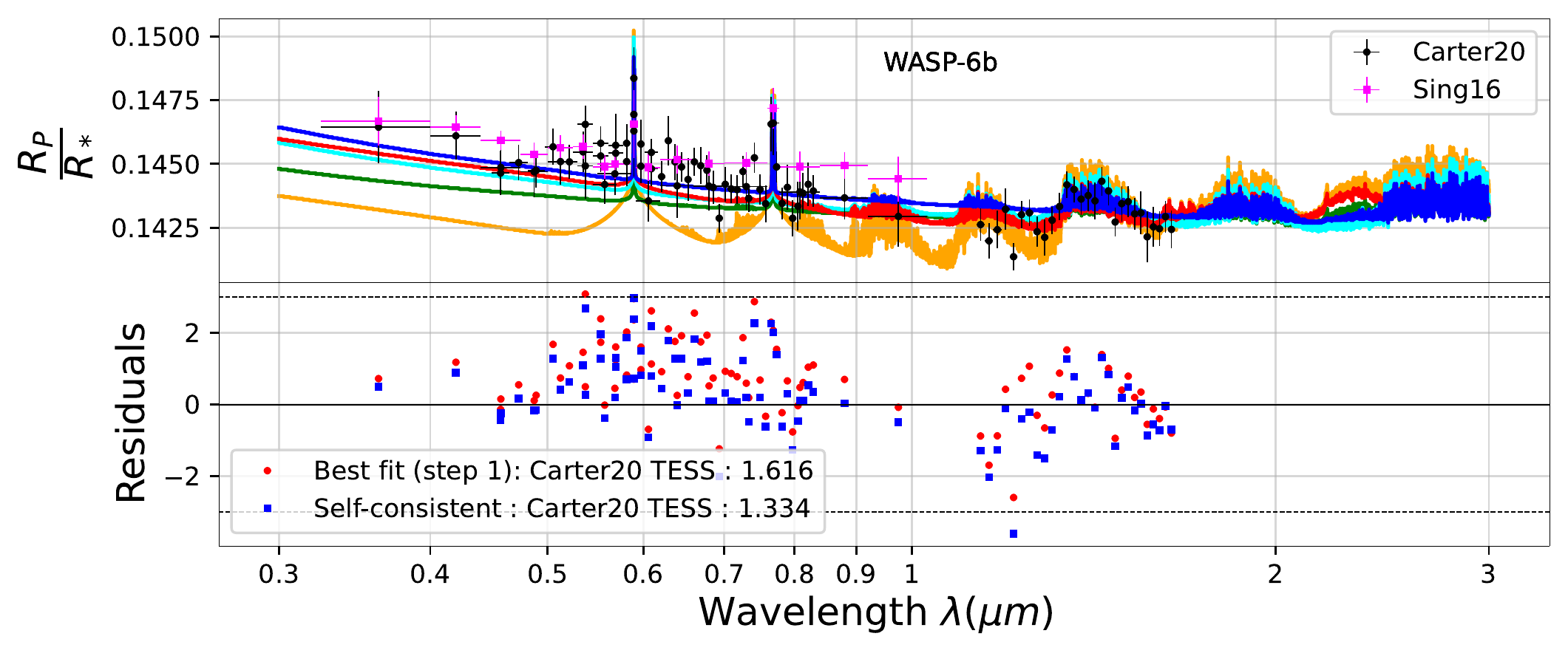}
\caption{WASP-6b: 
The orange and green lines correspond to a haze-free and a 10$^{-14}g.cm^{-2}.s^{-1}$ mass flux, respectively, with the nominal eddy and a solar metallicity.
The red line corresponds to a 10$^{-14}g.cm^{-2}.s^{-1}$ mass flux with a constant eddy profile and a sub-solar metallicity. 
The cyan and blue lines are the self-consistent model with a 10$^{-14}g.cm^{-2}.s^{-1}$ mass flux, the nominal eddy and with solar and sub-solar metallicities, respectively.}
\label{Fig:W6}
\end{figure*}

\subsection{WASP-6b}

\bf Step 1: \rm
For this planet, \cite{Sing16} do not provide water band observations that are required for our pressure referencing. We thus use data from \cite{Carter20} that observed WASP-6 with STIS, WFC3 and IRAC. Three sets of data are provided: uncorrected, star spot corrected with AIT and with TESS. The latter being based on more precise space-born photometric observations, we consider it to be more reliable, although these observations are not contemporaneous with HST's. We decided to use this one for both altitude mapping and $\chi^2$ computation.

Although the water band features are in good agreement with our haze-free model ($\chi^2$=6.102, orange line in \cref{Fig:W6}), the latter underestimates the steep UV slope.
It would therefore require the presence of haze, albeit none of our attempts with solar metallicity were able to correctly fit the UV range, providing more shallow slopes than the observed one, as demonstrated by our 10$^{-14}g.cm^{-2}.s^{-1}$ mass flux case ($\chi^2$=2.923, green line in \cref{Fig:W6}). On the other hand, a sub-solar metallicity can help increase the UV slope due to changes in the pressure referencing.
Using a $0.1\times$solar metallicity gives a best-fitting mass flux of 10$^{-14}g.cm^{-2}.s^{-1}$ with a constant eddy of 10$^8 cm^2.s^{-1}$ ($\chi^2$=1.616, red line in \cref{Fig:W6}). This constant eddy provides a steeper slope than the nominal profile ($\chi^2$=2.489, not shown).

\bf Step 3: \rm
The inclusion of haze radiative feedback through our self-consistent model, with sub-solar metallicity and the nominal eddy, leads to a steeper UV slope (blue line in \cref{Fig:W6}), resulting in a better fit of the observations with a $\chi^2$ of 1.334 for a haze mass flux of 10$^{-14}g.cm^{-2}.s^{-1}$. The solar metallicity case produces a smaller transit depth in the UV-visible compared to the sub-solar case for the same haze mass flux ($\chi^2$=1.853, cyan line in \cref{Fig:W6}). 
This reflects the changes in the reference pressure taken in the water band. As the water abundance increases, the observations are probing higher up in the atmosphere in the infrared, while in the UV-visible range, they are probing the same pressures. Therefore, the lower reference pressure obtained in the water band results in a shift of the UV-visible spectrum toward smaller transit depths, to keep the water band consistent with WFC3 observations.
The rather small haze mass flux retrieved using our sub-solar self-consistent calculation is supported by photochemistry that provides a haze precursors photolysis mass flux of 2.6x10$^{-13}g.cm^{-2}.s^{-1}$, which corresponds to a formation yield of 3.8\%. \cite{Sing16} and \cite{Carter20} observations are in agreement with our model about the presence of sodium and potassium.

Our simulated temperature profiles for solar and sub-solar metallicities present minor differences, while they are consistent with the \cite{Sing16} profile at pressures between $\sim$100 and 0.1 bar (\cref{Fig:FinalPTprofiles}). At higher pressures, modifications are due to different intrinsic temperature assumptions, while at lower pressures differences arise from the disequilibrium chemistry composition and haze heating. Differences between the two simulated metallicity cases arise due to reduced heating and cooling from the lower abundances of radiatively active species (dominated by H$_2$O and CO) in the sub-solar case. For all temperature profiles silicates and iron condense deep in the atmosphere ($\sim$100 bar), while the transit observations probe above 3 mbar (\cref{Tab:ProbedPressure}). Thus, such clouds are not likely to affect the observed spectrum. MnS clouds may form from 1 bar to 10 mbar, right below the region probed by observations in the water band, but should sublimate at higher altitudes.

\begin{figure*}
\includegraphics[width=\textwidth]{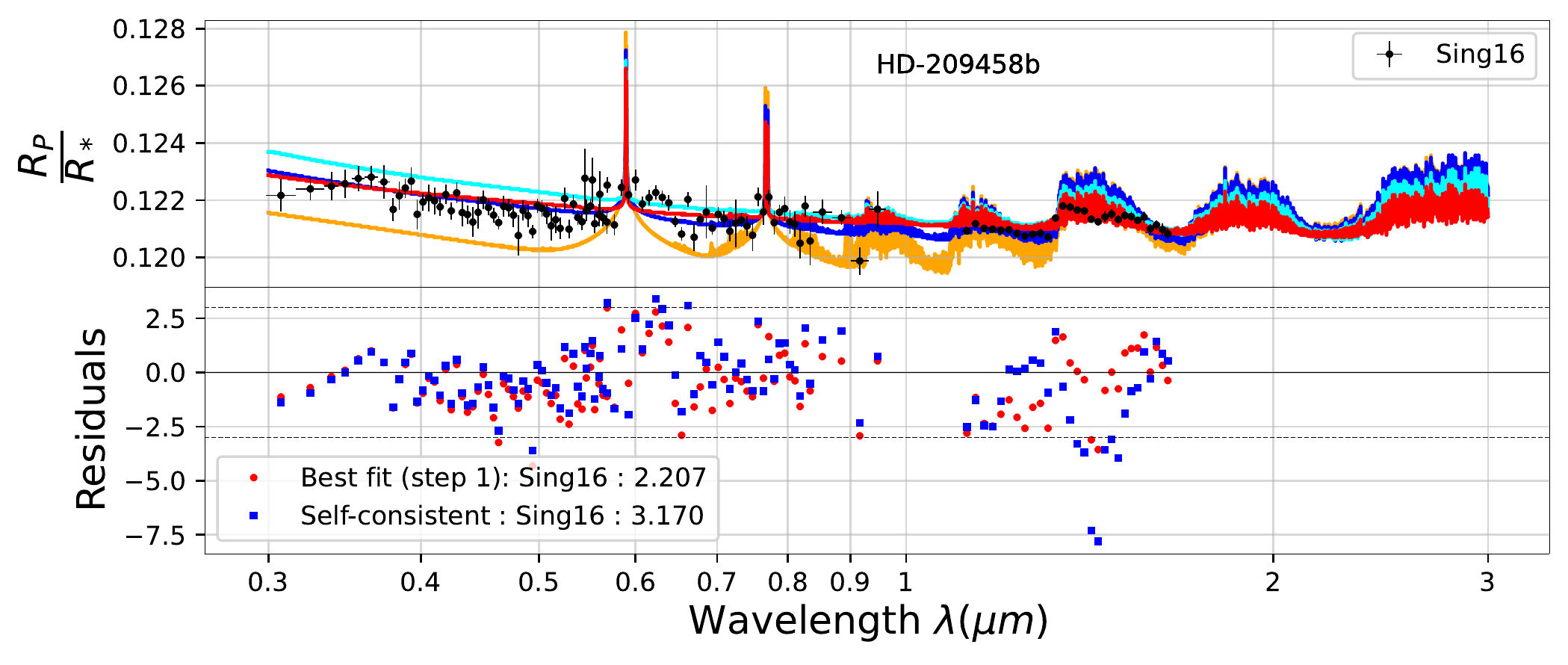}
\caption{HD-209458b: The red and cyan lines are the 10$^{-14}g.cm^{-2}.s^{-1}$ mass flux case from step 1 and step 3, respectively. The blue line is the 5x10$^{-15} g.cm^{-2}.s^{-1}$ mass flux case tried with the self-consistent model. The orange line is the haze-free model from step 1.}	\label{Fig:209}
\end{figure*}

\subsection{HD-209458b}

\bf Step 1: \rm
This planet is well-fitted by a 10$^{-14}g.cm^{-2}.s^{-1}$ mass flux and the nominal eddy profile ($\chi^2$=2.207, red line in \cref{Fig:209}), which provide a good fit to the WFC3 observations with the exception of the local minimum near 1.44 $\mu$m. A clear atmosphere case would provide a too low transit depth ($\chi^2$=10.94, orange line in \cref{Fig:209}).

\begin{figure*}
\includegraphics[width=\textwidth]{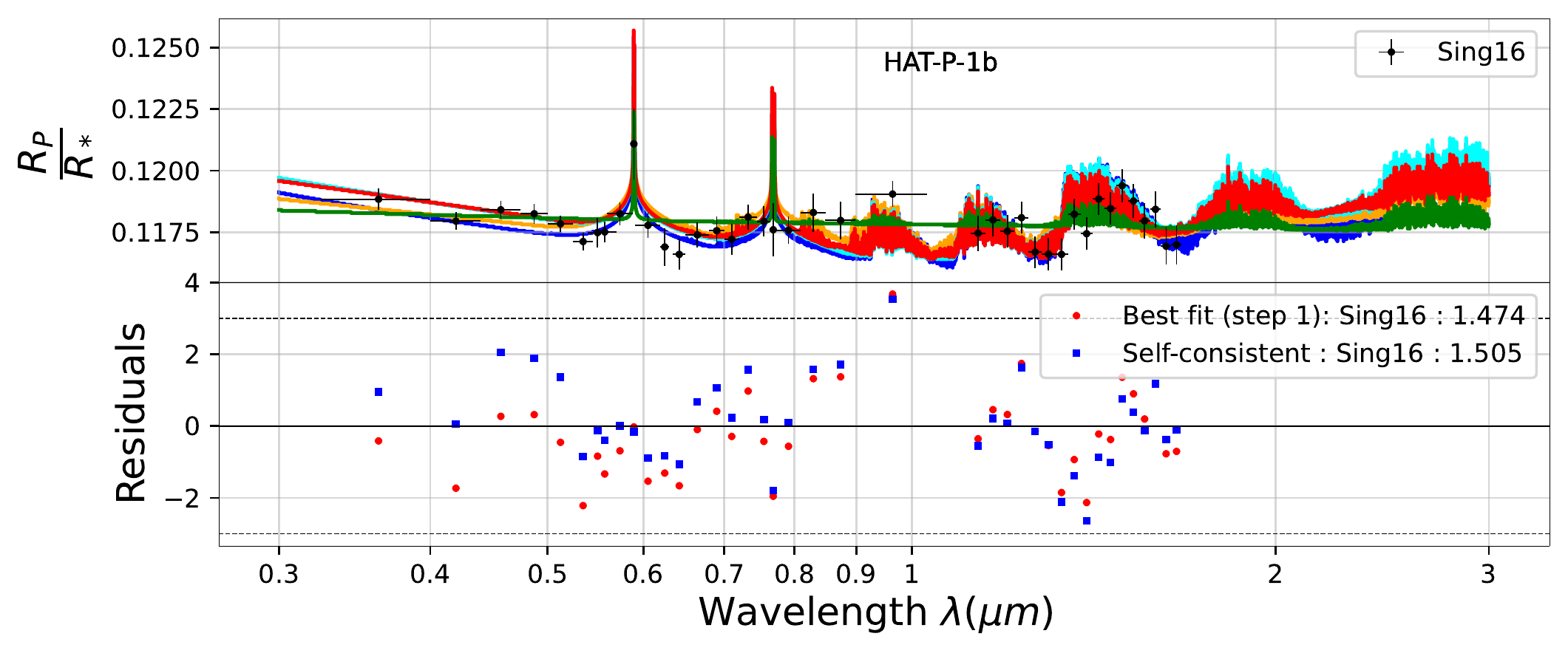}
\caption{The orange and green lines correspond to solar metallicity cases from step 1, for a haze-free atmosphere and for a hazy case with a 10$^{-12}g.cm^{-2}.s^{-1}$ mass flux, respectively. The blue line corresponds to a 10$^{-15}g.cm^{-2}.s^{-1}$ mass flux with solar metallicity from step 3. The red and cyan lines correspond to a 10$^{-16}g.cm^{-2}.s^{-1}$ mass flux with sub-solar metallicity from steps 1 and 3, respectively. }
\label{Fig:H1}
\end{figure*}

\bf Step 3: \rm
Our self-consistent temperature is slightly warmer in the upper atmosphere relative to the \cite{Sing16} profile resulting in an enhanced UV slope that worsens the fit in this wavelength range (cyan line in \cref{Fig:209}). We thus tested a lower haze mass flux case (5x10$^{-15} g.cm^2.s^{-1}$)  that provides a better fit (blue line in \cref{Fig:209}) and is in agreement with our previous evaluation \citep{Lavvas21}.
We must however note that due to haze sublimation, the effective mass flux obtained is $\sim$ 3.5x10$^{-15} g.cm^2.s^{-1}$.
Our calculated haze precursors mass flux of 1.6x10$^{-13} g.cm^2.s^{-1}$ is consistent with the effective mass flux and corresponds to a haze formation yield of 2.2\%.

The water band observations of HD-209458b are probing the pressure region from 0.2 mbar to 10 mbar (\cref{Tab:ProbedPressure}). Although silicate and iron clouds may be present in this region based on our calculated temperature profile, they should form deeper down, between 1000 and 10 bar for MgSiO$_3$ and even deeper for Mg$_2$SiO$_4$ and iron (\cref{Fig:FinalPTprofiles}), and their ability to expand up to the probed region remains to be assessed. We note that with \cite{Sing16} temperature profile, which is hotter than our profile in the deep atmosphere, mainly related to a different assumption on the intrinsic temperature, iron clouds can form only above the 10 bar level and silicate clouds above the 0.1 bar altitude. Such conditions may make the hypothesis of clouds expanding in the probed region more likely.

We may further note that the observed variations in the UV-visible are not reproduced by any of our simulations. An explanation for these features may arise from stellar activity, however, HD-209458 is not known to be a very active star. The features may also relate to metal absorption \citep{Lothringer20}, but such elements are expected to be lost in the cloud formation. On the other hand, our simulated temperature profile for the dayside conditions \citep{Lavvas21} suggest that such condensates may locally sublimate in the upper atmosphere if they can arrive there. Therefore variations along the line of sight during transit may be responsible for the observed features. A more detailed treatment of temperature and cloud microphysics would be required for a satisfactory interpretation of these observations. We note on the other hand, that the large error bars in the visible keep our best-fitting haze model almost within the 3$\sigma$ uncertainties of the observations (\cref{Fig:209}).

\subsection{HAT-P-1b}
\label{SSec:H1}

\bf Step 1: \rm
For HAT-P-1b, every solar metallicity model tested provides a spectrum within the 3$\sigma$ uncertainties of the observations, causing degeneracies in the retrieved parameters.
This behaviour can be explained by the large error bars and the shape of the spectra, with the flat hazy curves crossing through the haze-poor models.
Although the $\chi^2$ minimization is in favour of a large haze mass flux with a minimum $\chi^2$ of 1.314 obtained for the 10$^{-12} g.cm^{-2}.s^{-1}$ mass flux (green line in \cref{Fig:H1}), the strong water band as well as the sodium and potassium wings are consistent with the clear atmosphere model (orange line in \cref{Fig:H1}; $\chi^2$=1.478).
Moreover, the best-fitting hazy model demonstrates a UV slope that is flatter than the H$_2$ Rayleigh scattering slope, with the latter fitting better the observations.
We therefore consider that a clear atmosphere or a small amount of haze are the most likely solutions, with an upper limit on the haze mass flux at 10$^{-15}g.cm^{-2}.s^{-1}$.

On the other hand, the observed Na line wings seem more narrow compared to our theoretical clear atmosphere model that may indicate a lower atmospheric metallicity. We therefore attempted to evaluate a 0.1$\times$solar metallicity case through our step 1 minimization.
The best fit produced with this low-metallicity case suggests a haze mass-flux of 10$^{-16} g.cm^{-2}.s^{-1}$ ($\chi^2$=1.474, red line in \cref{Fig:H1}), though it is not significantly different from the sub-solar haze-free model ($\chi^2$=1.487, not shown). Thus, we consider this haze mass flux as the upper limit for the sub-solar metallicity case.

\bf Step 3: \rm
We test both solar and sub-solar metallicity cases in our self-consistent model using the corresponding upper limits on the haze mass fluxes derived in step 1. Our resulting temperature profiles suggest small temperature differences between the two metallicity cases (\cref{Fig:FinalPTprofiles}). However, our simulated temperatures are slightly warmer than the \cite{Sing16} profile above 1 bar and cooler below due to differences in the intrinsic temperature and the gas composition. 

Assuming the sub-solar metallicity with a mass flux of 10$^{-16}g.cm^{-2}.s^{-1}$, the photochemistry calculation partially works out the degeneracies as our simulated precursor photolysis rates provide a maximum haze mass flux of 1.4x10$^{-13} g.cm^{-2}.s^{-1}$, corresponding to a haze formation yield of 0.07$\%$. The simulated transit spectrum is consistent with the observed UV slope and the Na wings ($\chi^2$=1.535, cyan line in \cref{Fig:H1}). However, we note that our model slightly overestimates the Na wings, though the residuals are within 3$\sigma$ of the observations.

The solar metallicity case with a mass flux of 10$^{-15}g.cm^{-2}.s^{-1}$ (producing a smaller effective mass flux of 7.5x10$^{-16}g.cm^{-2}.s^{-1}$ due to sublimation) also provides a slightly better fit of the UV slope and the Na and K lines ($\chi^2$=1.505, blue line in \cref{Fig:H1}), with slightly weaker transit depths than the sub-solar case. As for WASP-6b, this is related to the lower reference pressure used for the solar case that shifts the UV-visible to smaller transit depths. A solar metallicity is therefore more likely considering the self-consistent results. The solar metallicity case provides a larger haze precursors photolysis mass flux of 2.9x10$^{-12}g.cm^{-2}.s^{-1}$ (\cref{Tab:PhotoFlux}), which would correspond to a small yield of 0.03\%. This larger haze precursors mass flux is mainly related to CO that presents larger upper atmosphere abundances, resulting in a CO photolysis mass flux fifty times larger compared to the sub-solar case. 

The solar metallicity we assume produces a good fit of the sodium line. However, Na and K lines are in disagreement: while a strong sodium core is observed, potassium is not observed in the probed region. These two elements are expected to present the same behavior: the presence of one involving the presence of the other \citep{Lavvas14}.
As temperature decreases, sodium is supposed to condense first, thus condensation can not explain such discrepancies. K could have been depleted, for example to K$^+$ via photoionisation and thermal ionisation. Indeed, our disequilibrium model shows that photoionization occurs above the 0.1 mbar level for Na and K, which leads to slightly weaker lines, but does not solve the discrepancy. We further note that, despite these important differences between the observed Na and K lines, the model is within the 3$\sigma$ of the observations at the two lines. More precise measurements are therefore required to address this issue. The observed strong water band is in agreement with our small haze mass flux everywhere in the spectrum except one location near 1 $\mu m$ that produces a residual off the 3$\sigma$ limit. We however consider this discrepancy not significant since it involves one single point at the edge of the detector.

\begin{figure*}
{\includegraphics[width=\textwidth]{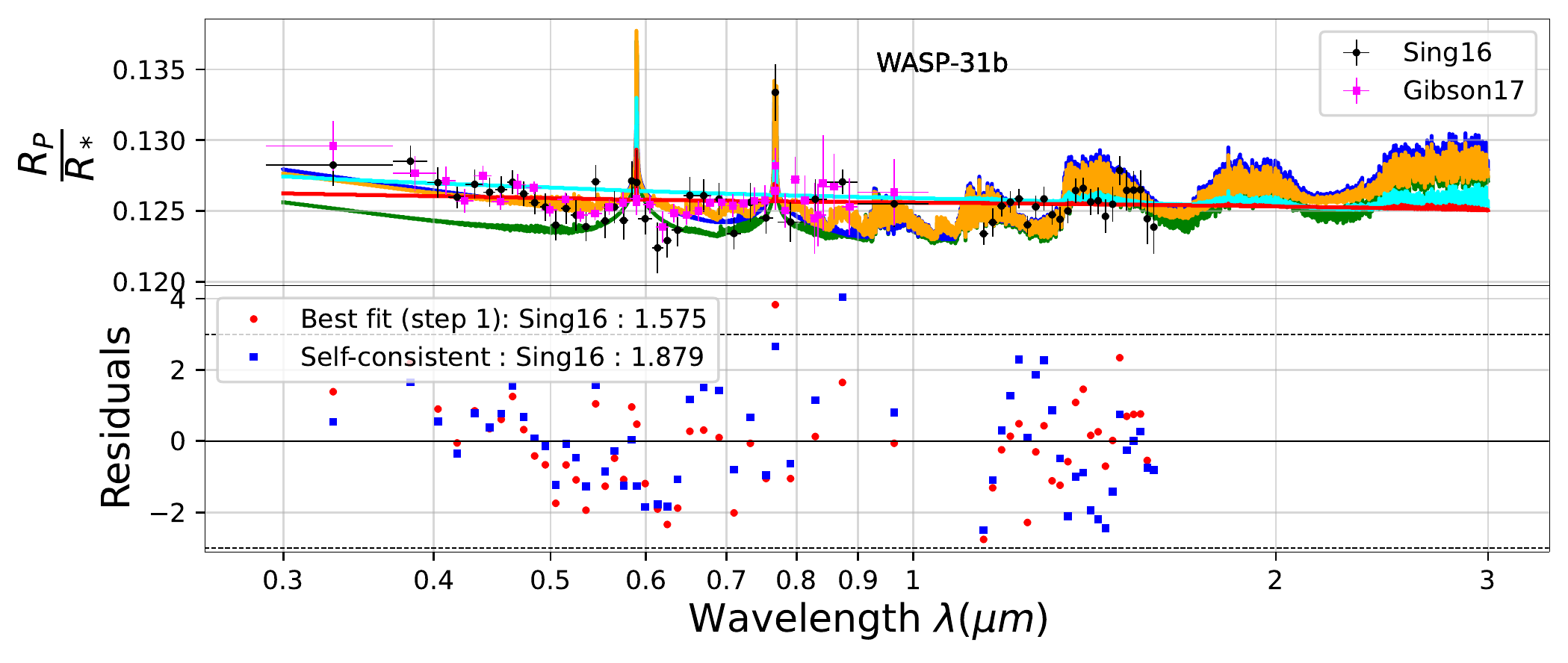}}
\caption{WASP-31b: The red line is the 10$^{-12}g.cm^{-2}.s^{-1}$ mass flux case from step 1. The orange and green lines are the haze-free 0.1xsolar and solar metallicity cases from step 1, respectively. The blue line is the 10$^{-16}g.cm^{-2}.s^{-1}$ with 0.1xsolar metallicity from step 3.
The cyan line is the 10$^{-12}g.cm^{-2}.s^{-1}$ mass flux case assuming a solar metallicity from step 3, though it corresponds to a smaller effective mass flux (see text).}
\label{Fig:W31}
\end{figure*}

\begin{figure}
	{\includegraphics[width=0.5\textwidth]{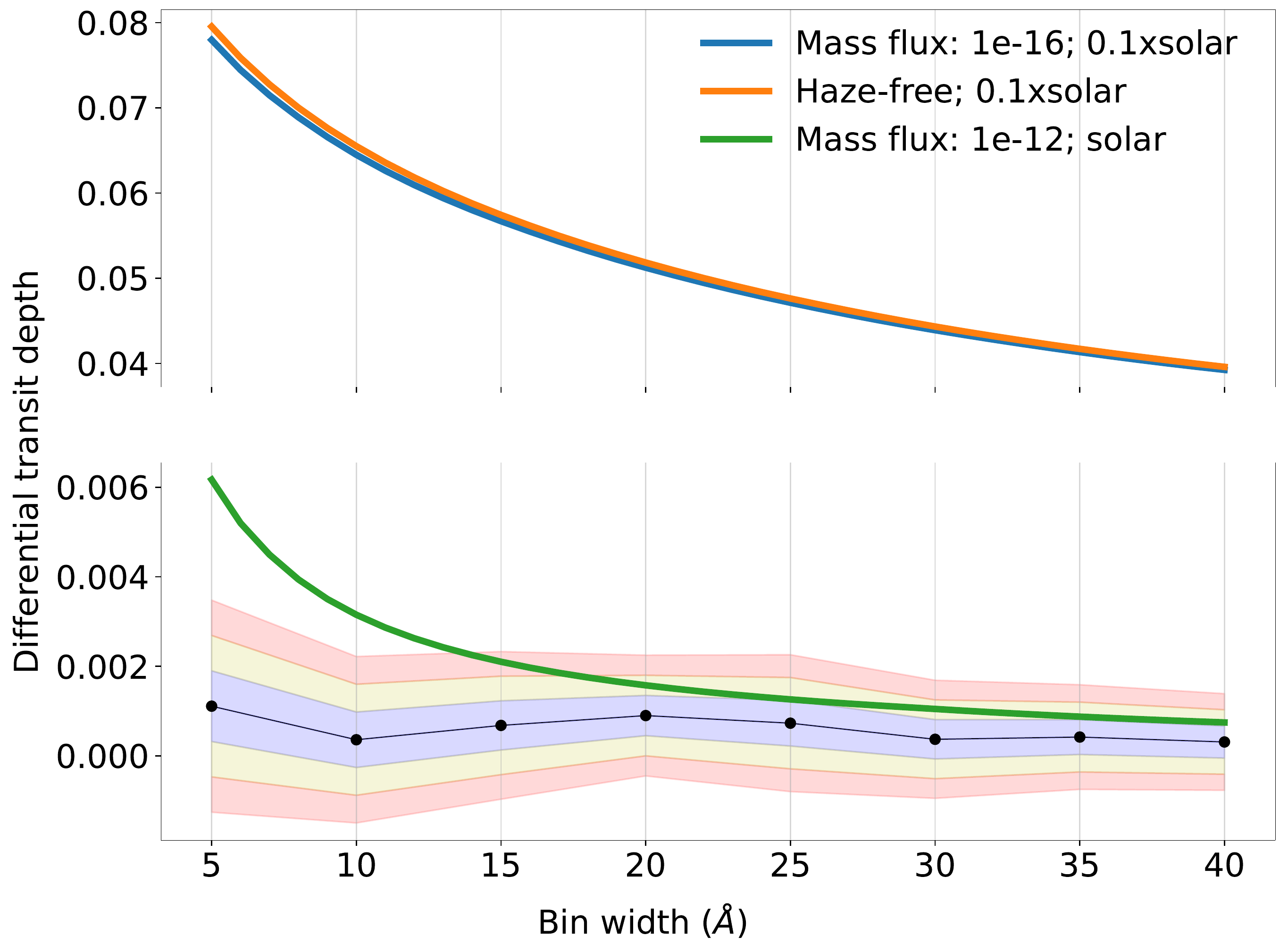}}
	\caption{Differential transit depth of WASP-31b K line from our self-consistent cases for different haze mass fluxes and metallicities.
	Black points are observations from \protect\cite{Gibson19}.
	The blue, yellow and red areas are the 1, 2 and 3 $\sigma$ limits of the observations, respectively.}
	\label{Fig:W31DTD}
\end{figure}

Our simulated temperature profiles for both metallicities suggest that silicate and iron condensates form in the deep atmosphere of HAT-P-1b (\cref{Fig:FinalPTprofiles}). For the solar-case, the temperature profile approaches asymptotically the condensation curve of Mg$_2$SiO$_4$, but does not cross this limit. Considering the probed pressures (pressures below 0.01 bar; \cref{Tab:ProbedPressure}), the impact of silicate and iron clouds depends on their stratification and how efficiently atmospheric mixing would allow cloud particles formed in the deep atmosphere to loft at such low pressures. The large pressure difference between the cloud formation region and transit probed region makes this scenario unlikely. MnS cloud formation is possible for the sub-solar metallicity case locally at $\sim$100 mbar and above 0.1 mbar. In the lower region particles should sublimate at pressures above the 10 mbar level thus should not affect the transit slope that probes pressures lower than 10 mbar, while cloud formation in the upper atmosphere will be limited by their potential formation in the deeper atmosphere. For either case, cloud contributions appear unlikely to affect the observations.

\subsection{WASP-31b}

\bf Step 1: \rm
For WASP-31b, we obtain the best fit in step 1 with a mass flux of 10$^{-12}g.cm^{-2}.s^{-1}$ assuming the nominal eddy profile ($\chi^2$=1.575; red line in \cref{Fig:W31}). The clear atmosphere model provides a $\chi^2$ of 3.162 (green line in \cref{Fig:W31}) that underestimates the atmospheric opacity at wavelengths below 0.5$\mu$m. We also evaluated a 0.1$\times$solar metallicity case for this planet that provides a better haze-free model ($\chi^2$=1.659, orange line in \cref{Fig:W31}). 
However, as for HAT-P-1b, WASP-31b presents degeneracies in the fitting parameters with all cases being within the 3$\sigma$ uncertainties of the observations, for both haze-free (with sub-solar metallicity) and very hazy (with solar metallicity) models. These degeneracies are inherent to the shape of the spectra and the large error bars and would be solved with more precise observations.

\begin{figure*}
{\includegraphics[width=\textwidth]{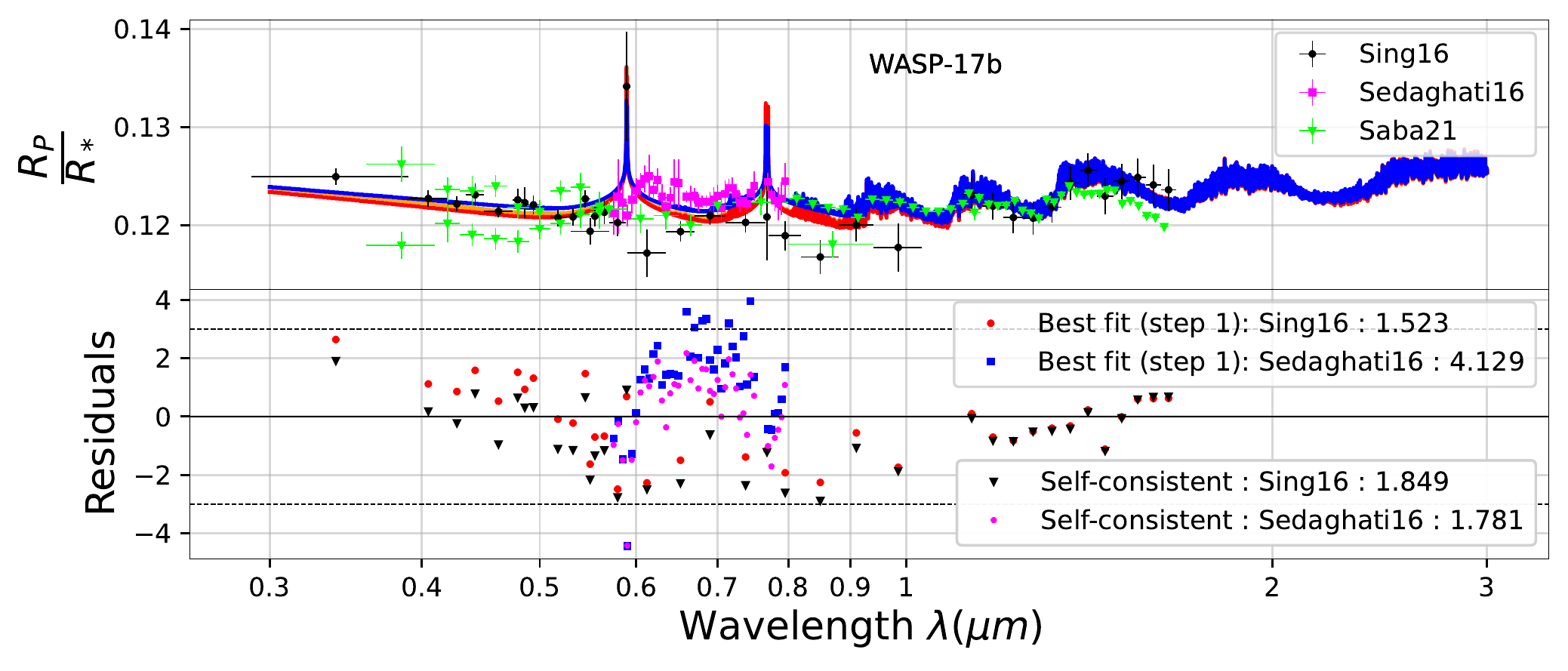}}
\caption{WASP-17b: Haze free spectra from step 1 (red line) and step 3 (blue line). The orange line is calculated with a small 10$^{-16}g.cm^{-2}.s^{-1}$ mass flux from step 1.
The red circles and black triangles are the residuals calculated based on the \protect\cite{Sing16} data for step 1 and 3 results, respectively.
The blue squares and pink dots are the residuals calculated based on the \protect\cite{Sedaghati16} data for step 1 and 3 results, respectively.}
\label{Fig:W17}
\end{figure*}

\bf Step 3: \rm
Considering the degeneracies observed in step 1, we tested both hazy solar and haze-free sub-solar metallicity cases with our self-consistent model.
The hazy solar case calculation for WASP-31b provides an homopause near 1$\mu$bar that is inside the assumed haze production region (\cref{Fig:FinalPTprofiles}).
As a consequence, the newly formed particles sublimate resulting in a clear atmosphere.
A deeper haze production altitude at 10 $\mu bar$ might allow the particles to survive due to the lower local temperatures. 
However, the strong heating by haze related to the large UV flux received by this planet leads to high particle temperatures in the upper atmosphere, resulting in a partial sublimation of the haze.
As a result, although we fix the haze mass flux at 10$^{-12} g.cm^{-2}.s^{-1}$, we obtain a much smaller effective haze mass flux of $\sim$ 7x10$^{-14} g.cm^{-2}.s^{-1}$.
The theoretical transit spectrum obtained with these parameters provides a small $\chi^2$ of 1.929 (cyan line in \cref{Fig:W31}).

The haze-free sub-solar metallicity case (not shown) provides a $\chi^2$ similar to the solar metallicity hazy case with a value of 1.927. This haze-free model is in better agreement with the observed UV-slope and Na and K features in the \cite{Sing16} data, though we note that the spectral variations seen in the observations do not allow for a clear conclusion. Specifically, some points in the observed transit depth between the two alkali lines seem consistent with the hazy case, while the near-IR region is quite difficult to assess as some variations seem to not follow the water band behavior. Moreover, both the hazy solar case with a strong muting of the water band and the haze-free sub-solar model with large water absorption features are within the 3$\sigma$ of the observations. This demonstrates the degeneracy for this planet with the two extreme models presenting the same $\chi^2$.
A small mass flux of 10$^{-16} g.cm^{-2}.s^{-1}$ for the sub-solar case further decreases the residuals ($\chi^2$  = 1.879; blue line in \cref{Fig:W31}) and provides a good fit of the different spectral features observed, though we note that, owing to sublimation, this case corresponds to an effective mass flux of 2x10$^{-17} g.cm^{-2}.s^{-1}$.
This case produces a haze precursors mass flux of 1.9x10$^{-13} g.cm^{-2}.s^{-1}$, corresponding to a 0.01\% formation yield.

On the other hand, the lack of potassium and sodium found by \cite{Gibson19} is in better agreement with a strong haze mass flux.
They used high resolution spectroscopy to construct the differential transit depth (defined as the relative difference between the integrated transit depth around a line center and in a surrounding region) of the potassium line and found no evidences for the presence of this element. We reproduce their calculations with our self-consistent spectra by computing the differential transit depth of the K line at different bin widths ranging from 5 to 40 \AA{}.
Our haze-free sub-solar metallicity case is in disagreement with \cite{Gibson19} observations (orange line in \cref{Fig:W31DTD}), while the same applies to the best fit sub-solar case found earlier with the 10$^{-16} g.cm^{-2}.s^{-1}$ mass flux (blue line in \cref{Fig:W31DTD}). We would need to assume a higher haze mass flux to bring the sub-solar model in agreement with the \cite{Gibson19} observation although this would result in particle sublimation in the production region as discussed above.
For the record, the hazy solar case is shown in green line in \cref{Fig:W31DTD} and is within the 3$\sigma$ limit for bins above 15 \AA{}.

Our simulated temperature profiles for both metallicity cases, as well as, the \cite{Sing16} temperature profile (\cref{Fig:FinalPTprofiles}), suggest that silicate clouds could form and affect the atmospheric region probed by the observations, i.e. above 0.1 (0.3) bar for a clear solar (sub-solar) atmosphere (\cref{Tab:ProbedPressure}). The contribution of iron in cloud formation depends on the intrinsic temperature assumed, with our calculations providing a formation deep in the atmosphere, while the higher intrinsic temperature used in \cite{Sing16} allows for iron cloud formation to start near 0.1 bar. Therefore we conclude that clouds could be affecting the transit spectrum of WASP-31b.

Based on the fit of the UV-slope, the water band behavior and the too large temperatures produced by haze heating, we conclude that photochemical hazes are not the most likely solution for this planet. The role of clouds and metallicity need to be further evaluated in the future. We further suggest that additional observations are required to accurately assess the atmospheric conditions of WASP-31b.

\subsection{WASP-17b}

\bf Step 1: \rm
The water band observed by \cite{Sing16} is in agreement with our haze-free model (red line in \cref{Fig:W17}) and suggests the absence of haze.
However, our haze-free model is slightly off in the visible based on STIS data, from 0.6 to 1 $\mu m$, suggesting our model overestimates the opacity in this range.
Additional observations by \cite{Sedaghati16} are in agreement with larger transit depths in the visible compared to the \cite{Sing16} data, standing for the need of accounting for an offset.
\cite{Saba21} reanalized the STIS data from 3 HST visits (the three datasets are combined in \cref{Fig:W17}) and used new WFC3 observations down to 800 nm covering the 1$\mu m$ observation by STIS.
The comparison of these two datasets at this wavelength shows a difference of 0.001 in transit depth.

Despite this discrepancy between \cite{Sing16} observations and our haze-free model, the latter remains almost within the 3$\sigma$ uncertainties of the observations (red circles in \cref{Fig:W17}) and provides an acceptable match of the shallow UV slope, indicating a rather clear atmosphere.
A low mass flux value of 10$^{-16}g.cm^{-2}.s^{-1}$ is also tested providing a slightly better fit with a $\chi^2$ of 1.394 against 1.523 for the haze-free model, although the differences in the spectra are negligible (orange line in \cref{Fig:W17}). We therefore consider this atmosphere as a low haze or haze-free case, in agreement with previous studies \citep{Sing16,Sedaghati16,Saba21}.

As for HAT-P-1b, the WASP-17b \cite{Sing16} observations present discrepancies between the Na and K cores.
While a strong Na core is observed, the core of the K line is at least weak, at most absent. The K wings appear to be present in the \cite{Sing16} observations, although offset to lower transit depth. These discrepancies seem to disappear with \cite{Sedaghati16} who do not show strong evidence for sodium and were not able to detect the core of the potassium feature, albeit they determined the presence of the K wings.  \cite{Saba21} were not able to detect the alkali lines either.

\bf Step 3: \rm
Our disequilibrium chemistry calculation is in agreement with the presence of both sodium and potassium.
The haze precursors photolysis provides a mass flux of 5.1x10$^{-13}g.cm^{-2}.s^{-1}$.
Considering the 10$^{-16}g.cm^{-2}.s^{-1}$ case as an upper limit for the haze mass flux possibly present in WASP-17b, we suggest a yield smaller than 0.02\%.

\begin{figure*}
\includegraphics[width=\textwidth]{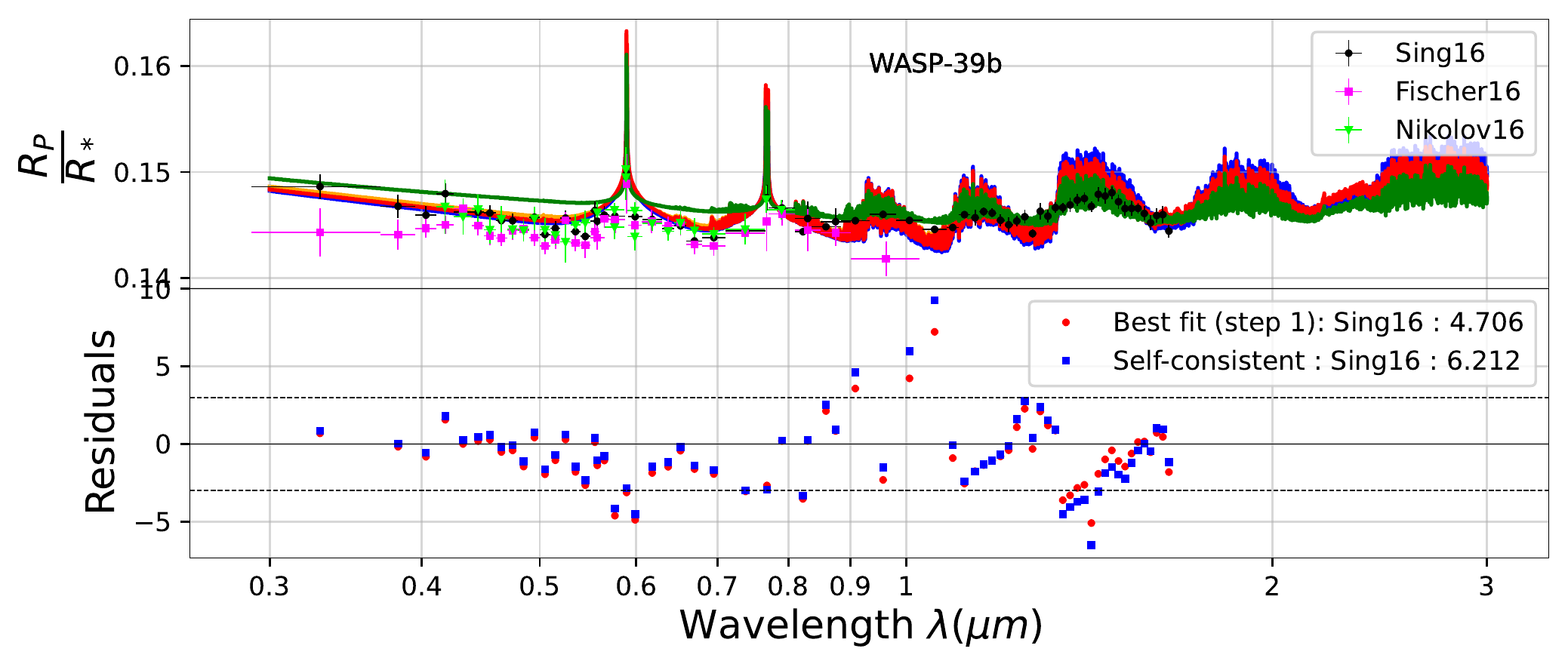}
\caption{WASP-39b: Haze free spectra from step 1 (red line) and step 3 (blue line).The orange and green lines correspond to 10$^{-16}g.cm^{-2}.s^{-1}$ and 10$^{-15}g.cm^{-2}.s^{-1}$, respectively.}
\label{Fig:W39}
\end{figure*}

The self-consistent calculation (blue line in \cref{Fig:W17}) shows larger transit depths in the visible range compared to our initial result from the first step, while the calculation provides colder temperatures than the \cite{Sing16} profile, which is expected to result in a collapse of the atmosphere.
This reflects the changes of reference pressure between the two cases.
Using the \cite{Sing16} temperature profile, the atmosphere expands increasing the length of the path traveled by the light through the atmosphere.
Therefore, the reference is shifted to lower pressures with the \cite{Sing16} profile compared to our profile.
As a result, the UV-visible range is shifted to smaller planet radius in our calculated spectrum compared to the step 1 to keep the water band in agreement with WFC3 observations.

Silicate and iron clouds are anticipated above 10$^{-2}$ bar based on the \cite{Sing16} temperature profile (\cref{Fig:FinalPTprofiles}), which would strongly impact the spectrum.
Especially, we can expect larger transit depths in the visible worsening the fit to \cite{Sing16} data, as well as, a strong muting of the water band that would be inconsistent with the WFC3 observations.
However, the globally colder temperatures presented by our profile allow the presence of silicate clouds deeper, between 1000 to 1 bar.
This is likely to result in a depletion of the necessary materials from the above atmosphere, and to prevent the formation of clouds in the region probed by the observations from 0.2 bar to 0.2 mbar (\cref{Tab:ProbedPressure}).

\subsection{WASP-39b}

\bf Step 1: \rm
Although WASP-39b is best-fitted by our haze-free theoretical model ($\chi^2$=4.706; red line in \cref{Fig:W39}), the observed water band is more shallow than our theoretical spectrum, meaning we are missing a source of opacity hiding the water lines. A small amount of haze could explain such a behavior, however, a haze mass flux of 10$^{-15}g.cm^{-2}.s^{-1}$ would be required to bring the water band in good agreement with the observations but such a value results in a bad fit of the UV slope ($\chi^2$=6.381; green line in \cref{Fig:W39}). A lower value of 10$^{-16}g.cm^{-2}.s^{-1}$ provides a smaller $\chi^2$ of 3.947 (orange line in \cref{Fig:W39}), albeit this case does not present substantial modification in the spectrum compared to the haze-free model.

We present additional datasets from \cite{Nikolov16} and \cite{Fischer16} (\cref{Fig:W39}). The former is in good agreement with the \cite{Sing16} data with a UV slope consistent with our haze-free model, while the latter presents a rather flat global UV slope. These discrepancies disappear in the visible range where both datasets are consistent with the \cite{Sing16} observations. Our model, is consistent with the presence of sodium and potassium in the atmosphere of this planet, although the Na and K lines near the cores present residuals off the 3$\sigma$ uncertainties of the \cite{Sing16} observations (\cref{Fig:W39}). We have tested a 0.1xsolar metallicity case to decrease the Na and K opacities, however, this drop in metallicity led to an increase of the UV slope, and thus to an enhanced transit depth signature worsening the residuals ($\chi^2$=6.666 for the haze-free sub-solar against $\chi^2$=4.706 for the haze-free solar). The \cite{Fischer16} observation near 1$\mu m$ is probing deeper in the atmosphere and is inconsistent with our model and with the water band observed by \cite{Wakeford18}.

\begin{figure*}
\begin{subfigure}{0.85\textwidth}{{\includegraphics[width=\textwidth]{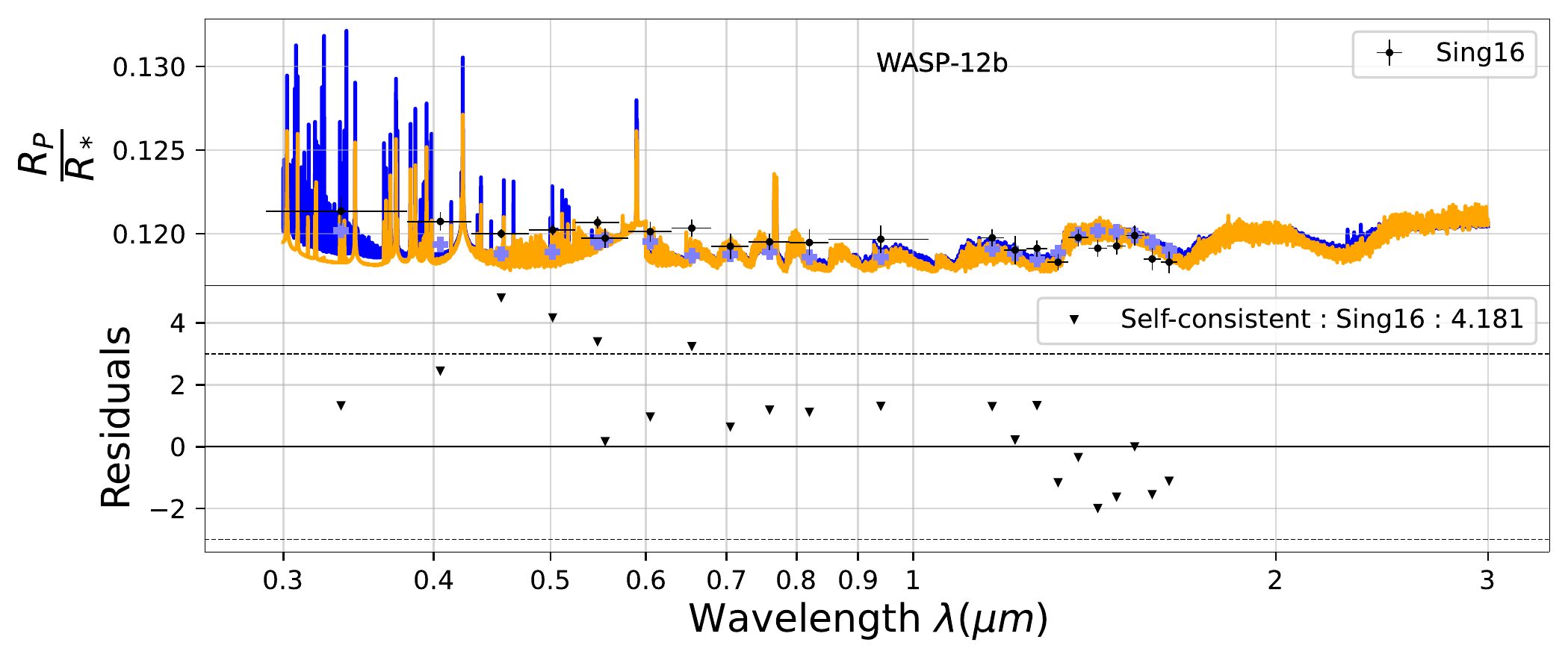}}
	\caption{WASP-12b: The orange and blue lines are the haze-free theoretical models with heavy metals from the first and third steps, respectively.}
	\label{Fig:W12}}\end{subfigure}
\begin{subfigure}{0.85\textwidth}{{\includegraphics[width=\textwidth]{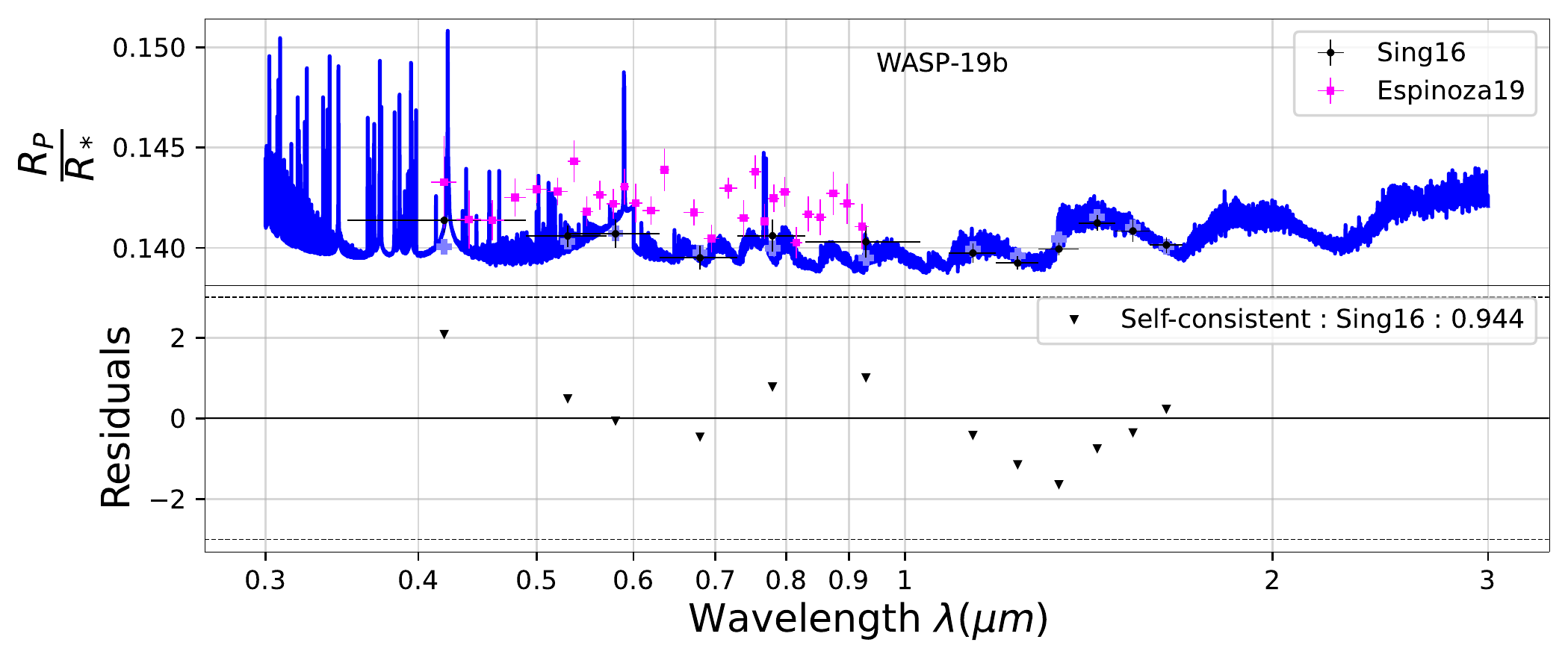}}
	\caption{WASP-19b: Self-consistent model including heavy metals and no haze.}
	\label{Fig:W19}}\end{subfigure}
\caption{Spectra of planets including heavy metals.
The black and pink crosses correspond to the observations.
The pale blue crosses are the self-consistent theoretical spectra degraded to the observation resolution.}
\label{Fig:WASP12b}
\end{figure*}

\begin{figure}
\includegraphics[width=0.48\textwidth]{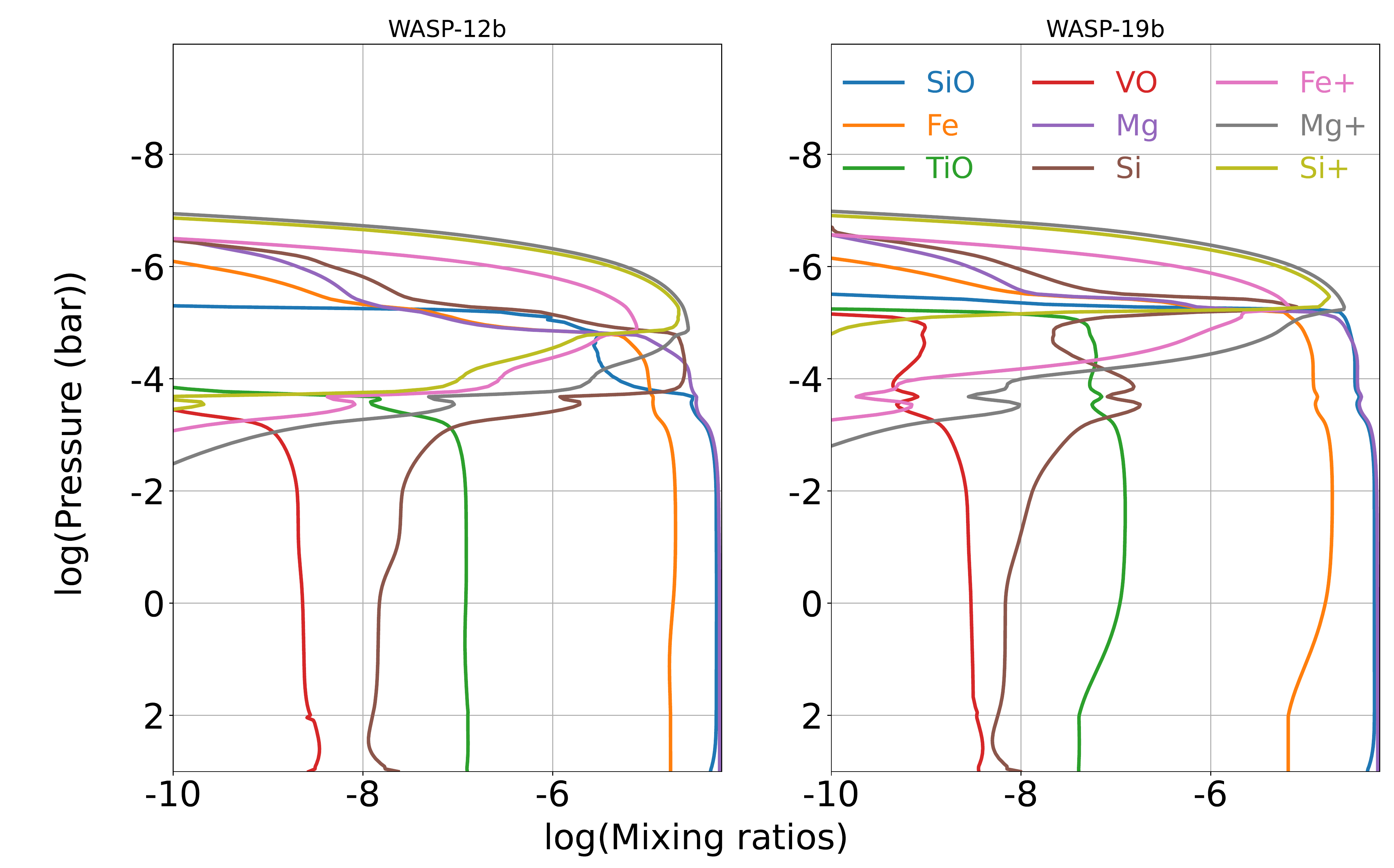}
\caption{Main heavy metal composition profiles for WASP-12b and WASP-19b.}
\label{Fig:TiOVO}
\end{figure}

\bf Step 3: \rm
The self-consistent calculation for a clear atmosphere results in a higher $\chi^2$ of 6.212 (blue line in \cref{Fig:W39}). This change relates to the temperature changes in the upper atmosphere due to the dis-equilibrium chemistry related to sulphur species, but does not cause any major modifications to the UV slope. Instead the higher $\chi^2$ is mostly due to the small changes imposed to the water band simulation and the small error bars reported for this region. The small haze precursors photolysis mass flux of 4.4x10$^{-14}g.cm^{-2}.s^{-1}$ is in agreement with the clear atmosphere observed. Considering the upper limit of 10$^{-16}g.cm^{-2}.s^{-1}$ we derived for the mass flux, we obtain an upper limit for the haze formation yield of 0.2\%.

Considering the large probed pressures (10$^{-2}$ bar), clouds lying deeper than haze in the atmosphere may bring a more satisfying answer to the muting of the water band.
According to both our simulated and \cite{Sing16} temperature profiles in \cref{Fig:FinalPTprofiles}, silicate clouds are anticipated to form in the deep atmosphere above 100 bar.
Although they may extend upward in the atmosphere, they are not anticipated in the probed region. On the other hand, sulphide clouds are likely to be present above $\sim$0.1 bar and to contribute to the transit depth in the water band that probes pressures from 0.3 bar to 0.2 mbar (\cref{Tab:ProbedPressure}).
If they are restrained to the region below 0.04 bar, they would not impact the UV-slope. Sulphide clouds in the region from 0.1 to 0.04 can therefore help enhancing the fit to the observations.
Based on \cite{Sing16} temperature profile, iron clouds may form above 100 bar, though our slightly colder deep atmosphere related to a smaller intrinsic temperature, is in agreement with an even deeper cloud production (below the 1000 bar level). Thus such clouds are not anticipated to play an important role in the transit spectrum.

\subsection{Heavy metals}

\subsubsection{WASP-12b}

Our self consistent simulation including heavy metal absorption produces a strong temperature inversion near 0.1 mbar, resulting in temperatures up to 3,000 K in this region.
These large temperatures are incompatible with the presence of soot particles that are expected to sublimate at temperatures above $\sim$1800 K. At pressures higher than 1mbar our simulated profile is quasi-isothermal ($\sim$2200K), contrary to the \cite{Sing16} profile that suggests an increasing temperature with higher pressure. This difference is attributed to the different values of intrinsic temperatures used between the two evaluations.

Despite these temperature differences our simulated transit spectra for these two cases are quite similar (orange for the \cite{Sing16} profile and blue for our simulated temperature profile in \cref{Fig:W12}). This agreement is related to the reference pressure that fixes the planet radius in the water band and to the rather small temperature difference between \cite{Sing16} and our self-consistent temperature profiles in the pressure range probed (from 30 to 1 mbar; \cref{Tab:ProbedPressure}). Moreover, due to the high temperatures, the differences between equilibrium and dis-equilibrium composition are small in the probed pressures. Therefore, the transit spectra are quite close and only the cores of the atomic lines, probing much higher in the atmosphere, are affected by higher temperatures in our simulation.

The simulated transit spectrum is consistent with the water band observations but falls short of the measured absorption depth at wavelengths shorter than 0.5$\mu$m (\cref{Fig:W12}).
Though the \cite{Sing16} temperature profile suggests that clouds cannot form in the atmosphere of WASP-12b, our colder deep atmosphere is consistent with the presence of iron clouds below the 100 bar level. However, such clouds are not anticipated to have an impact on the spectra considering the range of pressures probed by the observations (\cref{Tab:ProbedPressure}).

An answer to the bad fit of the UV-visible slope may arise from 3D effects \citep{Caldas19,Pluriel21}. Since WASP-12b is very hot and close to its host star, this planet may exhibit a strong day-night temperature shift that may have an impact on the terminator conditions and then on the spectra. Super-solar metallicity may also be plausible and have an impact on WASP-12b spectrum. Finally, tidal effects due to the close proximity of this planet to its host star may further contribute to the UV transit \citep{Koskinen22}.

The profiles in \cref{Fig:TiOVO} suggest Mg and SiO as the dominating heavy metal species up to the 0.1 mbar level.
At pressures lower than 1 mbar, SiO is lost to Si through its reaction with atomic hydrogen: $\rm SiO + H \rightarrow Si + OH $.
From 0.1 to 0.01 mbar, Si leads over SiO before being ionized in the above layer explaining the observed drop.
SiO is expected to play an important role in ultra-hot-Jupiter transit spectra \citep{Lothringer22}. Indeed, the presence of SiO in the atmosphere produces a strong absorption band below 0.35 $\mu$m, consistent with the observations. Though this feature appears in our theoretical spectrum (\cref{Fig:W12}), providing a rather good fit of the first STIS data point, SiO is not impacting the transit between 0.35 and 0.6 $\mu$m where our model is off the 3$\sigma$ limit of the observations.
Above the 0.01 mbar level, Mg and Fe are also ionized allowing Mg$^+$ to be the major heavy element species.
TiO and VO are dissociated above 0.1 mbar while above 0.01 mbar Ti and V are ionized.

\subsubsection{WASP-19b}

Our simulated temperature profile for WASP-19b in \cref{Fig:FinalPTprofiles} presents large temperatures above 2000 K everywhere in the atmosphere, therefore incompatible with the presence of photochemical hazes. Indeed, our temperature profile for WASP-19b follows the same behavior as for WASP-12b with an isothermal profile up to 1 mbar followed by a temperature inversion related to heavy metal absorption producing temperatures up to 2,500 K. Moreover, at such temperatures, clouds are not anticipated to affect the probed pressure range. The inclusion of heavy metals with our self-consistent model (blue line in \cref{Fig:W19}) provides a very good fit to the observations with a $\chi^2$ of 0.944. 
The \cite{Sing16} data are integrated over large wavelength bins preventing a clear detection of sodium and potassium lines.
However, additional ground-based observations by \cite{Espinoza19} with a higher resolution do not show clear evidence for the presence of Na and K, in agreement with our disequilibrium calculation that shows a depletion of alkalis due to ionisation above 0.1 mbar. \cite{Sedaghati21} also conducted ground based observations and were not able to detect these spectral features either, thus, supporting the anticipated absence of these elements from the probed region of the atmosphere. 
The heavy element profile are similar to WASP-12b with an ionization limit slightly higher up in the atmosphere.
TiO and VO remain the main Ti and V bearing species up to the 10 $\mu$bar altitude.

\section{DISCUSSION}
\label{Sec:Discussion}

Our results show an important diversity in the presence of haze with abundances ranging from a quasi absence for WASP-17b and WASP-39b to large values for HD-189733b.
We further note the case of WASP-12b and WASP-19b that present temperatures incompatible with the presence of haze.
For WASP-17b and WASP-39b, although they are assumed to be haze-free, an undetectable amount of haze can still be present.
We therefore provide an upper limit for them based on the smaller value we tested without significant changes compared to the clear atmosphere.

\label{Sec:Correlations}

\begin{figure*}
	\includegraphics[width=\textwidth]{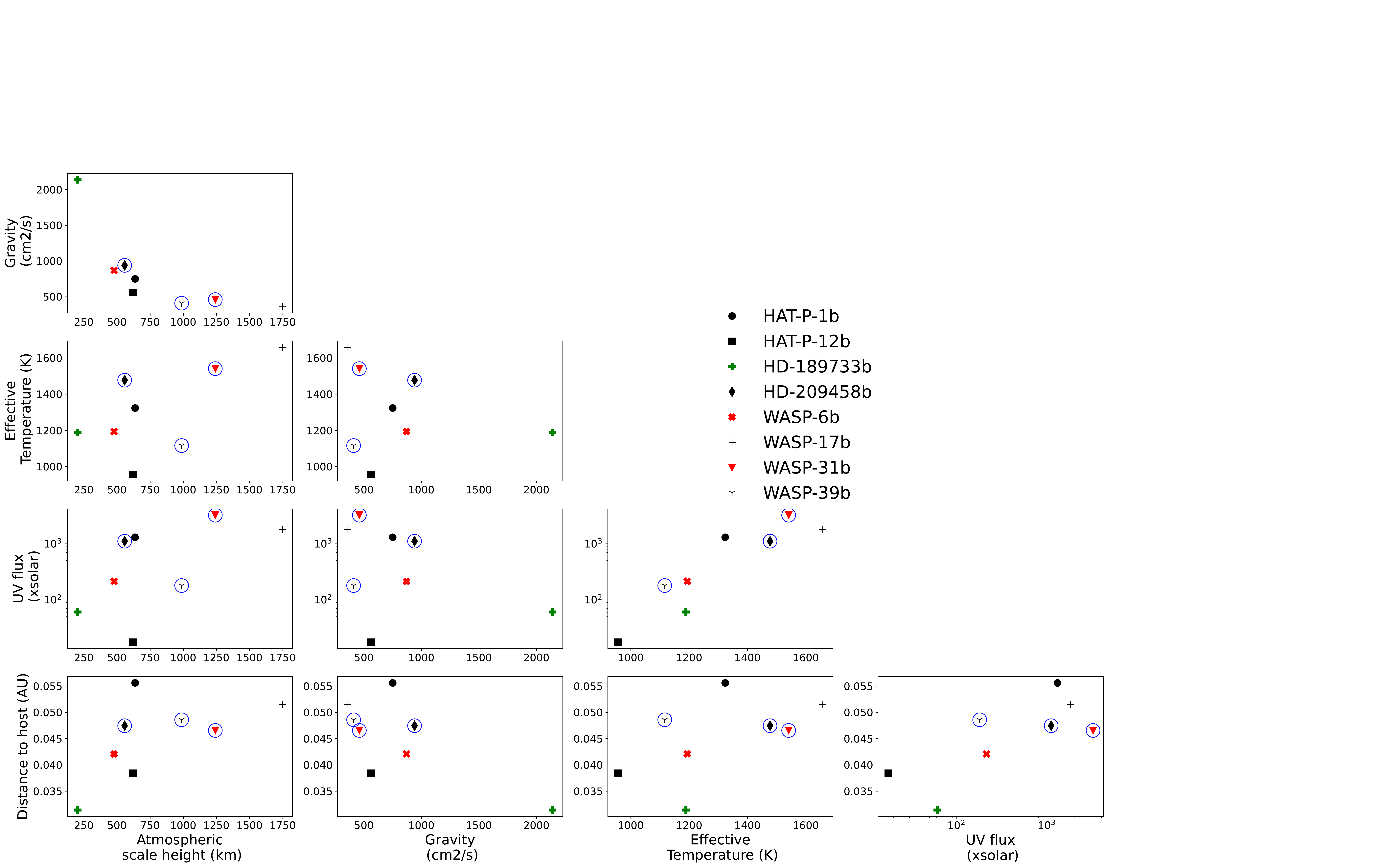}
	\caption{Correlations between the tested planetary parameters.
	The red markers are for planets best-fitted by a sub-solar metallicity, and the green for planets with a different eddy profile than the nominal.
	The blue circled planets are expected to present cloud opacities in their transit spectra.
	}
	\label{Fig:Relationships1}
\end{figure*}

\begin{figure*}
	{\includegraphics[width=\textwidth]{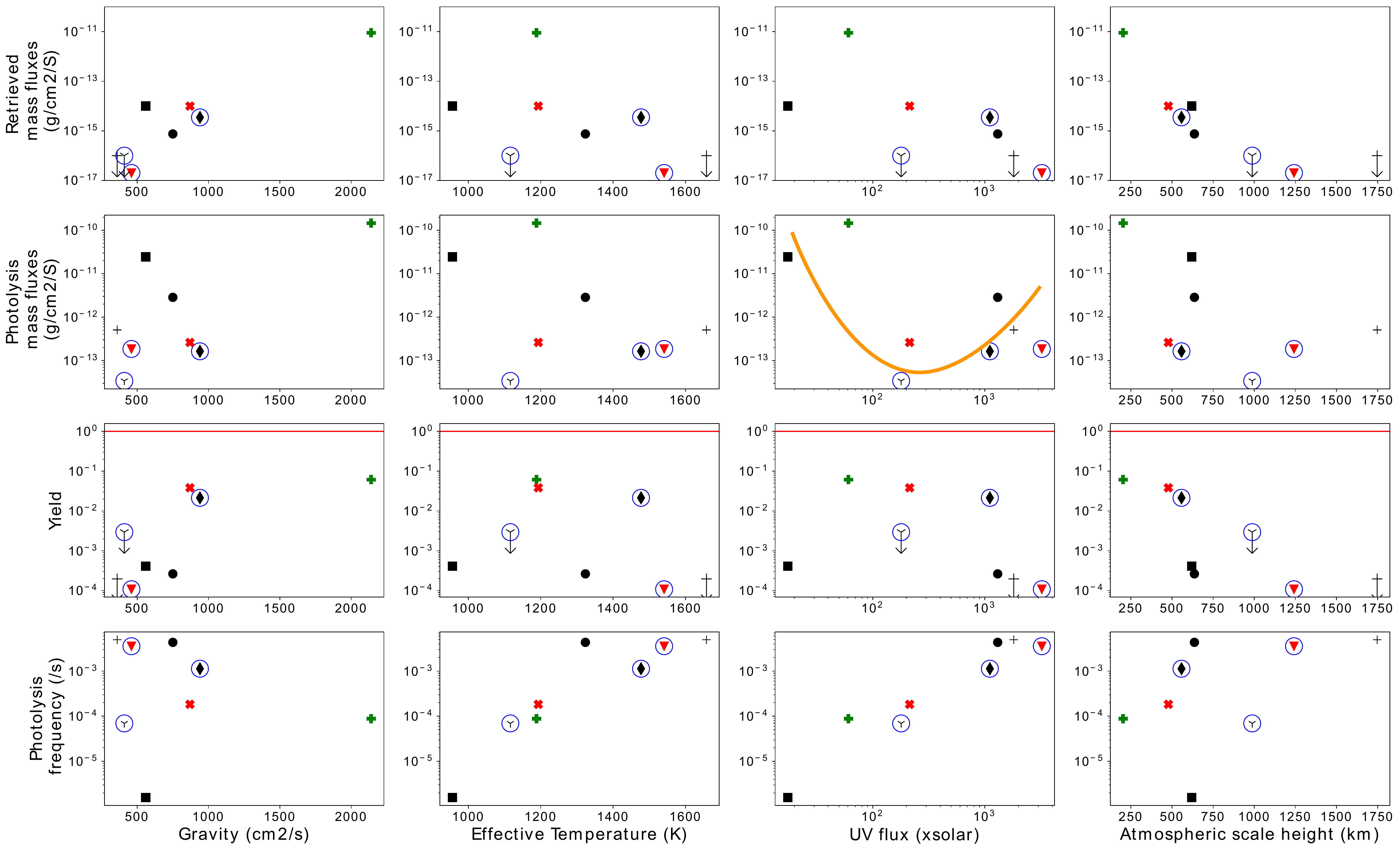}}
	\caption{Correlation plots of haze amount proxies and planet parameters. 
	The red markers are for planets best-fitted by a sub-solar metallicity, and the green for planets with a different eddy profile than the nominal.
	The blue circled planets are expected to present cloud opacities in their spectra.
	Downward arrows mean that the value provided is an upper limit to the haze mass flux.
	The orange line is meant to indicate the observed trend.}
	\label{Fig:Relationships2}
\end{figure*}

\begin{figure*}
	\includegraphics[width=\textwidth]{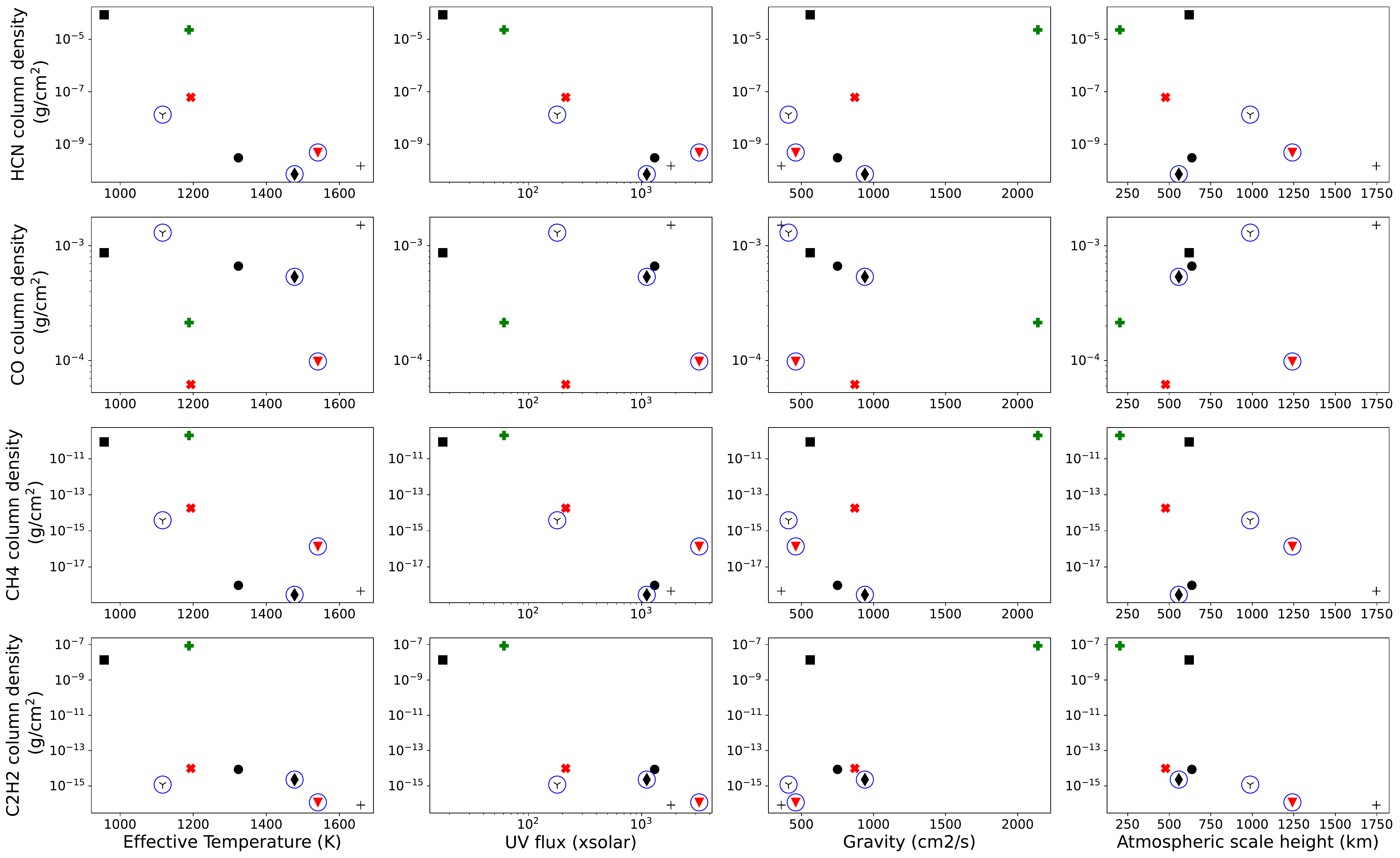}
	\caption{Haze precursors column densities against effective temperature, UV flux, gravity and scale height.
	The red markers are for planets best-fitted by a sub-solar metallicity, and the green for planets with a different eddy profile than the nominal.
	The blue circled planets are expected to present cloud opacities in their spectra.}
	\label{Fig:Relationships6}
\end{figure*}

\begin{figure}
	\includegraphics[width=0.5\textwidth]{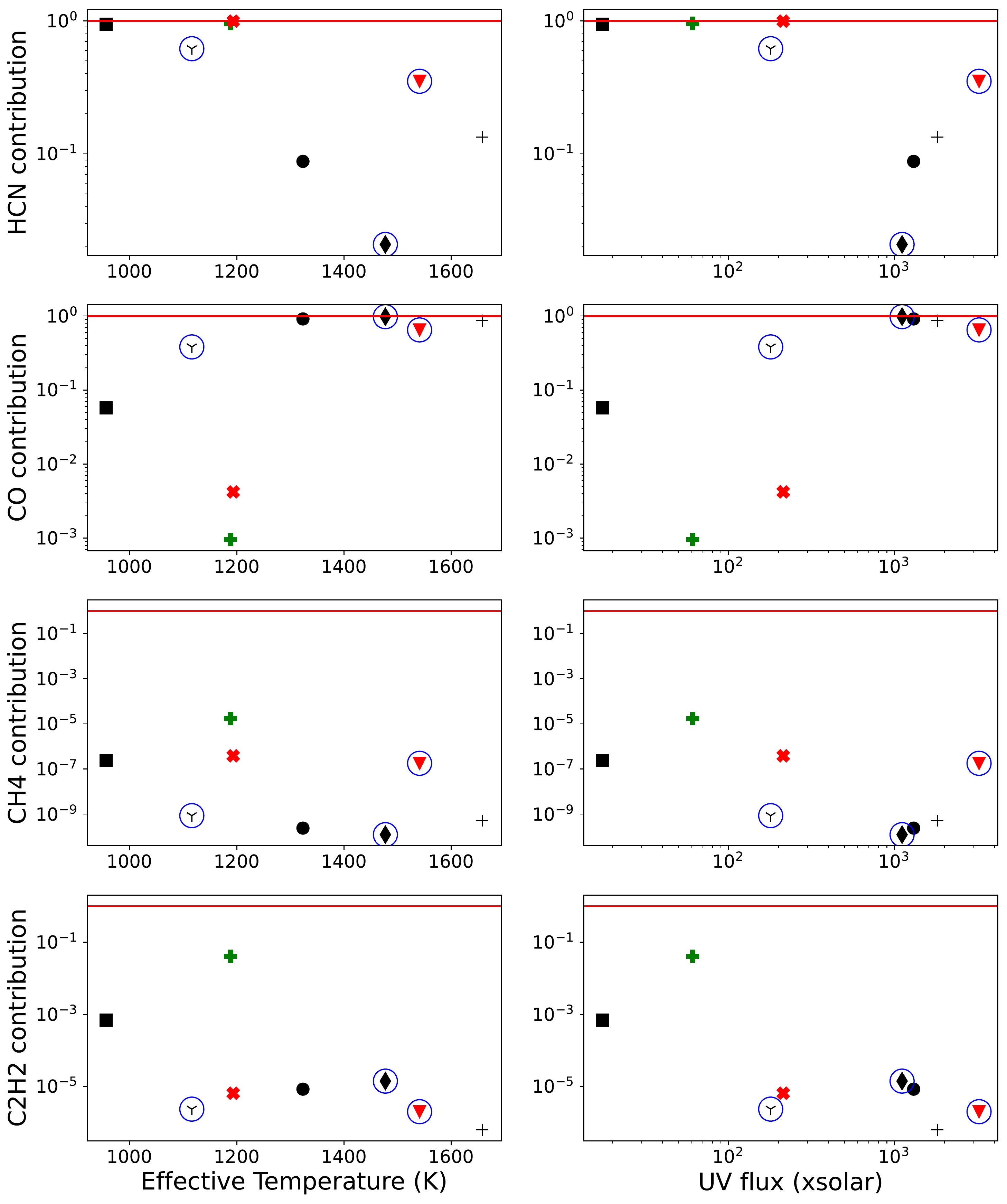}
	\caption{Haze precursors contribution against effective temperature and UV flux.
	The red markers are for planets best-fitted by a sub-solar metallicity, and the green for planets with a different eddy profile than the nominal.
	The blue circled planets are expected to present cloud opacities in their spectra.}
	\label{Fig:Relationships3}
\end{figure}

We try here to correlate the haziness conditions of the planets to their physical parameters.
First, we must note the trends observed among the planetary parameters (\cref{Fig:Relationships1}).
These planet parameters are the equilibrium temperature, gravity and atmospheric scale height, as well as the UV flux received by the planet. 
The UV flux is integrated from 0.5 to 300 nm and is provided in unit of the solar UV flux received at 1AU.
The scale height is calculated from the planet temperature and gravity and we logically see the corresponding trends in \cref{Fig:Relationships1}.
We also note a correlation between the UV flux and the planet effective temperature as they are both related to the planet-to-star distance.
As a result, a trend of increasing scale height with the UV flux also appears. 
Therefore, we need to keep in mind that a trend appearing with one of those parameters can propagate to another.

The retrieved haze mass flux, the haze precursors mass flux and the corresponding yield, are used as proxies for the planet haziness conditions and are plotted against the planet parameters (\cref{Fig:Relationships2}). The retrieved mass flux shows a positive trend with gravity and a negative trend with the atmospheric scale height.
We note that these trends are mainly due to HD-189733b and would not be that significant without this planet.
The remaining planets have mass fluxes ranging from 2x10$^{-17}$ to 10$^{-14}$ $g.cm^{-2}.s^{-1}$, which are in the range of values observed in the solar system.
The correlation of retrieved mass flux with temperature or UV flux does not present a clear trend. Nevertheless, considering only the planets without potentially significant cloud contribution to the transit spectrum, we observe a threshold near 1400 K beyond which no planet presents photochemical hazes, as for WASP-12b and WASP-19b. If we further consider the upper mass flux limit for WASP-17b that does not require haze opacities to provide an acceptable fit, the transit between hazy and clear atmospheres is somewhere between 1400 and 1740 K. This is roughly consistent with the particle sublimation expected above $\sim$1800 K based on soot saturation vapor pressure \citep{Lavvas17}. However, we would need to explore more planets in the temperature transition region to set better constraints on the limiting effective temperature.

To understand the precursor photolysis trends among the different exoplanets we need first to identify the general behavior of their abundance profiles.
The transition from H$_2$ to H dominated atmosphere related to photochemistry has important consequences on the chemistry of all species \citep{Moses11} and notably the haze precursors. 
Planets receiving a large UV flux present large atomic hydrogen abundances and then small HCN, CH$_4$ and C$_2$H$_2$ upper atmosphere column densities.
We therefore have decreasing trends of these species column densities (integrated above 10 $\mu$bar) with the UV flux (\cref{Fig:Relationships6}).
As the planet effective temperature correlates with the UV flux, similar trends are observable against the temperature.
On the other hand, CO is not strongly impacted by the H$_2$ to H transition and its column density remains almost constant with the UV flux.
Thus, CO can be an important haze precursor, though its photolysis requires high-energy radiation ($\lambda$ < 1100 \AA{}).
It therefore becomes really important for planets receiving a large flux of this high-energy photons.
Then, planets orbiting close to their host stars are expected to present strong CO photolysis mass fluxes.
CO is thus expected to be a major haze precursor for planets with high effective temperature.
We finally note that planets assuming a sub-solar metallicity present lower CO column densities (\cref{Fig:Relationships6}), while metallicity seems to have a relatively smaller impact for HCN, CH$_4$ and C$_2$H$_2$ because their upper atmosphere profiles are strongly affected by photochemistry and atmospheric mixing.

On \cref{Fig:Relationships3} we present the contribution of the assumed haze precursors, which is the ratio of their individual photolysis mass flux to the total photolysis mass flux.
As expected, HCN, CH$_4$ and C$_2$H$_2$ contributions tend to decrease with both effective temperature and UV flux.
On the other hand, the CO contribution tends to increase with the temperature and UV flux.
As the other species contributions decrease, the CO contribution increases and CO becomes the main haze precursor for temperatures above 1300 K and UV flux above 300 times the solar value.
We note the specific case of WASP-31b that shows an enhanced contribution of HCN compared to other very hot planets related to the sub-solar metallicity used.
A lower metallicity decreases CO abundance and then its contribution.
In addition, the lower water abundance leads to a weaker OH concentration and therefore to a slower transition from H$_2$ to H dominated atmosphere, resulting in an enhanced HCN upper atmosphere abundance.
We further note the small contributions of CH$_4$ and C$_2$H$_2$ that remain negligible in most cases owing to smaller column densities compared to HCN.
We therefore find HCN leading the haze precursors photolysis mass flux up to effective temperatures (UV flux) of 1300 K (300 times the solar value) and CO taking over above this temperature (UV flux).
This might have important and interesting ramifications about the haze particle composition and their physical and chemical properties.

With these properties in mind, the haze precursors photolysis mass flux seems to show a U-shape behavior against the UV flux (\cref{Fig:Relationships2}).
For UV fluxes lower than 300 times the solar value, HCN is leading the photolysis of haze precursors, therefore resulting in a decreasing trend of the latter that follows the behavior of HCN column density with the UV flux.
We mainly note HD-189733b off this trend related to the strong eddy profile used for this planet that produces enhanced HCN upper atmosphere abundances.
We further note that a sub-solar metallicity has the same effect of increasing HCN abundances as previously discussed, which results in slightly larger haze precursors photolysis mass flux for WASP-6b than what could be expected for a corresponding solar metallicity case.
For UV fluxes larger than 300 times the solar value, CO is leading the photolysis mass flux.
As CO column density is relatively constant with the UV flux, its photolysis increases with the latter and so do the haze precursors photolysis mass flux.
This produces the right limb of the U-shape behavior observed.
We note WASP-31b off this trend related to the sub-solar metallicity assumed that decreases the CO upper atmosphere column density.
To demonstrate the effect of different precursor abundances we also plot the photolysis frequency. This is the ratio of the haze precursors mass flux to their column density, therefore it is independent from species abundance effects and is only related to the UV flux and the efficiency of photolysis.
The clear positive trend of the photolysis frequency with the UV flux supports the hypothesis of haze precursors abundance effect to explain the observed behavior of the photolysis mass flux.

For planets demonstrating temperatures compatible with the presence of haze, we find yields ranging from 0.01 (or even lower) to 6.4\%.
The plot of the yield against temperature seems to present a positive trend up to 1300 K, that is in the region where HCN dominates the haze precursors.
Above 1300 K, where CO takes over, the behavior is more erratic, though we can note a drop of the yield.
In this region, HD-209458b presents a larger yield, however clouds are expected to play an important role for this planet, which would result in a smaller retrieved haze mass flux and then a smaller yield.
These results suggest that temperature plays an important role in the processes taking place between the precursors photolysis and the formation of the particles.
We further note that the yield is strongly related to the strength of the eddy diffusion.
Indeed, a large mixing allows for large haze precursors upper atmosphere abundances and then large photolysis mass fluxes.
This enhanced photolysis mass flux leads to smaller yields.
It is therefore necessary to further investigate in the future how the eddy magnitude affects the yield for each planet case.

\begin{figure}
	\includegraphics[width=0.5\textwidth]{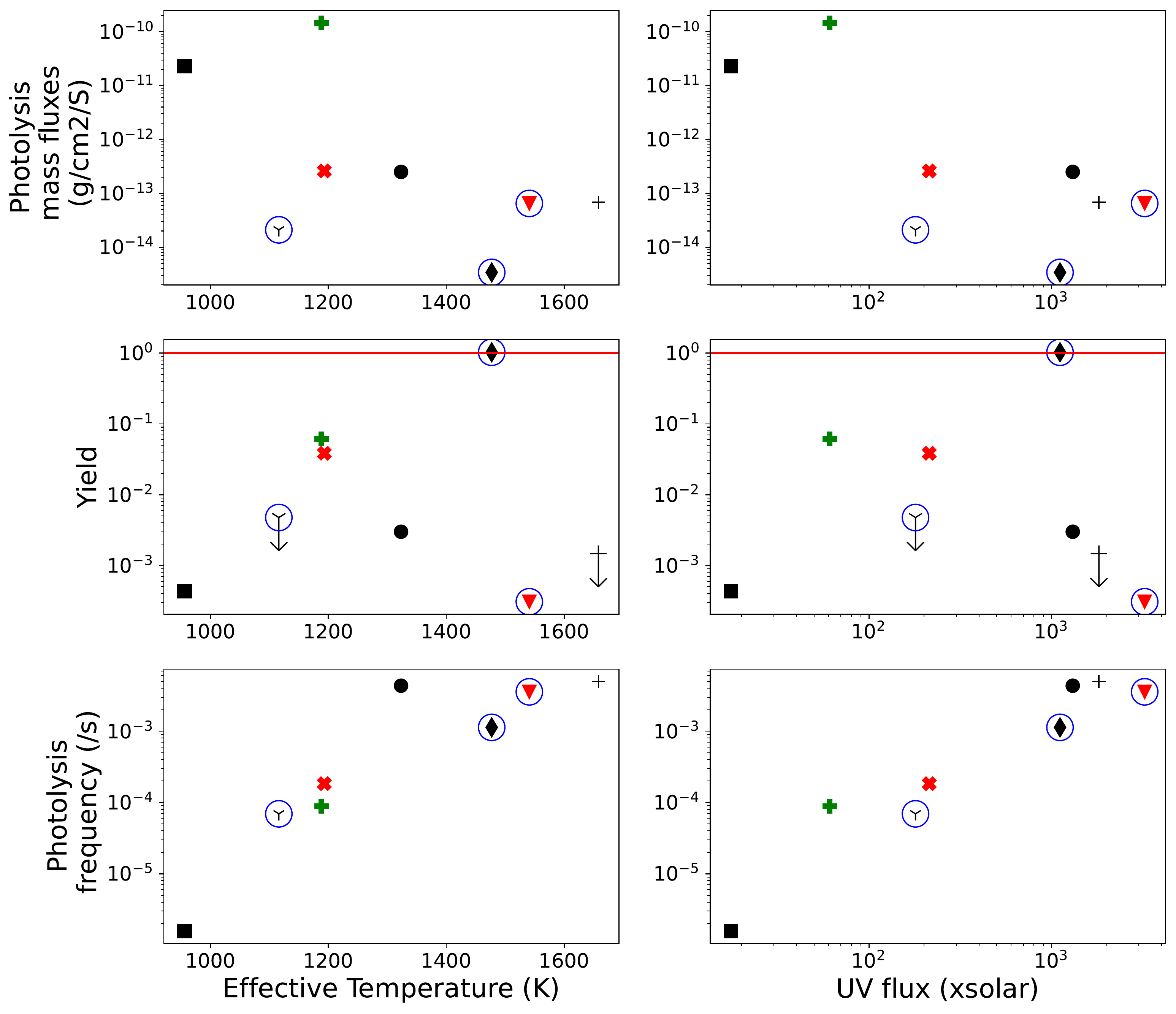}
	\caption{Correlation plots of the photolysis mass flux, yield and photolysis frequency against the effective temperature and UV flux without the CO contribution.
	The red markers are for planets best-fitted by a sub-solar metallicity, and the green for planets with a different eddy profile than the nominal.
	The blue circled planets are expected to present cloud opacities in their spectra.
	Downward arrows mean that the value provided is an upper limit to the haze mass flux.}
	\label{Fig:Relationships4}
\end{figure}

\begin{table}
\caption{Photochemical mass fluxes of sulfur-based haze precursors ($g.cm^{-2}.s^{-1}$) integrated above 10$^{-5}$ bar altitude with the self-consistent calculation. The last column provides the photolysis mass flux calculated based on carbon-bearing species alone (CH$_4$, C$_2$H$_2$, HCN and CO).}
\begin{adjustbox}{max width=0.5\textwidth}
\begin{tabular}{c|ccc|cc}
\hline
Planets		&	$H_2S$	&	$S_2$	&	$S_3$	&	Total		&	C-based		\\
\hline
HAT-P-1b		&	1.3(-12)	&	2.4(-10)	&	4.3(-13)	&	2.4(-10)	&	2.9(-12)		\\
HAT-P-12b	&	1.5(-15)	&	1.8(-11)	&	6.5(-11) 	&	8.3(-11)	&	2.4(-11)		\\
HD-189733b	&	9.3(-13)	&	9.6(-11)	&	1.9(-13)	&	9.7(-11)	&	1.4(-10)		\\
HD-209458b	&	2.7(-12)	&	1.8(-10)	&	7.3(-14)	&	1.8(-10)	&	1.6(-13)		\\
WASP-6b		&	2.1(-13)	&	5.3(-12)	&	6.1(-16)	&	5.5(-12)	&	2.6(-13)		\\
WASP-12b	&	1.7(-13)	&	5.7(-13)	&	4.0(-22)	&	7.4(-13)	&	9.5(-11)		\\
WASP-17b 	&	2.1(-12)	&	7.2(-11)	&	1.0(-12)	&	7.5(-11)	&	5.1(-13		\\
WASP-19b	&	9.4(-13)	&	3.7(-12)	&	5.9(-20)	&	4.6(-12)	&	5.1(-11)		\\
WASP-31b	&	2.5(-12)	&	1.5(-11)	&	8.4(-16)	&	1.8(-11)	&	1.9(-13)		\\
WASP-39b	&	1.8(-14)	&	1.1(-11)	&	6.9(-12)	&	1.8(-11)	&	4.4(-14)		\\
\hline
\end{tabular}
\end{adjustbox}
\label{Tab:PhotoFluxS}
\end{table}

So far we assumed all the precursors to share the same yield.
We may however consider the possibility of having a different efficiency among the haze precursors.
Specifically, laboratory measurements demonstrate that CO can have a different haze production yield \citep{He20,Horst14}. 
We therefore try to assess these correlations without accounting for CO in the photolysis mass fluxes, which obviously modifies the yields.
An important change brought by this consideration is a decreasing trend in the haze precursors photolysis mass flux against the planet effective temperature and the received UV flux (\cref{Fig:Relationships4}). These trends are consistent with the anticipated drop in the haze precursor abundances with increasing temperature.
The right limb of the U-shape previously observed was indeed related to the contribution of CO and it logically disappears as we remove CO from the haze precursors.
Moreover, the photolysis frequency correlations remain unchanged, as they are independent from abundance variation effects. 
The observed  trend propagates to the UV flux leading to a surprising behavior of decreasing haze precursors photolysis mass flux with increasing UV flux.
For the yield, we note the presence of HD-209458b slightly off the 100\% limit, though the smaller yield implied by additional opacities from clouds would solve this issue.
With the absence of CO, the yields are now larger above 1300 K (related to the lower haze precursors mass flux) but remain small compared to the yields of most planets with effective temperatures below 1300 K.

Several works consider the possibility of sulfur-based aerosols rather than or in addition to carbon \citep{Zahnle16,Gao17}.
If such particles could survive the hot temperatures found in hot-Jupiters, they indeed would present an additional source of particle.
However, sulfur-based aerosols are expected to sublimate for planet effective temperature above than 750 K \citep{Gao17,Adams19}, and might not be present in condense form in the atmosphere of the planets study here.
On the other hand, sulfur can be integrated in the carbon structure, increasing the haze mass flux without significant changes of the particle properties.
\cite{Chin81} was able to include 30\% of sulfur in soot particles via adsorption and did not notice sulfur loss up to $\sim$1000 K.
\cite{Vuitton21} ran lab experiments simulating photochemistry in high metallicity (10$^{4}$xsolar) exoplanet atmospheres with a 800 K temperature at 1 mbar, and assessed the composition of the produced hazes.
They found particles composed of carbon (40\%) with the inclusion of 21\% of sulfur.
If such a sulfur incorporation can be achieved at the higher temperatures and lower metallicities representative of hot-Jupiter atmospheres, it would certainly affect the chemical composition and properties of the resulting haze particles.
\cref{Tab:PhotoFluxS} presents the photolysis mass flux of the major sulfur-bearing species.
The comparison to the carbon-based value indicates that sulfur can have a non-negligible contribution to the total photolysis mass flux.
Additional lab experiments about the optical properties of such composites are required to go further in the exploration of the impact of sulfur in photochemical aerosols.

\section{CONCLUSIONS}

The planets we study present an important diversity in their haziness conditions, from the clear atmosphere of WASP-17b, to the large haze abundance of HD-189733b.
In increasing haze contribution we find:
\begin{itemize}
	\item{For HD-189733b the fit to the UV-visible observations was improved with our self-consistent modeling by assuming a higher haze production altitude and a large 9x10$^{-12} g.cm^{-2}.s^{-1}$ haze mass flux. However, we still cannot reproduce the observed steep UV slope. A larger mass flux would require a stronger eddy profile that is not physically anticipated. We note that the derived mass flux is significantly larger from those derived for the other planets, questioning the actual contribution of hazes in this atmosphere. Stellar activity may help to fit the steep UV slope observed and reduce the need for such a high haze mass flux.}
	\item{HAT-P-12b and WASP-6b present moderate amounts of haze with retrieved mass fluxes of 10$^{-14} g.cm^{-2}.s^{-1}$.
	The latter is better fitted with a 0.1$\times$solar metallicity.}
	\item{HD-209458b is fitted with a 5x10$^{-15} g.cm^{-2}.s^{-1}$ mass flux, though corresponding to an effective mass flux of 3.5x10$^{-15} g.cm^{-2}.s^{-1}$, while the value estimated from step 1 leads to a steeper slope than the observed one.
	This demonstrates that the inclusion of haze radiative feedback can decrease the need for haze opacities.}
	\item{HAT-P-1b demonstrates a very low haze abundance of 10$^{-15} g.cm^{-2}.s^{-1}$ and can be considered as a clear atmosphere.
	Indeed, this model is barely distinguishable from the haze-free with residuals within the 3$\sigma$ of the observations.
	Furthermore, haze heating by UV radiation leads to their partial destruction producing an effective mass flux of 7.5x10$^{-16} g.cm^{-2}.s^{-1}$.
	While our step 1 results are in agreement with a sub-solar metallicity, the self-consistent model suggests a solar metallicity.
	This shows that using a self-consistent model can decrease the need for using lower metallicities.}
	\item{WASP-31b remains uncertain owing to the degeneracies observed, although hazes, if present, require a deeper production altitude than the commonly assumed 1 $\mu bar$ level and very small mass flux to avoid particle sublimation.
	While the observed UV slope is in agreement with a small haze mass flux of 10$^{-16} g.cm^{-2}.s^{-1}$ (corresponding to an effective mass flux of 2x10$^{-17} g.cm^{-2}.s^{-1}$) and a 0.1$\times$solar metallicity, the lack of K feature observed by \cite{Gibson19} requires a sub-solar abundance of potassium and/or a strong haze opacity.}
	\item{For WASP-17b, we conclude to a rather clear atmosphere although we note that our model overestimates the opacity in the visible range based on \cite{Sing16} observations.
	The comparison of these observations in this range with the neighboring WFC3 data as well as with \cite{Sedaghati16} and \cite{Saba21} observations suggest the need for an offset of the \cite{Sing16} STIS observations.}
	\item{WASP-39b is best-fitted by a haze-free model although the water band and the Na and K lines seem to require additional opacities.
	Photochemical hazes however do not provide a satisfying answer and we suggest cloud opacities to be the most likely solution considering the relatively cold temperatures and the pressure probed by the observations.}
	\item{The large temperatures found for WASP-12b and WASP-19b are incompatible with the presence of haze.
	Our fit to WASP-12b transit spectrum is not consistent with the observed UV slope and we suggest 3D effects, super-solar metallicity or tidal effects to be a better explanation for the observations.
	The slightly cooler WASP-19b is accurately fitted with heavy metals and no haze.}
\end{itemize}

We demonstrate that the use of a self-consistent model enhances the UV slope via haze particle heating resulting in a larger planet radius.
This improvement leads to better fits for WASP-6b, WASP-19b and HD-189733b and can reduce the required haze mass flux for HD-209458b.
The photolysis mass fluxes from haze precursors are consistent with the retrieved haze mass fluxes required for fitting the observed transit spectra for all planets, with yields ranging from 0 to 6.4\%. Our results suggest that HCN is the main haze precursors at equilibrium temperatures below 1300 K, while CO dominates above, although the requirement for a haze opacity seems to drop at temperatures above 1400K. This change could point to a different efficiency of haze formation from CO compared to HCN as well as to limitations in the formation of photochemical hazes at very high temperatures. 
We evaluated the possible trends of the retrieved haze mass fluxes of the studied exoplanets with different planetary parameters such as effective temperature, insolation and gravity. Although deriving clear trends is inhibited due to different assumptions on atmospheric mixing and metallicity, our results indicate a positive trend with gravity that merits further investigation. Finally, our calculations suggest that clouds could impact the spectra of HD-209458b, WASP-31b and WASP-39b. A larger intrinsic temperature from that currently assumed may allow for their presence in the regions probed by some other planets within our study.
The effects of a larger intrinsic temperature will be investigated in a future publication.

\section*{ACKNOWLEDGEMENTS}

Based on data from the X-exoplanets Archive at the CAB.

\section*{DATA AVAILABILITY}

The data underlying this article will be shared on reasonable request to the corresponding author.



\bibliographystyle{mnras}
\bibliography{biblio} 




\appendix
%
%


\bsp	
\label{lastpage}
\end{document}